\documentclass[12pt]{article}
\usepackage[
backend=biber,
style=numeric-comp, %style of citations %numeric %alphabetic,
sorting=none % as they appear %ynt == year name title
]{biblatex}
\usepackage[utf8]{inputenc}
\usepackage{epsfig,amsfonts,amssymb,framed,amsmath,xcolor}
\usepackage{slashed}
\usepackage{ytableau}
\usepackage{hyperref}
\usepackage{mathtools}
\usepackage{cancel}
\input epsf.sty
\usepackage{hhline}
\usepackage{bm}
\usepackage{array,upgreek}

\usepackage{ytableau}
\usepackage{braket}

\usepackage[margin=2cm]{geometry}
\newcommand{\be}{\begin{eqnarray}}
	\newcommand{\ee}{\end{eqnarray}}   

\newcommand{\bi}{\begin{itemize}}
	\newcommand{\ei}{\end{itemize}}
\newcommand{\tk}{{\tilde k}}
\newcommand{\bse}{\begin{subequations}}
	\newcommand{\ese}{\end{subequations}}

	\newcommand{\del}{\partial}
	\newcommand{\eps}{\epsilon}
	\newcommand{\eq}[1]{
		\begin{align}
			#1
		\end{align}
	}
	\newcommand{\holdI}{{ \underline{\hat{\cal A}} } }
	
	\newcolumntype{P}[1]{>{\centering\arraybackslash}p{#1}}
	
	\newcommand{\N}{ \mathbb{N} }
	\newcommand{\Z}{ \mathbb{Z} }
	
	\newcommand{\R}{ \mathbb{R} }

	\newcommand{\oh}{ {\frac{1}{2}} }
	
	\newcommand{\uA}{{\underline A}}
	\newcommand{\uL}{{\underline L}}
	\newcommand{\uk}{{\underline p}}
	\newcommand{\up}{{\underline p}}
	
	\newcommand{\ualpha}{{\underline \alpha}}
	
	\newcommand{\cC}{{\cal C}}
	\newcommand{\cD}{{\cal D}}
	\newcommand{\cW}{{\cal W}}
	\newcommand{\cH}{{\cal H}}
	\newcommand{\cM}{{\cal M}}
	\newcommand{\cS}{{\cal S}}
	
	\newcommand{\sN}[1]{ { [ #1 ] } }
	
	\newcommand{\sdap}{ {\sqrt{2\alpha'}} } %% it was 2
	
	\newcommand{\samedap}{{2\alpha'}}
	\newcommand{\dap}{{2\alpha'}}
	\newcommand{\ap}{{\alpha'}}
	\newcommand{\ishap}{  } %% it was \sqrt{\frac{2}{\ap}}
	\newcommand{\oldap}{{\sdap}} %% it was \alpha' -> \sdap
	\newcommand{\dueoldap}{{\sdap}} %% it was 2\alpha' -> \sdap
	
	\newcommand{\upzero}{{\underline {\hat p}}}
	\newcommand{\uxzero}{{\underline {\hat q}}}
	
	\newcommand{\xr}{{x_b}}
	\newcommand{\xs}{{x_c}}
	\newcommand{\xt}{{x_a}}
	\newcommand{\snr}{{\sN b}}
	\newcommand{\sns}{{\sN c}}
	\newcommand{\snt}{{\sN a}}
	\newcommand{\rrr}{{b=}}
	\newcommand{\RRR}{{b}}

	\newcommand{\ttt}{{a=}}
	\newcommand{\TTT}{{a}}

	\newcommand{\xxrrtt}{ {x_{a b}} }

	\newcommand{\oldI}{{ \underline{\cal A} } }
	\newcommand{\oldJ}{{ \underline{\cal B} } }

\begin{document}

		\baselineskip 24pt
		
		\begin{center}
			{\Large \bf }
			THE REGGEON VERTEX FOR DDF STATES
		\end{center}
		
		\vskip .6cm
		\medskip
		
		\vspace*{4.0ex}
		
		\baselineskip=18pt

		\begin{center}
			{\large 
				\rm  Dripto Biswas$^a$, Raffaele Marotta$^b$ and Igor Pesando$^a$ }
		\end{center}

		\vspace*{4.0ex}
		\centerline{ \it \small $^a$ Dipartimento di Fisica, Universit\`{a} di Torino, 
			and I.N.F.N., Sezione di Torino}
		
		\centerline{ \it \small 
			Via P.\ Giuria 1, I-10125 Torino}
		
		\centerline{\it \small $^b$Istituto Nazionale di Fisica Nucleare (INFN), Sezione di Napoli, }
		\centerline{ \it \small Complesso Universitario di Monte S. Angelo ed. 6, via Cintia, 80126, Napoli, Italy.}

		\vspace*{1.0ex}
		\centerline{\small E-mail: dripto.biswas@to.infn.it,raffaele.marotta@na.infn.it,igor.pesando@to.infn.it}
		
		\vspace*{5.0ex}

		\centerline{\bf Abstract} \bigskip
		We provide a compact expression for the generating function of correlators involving an arbitrary number of bosonic open string DDF states. The explicit correlators for $M$ DDF states can then be obtained by differentiating this generating function with respect to the DDF polarization tensors.
		The generating function depends on single and double complex integrals centered around the punctures corresponding to the insertion points of the DDF vertices on the real axis in the upper half-plane. We have explicitly evaluated these integrals for an arbitrary number of DDF states. To check our results, we computed some massive scalars and  spin-2 amplitudes for $M=3$ and $M=4$, verifying that these amplitudes have the form expected from Lorentz invariance.

		Moreover, for the first excited levels, we also show that the DDF amplitudes with generic polarizations can be reassembled into Lorentz-covariant amplitudes, with emergent covariant polarizations that match the expressions obtained by comparing the string states in both the DDF and covariant formalisms.

		\vfill

		\vfill \eject

		\baselineskip18pt
		
		\tableofcontents

		\setcounter{section}{0}
		
		\section{Introduction}
		
		A hallmark of string theory is the presence of an infinite number of higher-mass spin excitations in its spectrum, which fill the higher-dimensional representations of the Poincaré group. String theory offers a consistent description of these massive modes. However, while the interactions of massless modes are well understood, the study of amplitudes involving higher-mass string modes remains relatively underexplored (see however \cite{Mitchell:1990cu,Manes:2001cs,Manes:2003mw,Manes:2004nd} for a discussion of string form factors,
		\cite{Iengo:2002tf,Iengo:2003ct,Chialva:2003hg,Chialva:2004ki,Chialva:2004xm,Iengo:2006gm,Iengo:2006if,Iengo:2006na} for the discussion on the stability of massive states which is relevant for the correspondence principle and
		\cite{Arduino:2020axy} for an explanation of divergences in temporal string orbifolds due to massive states).
		
		There are compelling reasons to address this gap. In the context of gravitational waves, and in line with the no-hair theorem, black holes are fully characterized by their mass, charge, and angular momentum. At large distances, their interactions can be described by the classical limit of on-shell amplitudes, where black holes are modeled as massive, charged, and spinning point particles. The study of scattering amplitudes involving higher-spin particles is an active area of research, due to its relevance in understanding binary black hole dynamic in the inspiral phase\cite{Bern:2020buy,Cangemi:2022abk,Bern:2022jnl}.
		
		Investigating higher-spin interactions is also of significant interest due to the black hole/string correspondence principle \cite{9507094,9612146,Damour:1999aw,1212.2606,2109.08563,2110.12617}, which suggests that perturbative string states may collapse into black holes when the closed string coupling constant reaches a critical threshold,
		$g_s  N^{1/4}\sim 1$, where $N$ represents the string’s excitation level \cite{9612146}. String theory interactions are encapsulated in correlators of BRST-invariant vertex operators. While these operators are well understood for highly massive states on the leading Regge trajectory, there have been few systematic attempts to formulate vertex operators for arbitrary massive string states\cite{Manes:1988gz,Bianchi:2010es,Markou:2023ffh}.
		
		In the 70s, the Del Giudice, Di Vecchia, and Fubini operators, commonly named  DDF operators, were introduced to describe excited massive string states in bosonic string theory \cite{DelGiudice:1971yjh,Ademollo:1974kz}.
		An important feature of these operators is that they commute with the generators of the Virasoro algebra. 
		This property allows for the generation of the complete Hilbert space of non null (non BRST exact) physical states by applying an arbitrary combination of them to the ground state. 
		Generic three point amplitudes involving such states of arbitrary string level  were computed \cite{Ademollo:1974kz} and subsequently, the framework was extended to the fermionic Neveu-Schwarz model \cite{Hornfeck:1987wt}.  
		DDF operators play a crucial role in constructing string coherent states, which are proposed as the Conformal Field Theory (CFT) description of macroscopic objects, potentially identified with cosmic strings\cite{1006.2559,1107.0730}.
		Recently, three level scattering amplitudes involving DDF- operators of exited strings have been explored 
\cite{Bianchi:2019ywd,Rosenhaus:2021xhm,Firrotta:2022cku,Hashimoto:2022bll,Das:2023xge,Savic:2024ock,Firrotta:2024qel,Firrotta:2024fvi,Bhattacharya:2024szw}.
		The erratic behavior of these  amplitudes  has  deepened our understanding of chaotic phenomena in the quantum field and string S-matrix
		\cite{2003.07381,2103.15301,2312.02217}.  %
		A quantitative indicator for chaotic quantum processes has been introduced in  \cite{2207.13112,2403.00713}, which draws upon the principles of Random Matrix Theory and the $\beta$-ensemble \cite{2303.17233}.
		DDF states have also been considered in connection with the S-matrix description of tidal excitations and absorption of energetic closed strings by $D$-branes\cite{1510.03837,DiVecchia:2023frv}.  %Three- and four-point DDF amplitudes are analyzed in Ref.s\cite{Firrotta:2022cku,Firrotta:2024qel,Firrotta:2024fvi}
		
		While DDF operators generate a complete set of physical states, their connection with BRST-invariant vertex operators describing single string excitations
		in the usual transverse and traceless gauge
		remains cryptic. 
		Significant progress in understanding this connection has been made recently, in the papers of Ref.\cite{Markou:2023ffh,2402.13066,2405.09987}. 
		In this paper, within the framework of open bosonic string theory, we explore DDF states from two distinct perspectives. 
		We begin by constructing a Sciuto-Della Selva-Saito like (SDS) vertex \cite{Sciuto:1969vz,DellaSelva:1970bj} to describe the interaction of DDF coherent states, as introduced in Ref. \cite{1006.2559,1107.0730}, in the context of cosmic strings. 
		Generic DDF vertex operators are subsequently derived by taking derivatives
		\cite{1006.2559,Pesando:2012cx,Pesando:2014owa} with respect to the polarizations of the SDS vertex.
		
		Next, we have constructed the generating functional that provides the correlation functions for an arbitrary number $M$ of such vertices, the $M$ Reggeon in the old parlance \cite{DiVecchia:1986jv,DiVecchia:1986uu,DiVecchia:1986mb,DiVecchia:1987ew}.
		
		We derive this generator both in the standard formulation of the DDF operators and in their framed version, as recently proposed in Ref. \cite{2402.13066}. 
		Due to the relationship between coherent and DDF states, the correlation functions involving DDF vertices are obtained by differentiating with respect to the polarizations of the external DDF states.
		
		A key feature of this generating functional is that all the information about the interactions is encoded in single and double complex integrals centered around the Koba-Nielsen points, where the SDS vertices are inserted on the upper complex plane. These integrals have been evaluated for the generic %three- and four
		$M$-point correlation functions of DDF  and they turn out to be polynomials in the anharmonic ratios of the Koba-Nielsen variables. 
		
		This approach provides a compact and efficient method for calculating correlators of an arbitrary number of coherent states.
		
		Different tests have been performed on the correctness of the functional generator.  Three point amplitudes with generic DDF states computed by using standard approaches and the proposed functional generator turn out to be in agreement. Furthermore, in the standard formulation of DDF amplitudes, we have obtained the expected relation between DDF and covariant polarizations from scattering amplitudes involving two tachyons and generic higher-spin states at the first four excited levels.
		
		In the framed DDF formulation of the generating functional, we have considered a generic linear combination of DDF states that describe $SO(24)$ spin-2 and massive scalars of the $N=4$ and $N=6$ string level.
		By imposing that the three point amplitudes have the expected structure under the full $SO(1,25)$ Lorentz group  we have checked that the coefficients of this linear combination are in agreement with the expected results proposed in the literature \cite{2405.09987} This analysis has also been extended to four-point interactions.
		
		This paper is organized as follows. In Section \ref{General}, we define the DDF states, showing their connection to perturbative string states, and provide a summary of their key properties in both the standard and framed formulations.
		In Section \ref{SDSV}, we introduce the coherent states and present the Sciuto-Della Selva-Saito (SDS) like vertex for the coherent states in both the standard and framed formulations.
		Section \ref{section4} is dedicated to constructing the generating functional that yields the correlation functions for an arbitrary number of SDS DDF vertices.
		Section \ref{Amplitude}, within the standard formulation of DDF operators, we test the generating function by computing generic three point amplitudes with arbitrary DDF external states, as well as amplitudes involving two tachyons and a massive string state at the $N=2,3,4$ excited level.
		In Section \ref{sec:3pt_scattering}, we use the correlators of coherent framed states to compute three point amplitudes involving scalars at the excited string levels $N=4$ and $N=6$, as well as spin-2 symmetric tensors. This section also demonstrates how sensitive these amplitudes are to the values of the complex integrals defining the generating functional.
		In Section \ref{sec:4pt_scattering}, we discuss the four-point amplitudes involving the excited string levels introduced in Section \ref{sec:3pt_scattering} and three tachyons. The paper ends with the conclusions given in section \ref{section7}.
		
		The paper includes several technical computations, the details of which are provided in the appendices. 
		Appendix \ref{app:notations} defines the notations used throughout the paper. 
		Appendix \ref{app:SDS_vertex} addresses the issues and subtleties involved in constructing the SDS vertex in the framed formulation of DDF operators and its connection with the standard formulation.
		Appendix \ref{app:correlator} provides details on obtaining the correlators for an arbitrary number of DDF states, while Appendix \ref{app:framed_DDF_correlator} offers the corresponding details within the framework of framed DDF vertices. 
		Appendix \ref{Covpol} discusses the first four excited string levels and the relationships between covariant and DDF polarizations. Finally, Appendix \ref{app:threeIntegral} evaluates the single and double complex integrals involved in defining the generating functional for the arbitrary correlation functions of SDS vertices.

		\newcommand{\ka}{{K_a}}
		\newcommand{\kb}{{K_b}}
		\newcommand{\pTa}{{P_{T a}}}
		\newcommand{\pTb}{{P_{T b}}}
		\newcommand{\pTc}{{P_{T c}}}
		
		\newcommand{\kka}{{k^a}}
		\newcommand{\kkb}{{k^b}}
		\newcommand{\ppTa}{{p^a_{T}}}
		\newcommand{\ppTb}{{p^b_{T}}}
		\newcommand{\ppTc}{{p^c_{T}}}

		\newcommand{\na}{{n}}
		\newcommand{\nb}{{m}}
		
		\newcommand{\COMMENTO}[1]{ { \bf #1} }
		\newcommand{\COMMENTOOK}[1]{OK: \bf #1}
		
		\section{Massive string states} 
		\label{General}
		One of the fundamental aspects of string theories is the presence in
		their spectrum of an infinite number of excited states forming
		$d$-dimensional massive representations of the Poincar\'e group.
		In the BRST-quantization of the open bosonic string in the critical
		dimension $d=26$, which is the model that we consider in these notes,
		these are obtained by applying on the lowest string state an arbitrary
		number of bosonic and ghost creation operators and imposing the
		BRST-conditions.
		The vertices associated with the physical states are constructed
		requiring their BRST invariance.  This procedure has been employed to
		determine the spectrum and vertex operators for the lowest massive
		string states and for those lying on the leading Regge trajectory
		\cite{Manes:1988gz}.
		However, it becomes cumbersome when applied to string states at
		arbitrary mass levels. The lightcone formulation of the string theory
		offers a solution to these difficulties but with the prize of losing
		the linear realization of the $d$-dimensional Lorentz covariance of
		the theory.  The DDF operators provide an alternative approach to
		constructing the physical string states that overcome all these
		difficulties \cite{DelGiudice:1971yjh,Green:1987sp}. In the case of the
		open bosonic string, they are defined as:
		\begin{eqnarray}
			A_n^i=
			i\,\epsilon_{(-n)\mu}^i\,
			\oint_{z=0}\frac{dz}{2\pi i} \,
			\partial_zL^\mu(z) 
			\,e^{i\,n \, \sqrt{2\alpha'} k \cdot L(z)}
			~~;~~
			n\in \mathbb{Z}
			~~,~~
			i=1,\dots,\,d-2~, \label{8}
		\end{eqnarray}
		with $k$ and $\epsilon_{(-n)\mu}^i$ the ``photon" momentum and
		polarization\footnote{In the following we write
			$\epsilon_{(-n) \mu}^i$ even if it is not necessary to consider
			different polarizations for any level $n$.
		}, respectively,
		satisfying the conditions $k^2=k\cdot \epsilon^i_{(n)}=0$. $L^\mu(z)$
		is the adimensional chiral part of the open string coordinate as
		defined in the App. \ref{app:notations} and the dot between two vectors
		denotes the contraction of spacetime indices with mostly plus
		Minkowski metric.  In the lightcone coordinates
		\begin{eqnarray}
			k^+=\frac{k^0+k^{d-1}}{\sqrt{2}}
			\qquad ;
			\qquad k^-=\frac{k^0-k^{d-1}}{\sqrt{2}}~,
			\label{photonmomentum}
		\end{eqnarray}
		where the metric takes the form
		$\eta^{+-}=\eta^{-+}=\eta_{+-}=\eta_{-+}=-1$,  the photon momentum  is
		usually  chosen in a frame where  takes  the form
		$k^\mu=(k^-,\,0,\dots)$ \cite{DelGiudice:1971yjh}.
		However, when possible we prefer to leave its expression arbitrary.  
		
		The DDF operators commute with the Virasoro generators and obey the
		algebra of the transverse string oscillators $\alpha_n^i$, provided
		that $\epsilon^i_{(-n)}\cdot\epsilon^j_{(n)}=\delta^{ij}$ and for all
		the physical states we take $2\alpha' k\cdot p_T=1$.
		Here, $p_T$ is the momentum of the lowest string state, i.e. the
		tachyon. From, these properties immediately follows that the action of
		a given number of DDF operators on the lowest tachyon state
		\begin{eqnarray}
			|{\rm Phys}\rangle 
			\equiv 
			\lambda_{i_1\dots i_m}\,
			:A^{i_1}_{-n_1}:\,\dots\,:A_{-n_m}^{i_m}:|0,\,p_T\rangle
			~~;~~
			p_T^2=-m^2=\frac{1}{\alpha'}~,
		\end{eqnarray}
		gives a physical state
		$|{\rm Phys}\rangle$
		of level $N=\sum_{i=1}^m n_i$
		with momentum $p=p_T -( \sum_{l=1}^m n_l)\, k$
		and polarization $\lambda_{i_1\dots i_m}$.
		Here we have written $:A:$ in order to stress that any $A$ is normal
		ordered but the product is not.

		The vertex operators associated with the emission of an open string at
		the point $x$ of the world-sheet gives the physical state through the
		correspondence states/operators:
		\begin{eqnarray}
			|{\rm Phys}\rangle = \lim_{x\rightarrow 0}V(x)|0,\,p=0\rangle~.
		\end{eqnarray}

		As example, the vertex operator associated with the following string
		state of the level $N=n$
		\begin{eqnarray}
			|n;\,\lambda\rangle
			\equiv
			\lambda_i\,A^i_{-n} |0,\,p_T\rangle~, \label{5}
		\end{eqnarray}
		turns out to be of the form %\cite{1902.07016}
		\begin{eqnarray}
			%V_{\lambda}(x)\equiv
			|n;\,\lambda\rangle
			=&
			\lambda_i \,\epsilon_{(n)\mu}^i 
			\oint_{z=x,\, |z-x|\rightarrow 0}\frac{dz}{2\pi i} 
			\frac{1}{(z-x)^n}
			:\,
			\left(
			i
			\partial_zL^\mu(z)
			+
			\frac{\sqrt{2\alpha'} p_T^\mu}{z-x}
			\right)
			\,e^{-i\sqrt{2\alpha'} n\,k\cdot L(z)}
			\,e^{i\sqrt{2\alpha'} p_T\cdot L(x)}
			\,.
			\label{6}
		\end{eqnarray}  
		Upon an integration by parts, this state can be rewritten as
		\begin{align}
			\lambda_i \,\varepsilon_{(n)\mu}^i \,V^\mu_{A_{-n}}(x)
			\equiv&
			\lambda_i \,\epsilon_{(n)\mu}^i  \Pi^\mu_{~\nu} 
			\oint_{z=x,\, |z-x|\rightarrow 0}\frac{dz}{2\pi i} 
			\frac{1}{(z-x)^n} :\partial_zL^\nu(z) 
			\,e^{-i\sqrt{2\alpha'} n\,k\cdot L(z)}
			\,e^{i\sqrt{2\alpha'} p_T\cdot L(x)}:\,,
			\label{8}
		\end{align}
		where we have introduced
		\begin{eqnarray}
			\Pi^\mu_{~\nu}= \delta^\mu_\nu -2\alpha'
			p_T^\mu\,k_\nu~~;~~\Pi^\mu_{~\nu}\,p_T^\nu=0
			~~;~~
			\epsilon_{(n)\mu}^i\,\Pi^\mu_{~\nu}\,k^\nu=0~
		\end{eqnarray}
		and the transverse polarization vector
		\begin{equation}
			\varepsilon_{(n)\nu}^i
			\equiv
			\epsilon_{(n)\mu}^i\,\Pi^\mu_{~\nu}
			~~;~~
			p_{T}^\mu\,\varepsilon^i_{(n)\mu}=k^\mu\,\varepsilon^i_{(n)\mu}=0 
			.
			\label{eq:covariant projector}
		\end{equation}

		The meaning of $|z-x|\rightarrow 0$ in Eq. \eqref{6}
		is that the contour must not encircle other vertices.
		This restriction is important when computing correlators of DDF vertices.

		The previous vertex satisfies the following OPE with the
		energy-momentum tensor
		\begin{eqnarray}
			T(u) V^\mu_{A_{-n}}(x)
			\sim
			\frac{\alpha' p_T^2}{(u-x)^2} V^\mu_{A_{-n}}(x)
			+\frac{\partial_xV_{A_{-n}}^\mu(x)}{u-x}+{\rm regular}\,.
		\end{eqnarray}
		This proves that the vertex operator has conformal dimension
		$\Delta=\alpha'\,p_T^2=1$ as required.
		
		The explicit expression of the DDF state, as provided in
		Eq. \eqref{5}, enables us to establish a direct connection with the
		physical states in the old-covariant formalism.
		Equation \eqref{5} can
		also be equivalently expressed in the following form:
		\begin{eqnarray}
			|n;\, \lambda\rangle
			= \lambda_i\,\epsilon_{(n)\mu}^i\,
			\sum_{m=0}^n \frac{1}{(n-m)!}\frac{\partial^{n-m}}{\partial z^{n-m}}
			\,
			e^{-n\sqrt{2\alpha'} \sum_{l=1}^\infty \frac{k\cdot
					\alpha_{-l}}{l}\,z^l}\Bigg|_{z=0}
			\alpha_{-m}^\mu|0,\,p\rangle
			~~;~~
			p=p_T-n\,k \,,\label{10L}
		\end{eqnarray}
		that shows that physical states of a generic level $n$ are obtained by
		taking suitable linear combinations of all the covariant creation
		operators associated with the given level.

		Recently a more compact reformulation of the usual DDFs has been
		proposed under the name of framed DDF. In this formulation, it is possible to completely decouple the DDF operators from the associated tachyon and show that they are good zero-dimensional conformal operators\cite{2402.13066}
		.
		Let us start by reminding the definition of the framed DDF\footnote{
			The map between the notations is
			$L_{there}(z)= \sqrt{\frac{\alpha'}{2}} L_{here}(z)$,
			$x_0=\sqrt{2\alpha'} \hat q$
			and
			$p_0=\hat p$.
			Moreover the $i$ range is $i_{there}=2,\dots D-1$ while
			$i_{here}=1,\dots d-2$.
		}
		\begin{equation}
			\uA^i_{n}(E)
			=
			i
			\oint_{z=0} \frac{d z}{ 2 \pi i}
			E^i_\mu \partial L^\mu(z)\,
			e^{i n \frac{E^+_\mu L^\mu(z)}{ \oldap E^+_\nu p_0^\nu} }
			=
			i
			\oint_{z=0} \frac{d z}{ 2 \pi i}
			\partial \uL^i(z)\,
			e^{i n \frac{\uL^+(z)}{ \oldap \upzero^+} }
			,
			\label{eq:app framed DDF}
		\end{equation}
		where $E^{\underline \mu}_{\nu}$ is a local orthogonal tangent frame, i.e.
		$ \eta_{\underline \mu \underline \nu} 
		E^{\underline \mu}_{\rho} E^{\underline \nu}_{\sigma} 
		=
		\eta_{\rho \sigma}
		$
		and
		$\uL^i = E^i_\mu\, L^\mu$,
		$\uL^+ = E^+_\mu\, L^\mu$
		(in particular $\upzero^+ = E^+_\mu \hat p^\mu$)
		are the flat coordinates fields.
		The advantages of this formulation are
		\begin{itemize}
			\item
			$\uA^i_{n}(E)$ are completely decoupled from the tachyon;
			\item
			$\uA^i_{n}(E)$ are good zero dimensional conformal operators;
			\item
			the notation is more intuitive given its resemblance with the light-cone.
		\end{itemize}
		When the operator $\hat p$ is evaluated on the tachyonic state
		the mapping between the usual notation and the one used for framed DDF is
		\begin{align}
			\epsilon^i_\mu
			&
			\rightarrow E^i_\mu
			,~~
			\varepsilon^i_\mu \rightarrow
			\underline{\Pi}^i_j\,E^j_\mu
			,
			\nonumber\\
			k_\mu
			&
			\rightarrow
			\frac{ E^+_\mu}{ \dap E^+_\nu\, p_T^\nu}
			,~~
			n \rightarrow n
			,
			\label{mapping framed usual}
		\end{align}
		where we have introduced
		the covariant projector Eq. \eqref{eq:covariant projector}
		written in flat coordinates
		\begin{equation}
			\underline{\Pi}^i_{\underline{\mu}}
			=
			%%      \left(
			\delta^{\underline{i}}_{\underline{\mu}}
			-
			\frac{\uk_T^i}{ \uk_T^+}
			\delta^{\underline{+}}_{\underline{\mu}}
			%%    \right)
			=
			%%      \left(
			\delta^{\underline{i}}_{\underline{\mu}}
			-
			\frac{E^i_\mu p_T^\mu}{E^+_\mu p_T^\mu}
			\delta^{\underline{+}}_{\underline{\mu}}
			%%    \right)
			.
			\label{eq:flat_coords_covariant projector}
		\end{equation}

		In the following for all results
		we give first the usual formulation and then the framed DDF one. 
		
		\section{ Sciuto-Della Selva-Saito vertex}
		\label{SDSV}

		The states introduced to describe cosmic strings \cite{Blanco-Pillado:2007hzh,Hindmarsh:2010if,Skliros:2011si} can be considered as the DDF realization of the Sciuto-Della Saito coherent state, which was introduced in the 60s to provide an efficient representation of string states at a generic massive level.

		These states are defined as the exponential of the DDF operators:
		\begin{eqnarray}
			\prod_{n=1}^\infty e^{\ \lambda_{n\,i} \,:A^i_{-n}:} |0_a,\,p_T\rangle
			= 
			e^{\sum_{n=1}^\infty \lambda_{n\,i} \,:A^i_{-n}:} |0_a,\,p_T\rangle
			,
		\end{eqnarray}
		where we have written $:A^i_{-n}:$  despite the fact there is no
		difference with $A^i_{-n}$ since all fields commute because of the
		choice of $\epsilon^i$ and $k$.

		Physical states are obtained by taking derivatives with respect to
		$\lambda_n$ at $\lambda_n=0$.
		
		The Sciuto-Della Selva-Saito vertex \cite{Sciuto:1969vz,DellaSelva:1970bj}, i.e. the vertex operator
		associated with the insertion of a DDF coherent state at point $x$ in the
		upper plane of the open bosonic string world sheet\footnote{
			See App. \ref{app:SDS_vertex} for details.}
		is defined {\sl naively }as follows \cite{Bianchi:2019ywd}
		\begin{eqnarray}
			{\cS}(x;\, \lambda_{i\,n},\, \,p_T)
			=
			\prod_{n=1}^\infty
			e^{\lambda_{n\, i}  : A^i_{-n}(x) :}\,
			:\,e^{i\sqrt{2\alpha'}  p_{T \mu}  L^\mu(x)}\,:\,
			,
		\end{eqnarray} 
		where
		\begin{align}
			A^i_n(x)
			=&
			i\,\epsilon_{(n)\mu}^i\,
			\oint_{z= x}\frac{d z}{2\pi i} \,
			\partial_zL^\mu(z) 
			\,e^{i\,n \, \sqrt{2\alpha'} k \cdot L(z)}
			\nonumber\\
			\equiv&
			i\,\varepsilon_{(n)\mu}^i\,
			\oint_{z= x}\frac{d z}{2\pi i} \,
			\partial_zL^\mu(z) 
			\,e^{i\,n \, \sqrt{2\alpha'} k \cdot L(z)}
			,
			\label{eq:A_w_eps_into_vareps}
		\end{align}
		is once again given by Eq. \eqref{8} but with the contour integral
		centered around the Koba-Nielsen variable $x$
		\footnote{
			The DDF operator in Eq.\eqref{8} is not really a zero-dimensional
			conformal operator which can be moved without care.
			This happens because its OPE has a cut due to
			$e^{\sdap k \cdot \hat p}$.
			The only DDF operators which are true zero-dimensional conformal
			operators are the one where $\hat p$ is used and not $k$, such as the
			framed DDF operator in Eq.\eqref{eq:app framed DDF}.
			Nevertheless, when it is applied to the tachyonic vacuum it becomes a
			good operator which can be moved.
		}.
		Notice that the two expressions are equivalent since the difference is
		an integral of a total derivative.
		As commented below the second form allows for a slight simplification
		of the final expression.

		The reason why we have written naively is that the expression is not
		well defined as it stands because of the integrals, in facts the
		integrations make some creators have bigger $|z|$ than some
		annihilators.
		This means that the normal ordering we are doing next requires
		some analytic continuation in order to be well defined.
		In App. \ref{app:SDS_vertex} we give a more rigorous treatment based
		on Sciuto-Della Selva-Saito vertex.
		Moreover the substitution $\epsilon \rightarrow \varepsilon$ is
		allowed because of an integration by parts and simplifies the computations.

		We now proceed with the normal ordering of the previous operator.
		The first step is to normal order
		$e^{\lambda_{n} \cdot :A_{-n}(x):}$ as\footnote{In this expression the normal order can also include the second factor because the two exponential operators commute.}  
		\begin{eqnarray}
			&&
			e^{\lambda_{n} \cdot :A_{-n}(x):}
			=
			:\, e^{\lambda_{n} \cdot A_{-n}(x)}\, :
			\nonumber\\
			&&
			\phantom{   e^{\lambda_{n} \cdot :A_{-n}(x):} }
			\exp\Bigg\{ \frac{1}{2} 
			\oint_{z_1=x\,|z_1|>|z_2|} \frac{dz_1}{2\pi i}\,
			\oint_{z_2=x} \frac{dz_2}{2\pi i}\,
			\frac{
				\lambda_{n} \cdot \varepsilon_{(n) }
				\,\lambda_{n} \cdot \varepsilon_{(n) }
			}
			{(z_1-z_2)^2}\, 
			e^{
				-i\sqrt{2\alpha'}\left[
				n\, k \cdot L(z_1)
				+n\, k \cdot L(z_2)
				\right]}
			\Bigg\}\,,\nonumber\\
			,
		\end{eqnarray}
		then we normal order the product\footnote{
			In the following 
			in order to easily track the origin of the different contributions
			we write $\varepsilon^a$, $\ppTa$ and $\kka$. Notice that $\kka$ have the same null directions for all vertexes but they are not equal but simply proportional
			as $k_a = \rho_{b a} k_b$. 
			This necessity follows from the normalization 
			$2\alpha' k_a \cdot p_{T a}
			=-2\alpha' k_a^- p_{T a}^+ 
			=-2\alpha' k_a^- p_a^+ 
			=1
			$ in the frame where the photon momenta are of the form
			$k_a= (k_a^-,\,0\dots 0)$
			and the fact that $p_{T a}$ are not equal.
			Then we define:
			$\rho_{ba}=\frac{p_b^+}{p_a^+}
			=\frac{k_a^-}{k_b^-}$.

			Moreover, $\ppTa$ differ in order to allow the momenta $p_a= \ppTa - N_a \kka$ to point in different directions.
			Because of this, the projectors $\Pi_a$ differ, and also
			$\varepsilon^{i a} = \epsilon^i \cdot \Pi_a$ differ.}
		
		\begin{eqnarray}
			&&
			:\,e^{\lambda_{n_a} \cdot A_{-n_a}(x;\, \varepsilon^a,\, \kka)}\,:\,
			:\, e^{\lambda_{n_b} \cdot A_{-n_b}(x;\, \varepsilon^b,\, \kkb)}\, :
			= 
			:\, e^{\lambda_{n_a} \cdot  A_{-n_a}(x)
				+\lambda_{n_b} \cdot A_{-n_b}(x)}\, :
			\nonumber\\
			&&
			\phantom{   e^{\lambda_{n} \cdot :A_{-n}(x):} }
			\times
			\,\exp\Bigg\{
			%% \frac{1}{2} 
			\oint_{z_1=x\,|z_1|>|z_2|} \frac{dz_1}{2\pi i}\,
			\oint_{z_2=x} \frac{dz_2}{2\pi i}\,
			\frac{
				\lambda_{n_a}\cdot \varepsilon^a_{(n_a)}
				\,\lambda_{n_b}\cdot \varepsilon^b_{(n_b)}
			}
			{(z_1-z_2)^2}\, 
			e^{
				-i\sqrt{2\alpha'}\left[
				n_a\,\kka\cdot L(z_1)
				+n_b\,\kkb\cdot L(z_2)
				\right]}
			\Bigg\}\,
			,\nonumber\\
		\end{eqnarray}
		and finally, we can assemble all the pieces and consider the tachyonic vertex
		\begin{eqnarray}
			&&
			:e^{\lambda_{n_1}\cdot A_{-n_1}(x)}: \,
			:e^{\lambda_{n_2}\cdot A_{-n_2}(x)}:\,
			: e^{i\sqrt{2\alpha'}  p_T\cdot L(x)}:
			\nonumber\\
			=&&   
			:\,
			e^{\sum_{a,b=1}^2\lambda_{i\,n_a}\,\lambda_{j\,n_b}\,Q^{ij}_{n_a\, n_b}(x)}\,
			e^{\lambda_{n_1\, i}\, Q^i_{n_1}(x)}\,
			e^{\lambda_{n_2\, i}\, Q^i_{n_2}(x)}\,
			e^{i\sqrt{2\alpha'} p_T\cdot L(x)}
			:\,
			.
		\end{eqnarray}
		Given the conditions, $k^2=k\cdot\epsilon=0$,  the normal ordering may also incorporate the last exponential factor in these expressions. In the previous equation, we have used the operator
		\begin{eqnarray}
			Q_{n}^{i}(x;\, \varepsilon,\, k)
			=&&
			\oint_{z=x} \frac{dz}{2\pi i} \,
			\frac{1}{ (z- x)^n }\,
			\epsilon_{(\na)}^i
			\cdot
			\left( i \partial_z L(z) + \frac{ \sdap p_T}{ z-x} \right)\,
			e^{-i \na\sqrt{2\alpha'} \,k \cdot L(z)}
			\nonumber\\
			\equiv&&
			\oint_{z=x} \frac{dz}{2\pi i} \,
			\frac{1}{ (z- x)^n }\,
			\varepsilon_{(\na)}^i
			\cdot
			\left( i \partial_z L(z) + \frac{ \sdap p_T}{ z-x} \right)\,
			e^{-i \na\sqrt{2\alpha'} \,k \cdot L(z)}
			\nonumber\\
			=&&
			\oint_{z=x} \frac{dz}{2\pi i}\,
			\frac{
				\varepsilon_{(\na)}^i\cdot \partial_z L(z)
			}
			{(z- x)^n       }\,
			e^{-i \na\sqrt{2\alpha'} \,k \cdot L(z)}
			,
			\nonumber\\
		\end{eqnarray}
		where the second expression is due to the second expression in
		\eqref{eq:A_w_eps_into_vareps} and the third equation shows the
		simplification occurring because of the substitution $\epsilon
		\rightarrow \varepsilon$ .
		We use also the operator
		\begin{eqnarray}
			Q_{n,\, m}^{i j}(x;\, \varepsilon,\, k)
			=&&
			\frac{1}{2} \oint_{z_1=x, |z_1|>|z_2|}
			\frac{d z_1}{2\pi i}\frac{1}{(z_1-x)^{\na}}\,
			\oint_{z_2=x} \frac{d z_2}{2\pi i}\frac{1}{(z_2-x)^{\nb}}\,
			\frac{\varepsilon_{(\na)}^i  \,\cdot\varepsilon_{(\nb)}^j}{(z_1-z_2)^2}\, 
			e^{-i\sqrt{2\alpha'}\left[\na\,k \cdot L(z_1)+\nb\,k \cdot L(z_2)\right]}\,
			,
			\label{eq:integrals in usual DDF vertex}
		\end{eqnarray}
		where we have used the transverse polarization
		defined in Eq. \eqref{eq:covariant projector}.

		We then get the following expression for the vertex operator
		describing the emission of coherent states at $x=x_a$
		from the boundary of world sheet of an open string:
		\begin{eqnarray}
			{\cS}(x_a;\, \lambda^a_{n},\, p^a_{T} )
			=&
			%%%%   e^{\sum_{n=1}^\infty\lambda_n^a\cdot Q_{n}(x_a)}\,
			: e^{\sum_{n_1,n_2=1}^\infty  \lambda^a_{n_1\, i_1}\,
				\lambda^a_{n_2\, i_2}
				Q^{i_1i_2}_{n_1\,n_2}(x_a;\, \varepsilon^a,\, \kka) }\,
			%%   \nonumber\\
			%%   &
			\,
			e^{\sum_{n=1}^\infty \lambda^a_{n\,i} \, Q^i_{n}(x_a;\, \varepsilon^a,\, \kka) }\,
			e^{i\sqrt{2\alpha'} p^a_{T\, \mu} L^\mu(x_a)}:\,
			.
			\label{19}
		\end{eqnarray}
		Notice that $\ppTa$ differ while $\kka$ are proportional in order to allow the momenta $p_a= \ppTa - N_a k_a$ to point in different directions, with $N$ the level of the state.

		The vertex associated with the generic DDF state, with polarization
		$\lambda^a_{i_1\dots i_n}$, is obtained by acting on Eq. \eqref{19}
		with an arbitrary number of derivatives with respect to $\lambda^a_n$:
		\begin{eqnarray}
			V_N(x_a;\, \lambda^a_{i_1\dots i_n},\, p^a)
			=
			\lambda^a_{i_1\dots i_n}\,
			\frac{\partial^n}{
				\partial \lambda^a_{n_1 {i_1}} \dots\partial \lambda^a_{n_n{i_n}}
			}
			{\cal S}(x_a;\, \lambda^a_{n},\,  \ppTa)\Bigg|_{\lambda^a_{n_i}=0}
			,
			~~~~
			p^a=\ppTa + ( \sum_{l=1}^n n_l)\,k^a
			,
			\label{Genexp}
		\end{eqnarray} 
		where, in general, some $n_i$ can also be equal to some $n_j$ for
		$i\neq j$.
		The level of the state is
		$ N_a=\sum_{i=1}^n n_i$.

		Due to this identity, the correlation function with an arbitrary
		number of coherent  vertices in
		Eq. \eqref{eq:corr_func_coher_states}
		can be thought of as the functional
		generator of amplitudes involving an arbitrary number of DDF states,
		describing the open string excited states associated with a generic
		massive level.

		The corresponding expression for the Sciuto-Della Selva-Saito vertex
		for the framed DDF reads
		\begin{align}
			%%  \widetilde \cS(x)
			%%  &
			%%  | \{\lambda^i_n\}, \uk_T )
			%%  \equiv
			\cS( x_a;\, \lambda^a_{n},\, \uk^a_T)
			%%  \nonumber\\
			&=
			:
			\exp \left(
			\sum_{i=1}^{d-2}
			\sum_{n=1}^\infty
			\lambda^a_{n\, i}
			\oint_{z=x_a,\, |z-x_a|\rightarrow 0} \frac{d z}{ 2\pi i}
			\frac{1}{z^n}
			\left(
			i
			%%    \ishap
			\partial_z \uL^{i}(z)
			+
			\frac{
				\sdap \uk^i_{T a}
			}{z-x_a}
			\right)
			e^{-i n \frac{ \uL^{+} (z) }{\oldap \uk^+_{T a}} } 
			\right)\,
			e^{i \sdap \uk^a_{T \mu} \uL^\mu(x_a)}
			\nonumber\\
			\times
			\exp
			&
			\left(
			\oh
			\sum_{i=1}^{d-2}
			\sum_{n_1, n_2=1}^\infty
			\lambda^a_{n_1\, i}
			\lambda^a_{n_2\, i}
			\oint_{z_1=x_a, |z_1-x_a| > |z_2-x_a|} \frac{d z_1}{ 2\pi i}
			\oint_{z_2=x_a} \frac{d z_2}{ 2\pi i}
			\frac{
				e^{
					-i
					\frac{
						n_1  \uL^+(z_1)
						+
						n_2\uL^+(z_2)
					}{\oldap \uk^+_{T a}}
				}
			}{
				(z_1-x_a)^{n_1} (z_2-x_a)^{n_2} (z_1 -z_2)^2
			}
			\right):
			,
			\label{DDF_S_vertex_1}
		\end{align}
		where $|z-x_a|\rightarrow 0$ means that the contour must not
		encircle other vertices and
		the double integral in the second line is the flat coordinate
		expression for $ Q_{\na,\nb}^{ij}(x)$.

		Upon an integration by parts in the single integral we get a more
		compact form
		\begin{align}
			\cS( x_a;\, \lambda^a_{n},\, \uk^a_T)
			&=
			:
			\exp \left(
			\sum_{i=1}^{d-2}
			\sum_{n=1}^\infty
			\lambda^a_{n\, i}
			\oint_{z=x_a,\, |z|\rightarrow x_a} \frac{d z}{ 2\pi i}
			\frac{1}{z^n}
			i
			%%    \ishap
			\underline{\Pi}^i_{\underline{\mu}}
			\partial_z \uL^{\underline{\mu}}(z)
			e^{-i n \frac{ \uL^{+} (z) }{\oldap \uk^+_{T a}} } 
			\right)\,
			e^{i \sdap \uk^a_{T \mu} \uL^\mu(x_a)}\,
			:
			\nonumber\\
			\exp
			&
			\left(
			\oh
			\sum_{i=1}^{d-2}
			\sum_{n_1, n_2=1}^\infty
			\lambda^a_{n_1\, i}
			\lambda^a_{n_2\, i}
			\oint_{z_1=x_a, |z_1-x_a| > |z_2-x_a|} \frac{d z_1}{ 2\pi i}
			\oint_{z_2=x_a} \frac{d z_2}{ 2\pi i}
			\frac{
				e^{
					-i
					\frac{
						n_1  \uL^+(z_1)
						+
						n_2\uL^+(z_2)
					}{\oldap \uk^+_{T a}}
				}
			}{
				(z_1-x_a)^{n_1} (z_2-x_a)^{n_2} (z_1 -z_2)^2
			}
			\right)
			\,
			&
			\label{DDF_S_vertex_2}
		\end{align}
		This form reveals the covariant projector written in flat coordinates
		introduced in Eq \eqref{eq:flat_coords_covariant projector}.
		\section{The DDF Reggeon: the generating function of DDF amplitudes}
		\label{section4}
		The amplitudes involving DDF states are obtained by starting with the correlators of coherent state vertices defined in Eq. \eqref{8} and then acting with the appropriate number of derivatives, as discussed in Eq. \eqref{Genexp}. The complex integrals around the cycles centered at the punctures $x$ can be evaluated, and the resulting DDF vertex is expressed as a sum of $N$ terms, where $N$ denotes the string level. These terms can be written in terms of Schur polynomials and string coordinate fields \cite{Gross:2021gsj}.  Scattering amplitudes are then obtained by considering correlation functions of these polynomial operators.
		
		In this section analogously to  the $M$ vertex operator constructed in the 1980s for computing correlation functions of perturbative string states, we aim to construct a generating function that yields all correlation functions of $M$
		Sciuto-Della Selva-Saito vertex: 
		\begin{eqnarray}
			{\cal W}_M(x_1,\,\lambda^1,\,p_T^1; \dots x_M,\, \lambda^M,\,p_T^M)
			=
			\langle p=0| \prod_{a=1}^M {\cal S}(x_a;\,\lambda^a,\, \ppTa)|p=0\rangle
			~~;~~
			x_1>x_2\dots>x_M\,,
			\label{eq:corr_func_coher_states}
		\end{eqnarray}
		Moreover, we have introduced two different but equivalent expressions for the DDF vertex and in  both formulations 
		we present the expression for the generator of the correlation functions collecting the details of the calculation in App. \ref{app:correlator} for the standard formulation 
		and in App. \ref{app:framed_DDF_correlator}  for the framed one.
		
		In the formalism where the vertex operator is represented as given in  Eq. \eqref{19} the resulting expression for the generating function is:
		\begin{eqnarray}
			&&  {\cal W}_M(\{x\},\,\{\lambda\},\,\{p_T\})
			=
			(2\alpha')^{\frac{d}{2}}\int \frac{d^d q}{(2\pi)^d}
			\,
			\Bigg[
			\prod_{a=1}^M
			e^{i\sqrt{2\alpha'}\ppTa \cdot q}\,
			\prod_{1\le a<b\le M} (x_a-x_b)^{2\alpha' \ppTa \cdot \ppTb}
			\nonumber\\
			&&
			\times\,
			\prod_{a=1}^{M}
			e^{\sum_{n_a=1}^\infty{\cal  A}_a(\{x\};\, \lambda^a_{n_a};\, \{k\};\, \{p_T\};\, q) }
			\,
			\prod_{a,  b=1}^M
			e^{\frac{1}{2}\sum_{n_a,\,n_b=1}^\infty
				{\cal B}_{ab}(\{x\};\,\lambda^a_{n_a};\,
				\,\lambda^b_{n_b};\, \{k\};\{p_T\};\,  q) }
			\Bigg]
			.
			\nonumber\\
			\label{23}
		\end{eqnarray}
		where we have denoted with
		$p_T=\{p_T^a\}_{1\leq a\leq M}$ the set of all the tachyonic momenta associated with the external states and a similar notation for the other quantities introduced in Eq. \eqref{23}.

		Here, we have introduced the quantities:
		\begin{eqnarray}
			{\cal  A}_a(\{x\};\, \lambda^a_{n_a};\,\{k\};\, \{p_T\};\, q)
			&&=
			\sum_{n_a=1}^\infty\,
			\sum_{i=1}^{d-2}
			\lambda^a_{n_a\, i}\,
			\,e^{-i\sqrt{2\alpha'}n_a \kka\cdot q}\,
			{\cal  A}_{a,n_a}^i(\{x\};\, \varepsilon^a_{(n_a)},\, \{k\};\, \{p_T\})
			,
			\nonumber\\
			{\cal  A}_{a,n_a}^i(\{x\}; \varepsilon^a_{(n_a)},\, \{k\};\, \{p_T\})
			&&=
			\oint_{x_a} \frac{dz_a}{2\pi i} \,
			\frac{
				\prod_{b >a}(z_a-x_b)^{-2\alpha'  n_a \kka \cdot \ppTb } \,
				\prod_{b <a}(x_b-z_a)^{-2\alpha'  n_a \kka \cdot \ppTb }
			}{(z_a-x_a)^{n_a}}
			\nonumber\\
			&&
			\times
			\left[\sum_{b\neq a }
			\frac{
				\sqrt{2\alpha'}\,
				\varepsilon_{a (n_a)}^i\cdot \ppTb
			}{z_a-x_b}\right] \,,
			\nonumber\\
			\label{25}
		\end{eqnarray} 
		and
		\begin{eqnarray}  
			{\cal    B}_{ab}(
			\{x\};\, \{k\};\, \lambda^a_{n_a};\,
			\,\lambda^b_{n_b};\, 
			\{p_T\};\, \,q )
			&& =
			\sum_{n_a, n_b=1}^\infty\,
			\sum_{i, j=1}^{d-2}
			\lambda^a_{n_a\, i}
			\,\, \varepsilon_{(n_a)}^i\cdot\varepsilon_{(n_b)}^j
			\,\,
			\lambda^b_{n_b\, j}\, e^{-i\sqrt{2\alpha'}(n_a\kka+n_b\kkb)\cdot q}\nonumber\\
			&&\times\,
			{\cal    B}_{a,n_a\, b,n_b} %_{n_a\, i; n_b\, j}
			(\{x\};\,\{k\};\, \{p_T \})
			,
			\nonumber\\
		\end{eqnarray}
		with:
		\begin{eqnarray}
			&&
			{\cal    B}_{a,n_a,b\, n_b}
			( \{x\};\, \{k\};\,\{p_T\} )
			=
			\oint_{x_a} \frac{dz_a}{2\pi i}
			\frac{  \prod_{c >a}(z_a-x_c)^{-2\alpha'  n_a \,k_a\cdot p_{T}^c} \,
				\prod_{c <a}(x_c-z_a)^{-2\alpha' n_a \,k_a\cdot p_{T}^c}
			}{(z_a-x_a)^{n_a}}
			\nonumber\\
			&&
			\times \oint_{x_b} \frac{dz_b}{2\pi i}
			\frac{ \prod_{c >b}(z_b-x_c)^{-2\alpha'  n_b \kkb \cdot \ppTc } \,
				\prod_{c <b}(x_c-z_b)^{-2\alpha'  n_b \kkb \cdot \ppTc }
			}{(z_b-x_b)^{n_b}}\,
			%%   \left[
			\frac{1
			}
			{ (z_a - z_b)^2 }
			%%     \right]
			\,
			.
			\nonumber\\
			\label{26a}
		\end{eqnarray}
		In getting these expressions 
		we have used the conditions:
		\begin{eqnarray}
			\label{ktransv}
			\kka \cdot \kkb
			= \kkb \cdot \varepsilon_{(n_a)}=0
			\qquad \forall a,b\,.
		\end{eqnarray}

		The corresponding result for framed DDF reads
		\begin{align}
			\cW_M
			=&
			\langle \up^i=\up^-=0|
			\nonumber\\
			%
			% exp number t linear
			%
			\Bigg\{
			\exp
			&
			\Bigg(
			\sum_{a=1}^M
			\sum_{i=1}^{d-2}
			\sum_{n=1}^\infty
			\lambda^a_{n\, i}
			\,
			\oldI^{i}_{a, n}( \{x\};\, \{\underline{p^i}\};\,\{\underline{p}^+\})
			e^{-i n \frac{ \uxzero^+ }{\dueoldap \uk^+_{T a}} } 
			\Bigg)
			%\end{align}
			%%%%%%%
			%  %
			\nonumber\\  
			\times\, \exp
			&
			\Bigg(
			% +
			\oh
			\sum_{a,b= 1}^{M}
			\sum_{i=1}^{d-2}
			\sum_{n_1, n_2=1}^\infty
			\lambda^a_{n_1\, i}\,
			\lambda^b_{n_2\, i}\,  e^{
				-i
				\frac{
					n_{ 1}  \uxzero^+
				}{\dueoldap \uk^+_{T a} }
				-i
				\frac{
					n_{ 2} \uxzero^+
				}{\dueoldap \uk^+_{T b} }
			}
			\oldJ
			_{a,n_1;b, n_2}( \{x\};\,\{\underline{p}^+\} )
			\Bigg)
			\Bigg\}
			\nonumber
		\end{align}
		%%%%%%%%%
		\begin{align}
			\times
			&
			\prod_{a=1}^M
			e^{i  \sdap (\uk^{a i}_{T} \uxzero^i 
				-\uk^{a -}_{T} \uxzero^+ )}
			\,
			| \up^i=\up^-=0 \rangle
			\,
			%
			%% \nonumber\\
			%% \times
			%% &
			\prod_{a= 1}^{M-1} \prod_{b=a+1}^M
			\xxrrtt ^{\dap \uk^a_{T} \cdot \uk^b_{T} }
			\,
			\delta\left( \sum_{a=1}^M \uk^+_{a} \right)
			,
		\end{align}
		where we have introduced the notation 
		$\underline{p}^+\equiv\underline{p}_T^+
		=\{\underline{p}_{ a}^+ \}
		=\{\underline{p}_{T a}^+ \}$
		where we have introduced the analogous integrals of Eq.s \eqref{25},\eqref{26a}
		in flat coordinates
		\begin{align}
			\oldI^{i}_{a,n}(\{x\};\,\{\underline{p}^i\};\,\{\underline{p}^+\})
			=&
			\oint_{z=x_a} \frac{d z}{ 2\pi i}
			\frac{1}{(z - x_a)^n}
			\sum_{b \ne a}
			\frac{\sdap  }{z - x_{b}}
			\left(
			\uk^i_{T b}
			-
			\frac{ \uk^+_{T b} }{ \uk^+_{T a} }
			\uk^i_{T a}  
			\right)
			\nonumber\\
			&
			\phantom{\exp}
			\prod_{b=1}^{a-1}
			(x_{b}  - z)^{ -n \frac{\uk^+_{T b} }{\uk^+_{T a}} }
			\,
			\prod_{\rrr \TTT+1}^{N}
			(z - x_{b}  )^{ -n \frac{\uk^+_{T b} }{\uk^+_{T a}} }
			,\label{34}
		\end{align}
		and
		\begin{align}
			%%  \oldJ_{\snr n_1; \snt n_2}( \uk_{\sns},\, x_\SSS)
			\oldJ_{a,n_1;\,b, n_2}( \{x\};\{\underline{p}^+\}\, %\uk^a;\, x_b,\, \uk^b
			)
			=&
			\oint_{z_1=x_a, |z_1-x_a| > |z_2-x_b|} \frac{d z_1}{ 2\pi i}
			\oint_{z_2=x_b} \frac{d z_2}{ 2\pi i}
			\frac{ 1
				% e^{-i\frac{ n_{ 1}  \uxzero^+}{\dueoldap \uk^+_{T a} }-i\frac{n_{ 2} \uxzero^{\dueoldap \uk^+_{T b} }}
				}{
					(z_1-x_a)^{n_1} (z_2-x_b)^{n_2} (z_1 -z_2  )^2
				}
				\nonumber\\
				&
				\phantom{\exp}
				\prod_{c=1}^{a-1}
				(x_{c}   - z_1)^{ -n_1 \frac{\uk^+_{T c} }{\uk^+_{T a}} }
				\prod_{c=a+1}^{M}
				(-x_{c}   + z_1)^{ -n_1 \frac{\uk^+_{T c} }{\uk^+_{T a}} }
				\nonumber\\
				&
				\phantom{\exp}
				\prod_{c=1}^{b-1}
				(x_{c}    - z_2)^{ -n_2 \frac{\uk^+_{T c} }{\uk^+_{T b}} }
				\,
				\prod_{c=b+1}^{M}
				(-x_{c}    + z_2)^{ -n_2 \frac{\uk^+_{T c} }{\uk^+_{T b}} }
				.\label{35}
			\end{align}
			
			Starting from this expression is then possible to show that on shell DDF amplitudes are the same as light-cone amplitudes and that Mandelstam maps appear automatically \cite{Biswas:2024epj}.
			\newcommand{\raapuno}{\rho_{a+1\,a}}
		\newcommand{\raamuno}{\rho_{a-1\,a}}
		\newcommand{\rab}{\rho_{b\,a}}
		\newcommand{\rac}{\rho_{c\,a}}
		\newcommand{\rba}{\rho_{a\,b}}
		\newcommand{\rbc}{\rho_{c\,b}}

		\section{Three-points amplitudes of DDF-states} 
		\label{Amplitude}
		In the previous section, we have computed the functional generator of
		all $M$ point amplitudes of DDF states. The expression, however, is
		still formal because requires the explicit evaluation of the residues
		around the worldsheet points where the external states are inserted
		on the open string worldsheet. In the case of the three point
		amplitudes, these integrals have been computed in  \cite{DelGiudice:1971yjh}. Here, we give their expression for the
		insertion of the vertex operators at general points $x_a$ of the
		world-sheet, leaving the details on their derivation in
		App. \ref{app:threeIntegral}.  The results of the single integrals are:
		\begin{eqnarray}
			{\cal A}^i_{a,n_a}(\{x\}; \{k\}; \{p_T\})
			&&=\,P_{a\,a+1}^i%\lambda_{n_a\,i}^a \varepsilon_{(n_a)}^i\cdot p_{a+1}\,e^{-i\sqrt{2\alpha'}n_a k_a\cdot q} 
			\left(\frac{x_{a-1 ;a+1}}{x_{a-1;a}\,x_{a;a+1}}\right)^{n_a}(x_{a-1;a})^{-2\alpha' \,n_a\,\raapuno%k_a\cdot p_{a+1}
			}\nonumber\\
			&& \times(x_{a;a+1})^{-2\alpha'\,n_a\,\raapuno
				%k_a\,\cdot p_{a+1}
			}
			\,(-1)^{n_a-1}\left( \begin{array}{c}
				2\alpha' n_a\,\raapuno %k_a\cdot p_{a+1}
				+n_a-1\\
				n_a-1\end{array}\right)\,,\nonumber\\
			\label{26}
		\end{eqnarray}
		with $a=1,2,3$ and the cyclic identification  $3+a\equiv a$ and we have denoted with 
		$x_{a;b}=x_{{\rm min}(a,\,b)}-x_{{\rm max}(a,\,b)}$. 
		
		Here we have introduced, in both formulations of DDF-amplitudes discussed in the previous sections,  the quantities: 
		\begin{eqnarray}
			\rab
			\equiv 
			2\alpha'\,k_a\cdot p_b
			~~&,&~~
			P_{ab}^i
			\equiv
			\sqrt{2\alpha'}\,
			e^{-i\sqrt{2\alpha'}n_a k_a\cdot q} \,  \varepsilon_{(n_a)}^i\cdot p_{b}
			\nonumber\\
			\rab
			\equiv 
			\frac{E_{[a]}^+\
				\cdot p_{T[b]}}{ E_{[a]}^+ \cdot p_{T[a]}} 
			\equiv  
			\frac{\underline{p}_{T[b]}^+}{\underline{p}_{T[a]}^+} 
			~~&,&~~
			P_{ab}^i
			\equiv
			\sqrt{2\alpha'}\, e^{-i n \frac{ \uxzero^+ }{\dueoldap \uk^+_{T a}} } \,\left(
			\uk^i_{T b}
			- \rab\,
			%\frac{ \uk^+_{T b} }{ \uk^+_{T a} }
			\uk^i_{T a}  
			\right)
			\label{eq:rhoab}
		\end{eqnarray}
		There are two different typologies of double integrals. Those with the two contour integrals centered around the same Koba-Nielsen variable $x_a$ and those where the integration is performed around different points. The former provides the auto-interaction contributions to the amplitudes, and the result of the integration is as follows
		\begin{eqnarray}
			&&
			{\cal  B}_{a,n_a;\,a, m_a}(
			\{x\},\, \{p\})
			=(-1)^{n_a+m_a}
			x_{a;a+1}^{-2 n_a\,
				\raapuno %k_a\cdot p_{a+1}
			}\,x_{a-1;a}^{-2 n_a\,\raamuno %k_a\cdot p_{a-1}
			}\,\nonumber\\
			&&\times \left( \frac{x_{a+1;a-1}}{\,x_{a;a+1}\,x_{a-1;a}}\right)^{2n_a}\,\left( \begin{array}{c} - n_a\,\raamuno %k_a\cdot p_{a-1}
				\\n_a\end{array}\right) \left( \begin{array}{c} - m_a \,\raamuno %k_a\cdot p_{a-1}
				\\ m_a \end{array}\right)\, \frac{n_a\,m_a}{n_a+m_a}\,\frac{(1+\,\raamuno%k_a\cdot p_{a-1}
				)}{\,\raamuno %k_a\cdot p_{a-1} 
			}\,.\nonumber\\ 
			&&\label{27}
		\end{eqnarray} 
		The result of the integration of the double integrals, 
		evaluated around two different Koba-Nielsen variables, turns out to be:
		\begin{eqnarray}
			&&{\cal  B}_{a,n_a;\,b, n_b}
			(\{x\},\,\{p\})
			=
			%e^{-i\sqrt{2\alpha'}(n_ak_a+n_bk_b)\cdot q}\,\lambda_{n_a\,i}\,\varepsilon^i_{(n_a)} \cdot \varepsilon_{(n_b)}^j\,\lambda_{n_b}^j\,
			x_{a;b}^{n_a\rab%k_a\cdot p_b
				- n_b\rba%k_b\cdot p_a
			}\, x_{a;c}^{-%2\alpha'
				n_a\,\rac}\,x_{b;c}^{-%2\alpha' 
				n_b\,\rbc}\nonumber\\
			&&\times\,\left(\frac{x_{b;c}}{x_{a;b} \,x_{a;c}}\right)^{n_a}\left( \frac{x_{a;c}}{x_{b;c}\, x_{a;b}}\right)^{n_b} (-1)^{n_b-1}\left( \begin{array}{c} -n_a\rab \\ n_a\end{array}\right)
			\left( \begin{array}{c} -\frac{n_b}{\rab}\\n_b \end{array} \right)\frac{n_an_b(1+\rab)}{n_a\rab +n_b}\,,\label{28}
		\end{eqnarray}
		with  $c\neq a,b$.  
				Eq. \eqref{28} gives the expressions of the cyclic double integrals   $ ({\cal B}(1;2),\,{\cal B}(2;3) ,\,{\cal B}(3;1))$, the other integrals are obtained from the identity ${\cal B}(a,b)={\cal B}(b;a)$%\footnote{Here and in the rest of the paper,  we will concisely denote with 
					%${\cal B}(a;b)\equiv {\cal B}_{a, n_a;\, b, n_n}(\{x\},\, \{p\})$. 
					%Similarly, we will denote with 
					%${\cal A}(a)\equiv {\cal A}_{a,n_a}(\{x\},\,\{k\},\,\{p_T\})$.}.
				
				\subsection{A direct check of $M=3$ Reggeon on an explicit non trivial amplitude}
				
				As a direct check on the validity of the generator of all DDF-correlators given in Eq. \eqref{23}, we will now compute the correlator of three DDF-states as defined in Eq. \eqref{5}. These states represent physical states belonging to three different levels, $n_a$, where $a=1, 2, 3$. We will do this using two different approaches. First, we will consider the correlator of three vertices, as defined in Equation \eqref{6}. Then, we will compute the corresponding expression using the functional generator of the correlators, which is defined in Equation \eqref{23}. We will show that the results are the same.
				
				Further checks of the validity of these expressions are obtained in section \ref{sec:3pt_scattering} where the amplitudes of one massive scalars with two tachyons are computed and shown to be extremely sensitive to the coefficients.
				
				The correlator of the three vertices given in Eq.\eqref{5} 
				stripped by the delta over the momenta $(2\pi)^d \delta^d\left(\sum_{i=1}^3p_i\right)$
				is:
				\begin{eqnarray}
					&&C^{123}_{3,n_1,n_2,n_3}(\lambda^{1},  k_1,p_1; \lambda^{2 },k_2, p_2; \lambda^{3}, k_3,\,p_3)=\lambda_{n_1,\,i_1}^{1}\,\lambda_{n_2,i_2}^{2}\,\lambda_{n_3, i_3}^{3}
					\langle V_{A_{-n_1}}^{i_1}(x_1)\,V_{A_{-n_2}}^{i_2}(x_2)\,V_{A_{-n_3}}^{i_3}(x_3)\rangle
					\nonumber\\
					&&=\oint_{x_1} \frac{dz_1}{2\pi i}\frac{\lambda_{n_1,\,i_1}^{1}}{(z-x_1)^{n_1}}\oint_{x_2}\frac{dz_2}{2\pi i} \frac{\lambda_{n_2,i_2}^{2} }{(z_2-x_2)^{n_2}}\oint_{x_3} \frac{d z_3}{2\pi i} \frac{ \lambda_{n_3, i_3}^{3}}{(z_3-x_3)^{n_3}} \nonumber\\
					&&\times \prod_{a=2}^3(z_1-x_a)^{-2\alpha' n_1k_1\cdot p_T^{(a)}}\prod_{a=1}^2(x_a-z_3)^{-2\alpha'n_3k_3\cdot p_T^{(a)}}(x_1-z_2)^{-2\alpha' n_2k_2\cdot p_T^{(1)}}\,(z_2-x_3)^{-2\alpha' n_2 k_3\cdot p_T^{(3)}}\nonumber\\
					&&\times \sqrt{2\alpha'}\Bigg[\frac{\varepsilon^{i_1}_{n_1}\cdot \varepsilon^{i_2}_{n_2}}{(z_1-z_2)^2} I_{3}^{i_3}+\frac{\varepsilon^{i_1}_{n_1}\cdot \varepsilon^{i_3}_{n_3}}{(z_1-z_3)^2} \,I_{2}^{i_2}+\frac{\varepsilon^{i_2}_{n_2}\cdot \varepsilon^{i_3}_{n_3}}{(z_2-z_3)^2} I_{1}^{i_1}+2\alpha'\,I_1^{i_1}\,I_2^{i_2}\,I_3^{i_3}\Bigg]\,,\label{stripped}
				\end{eqnarray}
				where we have introduced the quantity:
				\begin{eqnarray}
					I_a^{i_a}= \frac{\varepsilon_{n_a}^{i_a} \cdot p_T^{(b)}}{z_a-x_b}+\frac{\varepsilon_{n_a}^{i_a} \cdot p_T^{(c)}}{z_a-x_c}~~;~~a\neq b\neq c\,.
				\end{eqnarray}
				The integrals that appear in such a correlator are deeply related to those defined in Eq.s \eqref{25} and \eqref{26a}. Indeed,   %by writing:
				%\begin{eqnarray}
					%{\cal A}_a= e^{-i\sqrt{2\alpha'} n_a k_a\cdot q} \,\lambda_{n_a,\,i}^a \, {\cal \hat A}^i(a) ~~;~~{\cal B}^{i\,j}_{a,b}=e^{-i\sqrt{2\alpha'}(n_a\kka+n_b\kkb)\cdot q}\,
					%\,{\cal \hat B}^{ij}(a;\,b)\,,\nonumber\\
					%\label{46}
				%\end{eqnarray}
				we can write Eq. \eqref{stripped} in the form:
    \begin{eqnarray}
					\langle V_{A_{-n_1}}^{i_1}(x_1)\,V_{A_{n_2}}^{i_2}(x_2)\,V_{A_{n_3}}^{i_3}(x_3)\rangle &&=\,
					\varepsilon_{(n_1)}^{i_1}\cdot \varepsilon_{(n_2)}^{i_2} {\cal  B}_{1,n_1;2,n_2}\,{\cal  A}^{i_3}_{3,n_3}+\varepsilon_{(n_2)}^{i_2}\cdot \varepsilon_{(n_3)}^{i_3} {\cal  B}_{2,n_2;3,n_3}\,{\cal  A}^{i_1}_{1,n_1}\nonumber\\
					&& +\,\varepsilon_{(n_3)}^{i_3}\cdot \varepsilon_{(n_1)}^{i_1} {\cal  B}_{3,n_3;1,n_1}\,{\cal  A}^{i_2}_{2,n_2}
     +{\cal  A}^{i_1}_{1,n_1}\,{\cal  A}^{i_2}_{2,n_2} \,{\cal  A}^{i_3}_{3,n_3}\,.\nonumber\\
					&&\label{stripped1}
				\end{eqnarray}
				%\begin{eqnarray}
					% \langle V_{A_{-n_1}}^{i_1}(x_1)\,V_{A_{-n_2}}^{i_2}(x_2)\,V_{A_{-n_3}}^{i_3}(x_3)\rangle &&=\,
					%{\cal \hat B}^{i_1\,i_2}(1;2)\,{\cal \hat A}^{i_3}(3)+{\cal \hat  B}^{i_2\,i_3}(2;3)\,{\cal \hat A}^{i_2}(2)\nonumber\\
					%&&+\,{\cal \hat B}^{i_3\,i_1}(3;1)\,{\cal \hat A}^{i_2}(2)+{\cal \hat A}^{i_1}(1)\,{\cal \hat A}^{i_2}(2) \,{\cal \hat A}^{i_3}(3)\,.\nonumber\\
					%&&\label{stripped1}
				%\end{eqnarray}
				We observe that by computing the derivatives of Eq. \eqref{23} with respect to $\lambda_{n_a,\,i_a}^a$, $a=1,\,2,\,3$ and performing the integral over $q$, that gives the delta over the momenta, we get, exactly,  the stripped correlator written in Eq.\eqref{stripped1}. Finally, the integrals can be explicitly evaluated with the use of Eq.\eqref{26},\eqref{27} and \eqref{28}, getting:
				\begin{eqnarray}
					&&C^{123}_{3,n_1,n_2,n_3}(\lambda^{(1)},  k_1,p_1; \lambda^{(2) }, k_2,\,p_2; \lambda^{(3)}, k_3,\,p_3)=\frac{\sqrt{2\alpha'}\lambda_{n_1,\,i_1}^{(1)}\,\lambda_{n_2,i_2}^{(2)}\,\lambda_{n_3, i_3}^{(3)}}{(x_1-x_2)(x_1-x_3)(x_2-x_3)}\,\frac{1}{\Gamma[n_1]\,\Gamma[n_2]\,\Gamma[n_3]}\nonumber\\
					&&\times \Bigg[-(-1)^{n_2+n_3} \,\frac{\varepsilon^{(1)i_1}_{n_1}\cdot \varepsilon^{(2)i_2}_{n_2}\,\varepsilon^{(3)i_3}_{n_3}\cdot p_2}{(n_2+2\alpha' n_1 k_1\cdot p_2)}\,\frac{ \Gamma\left[-\frac{n_2}{2\alpha' k_1\cdot p_2}\right]\Gamma \left[1-2\alpha' n_1 k_1 \cdot p_2\right]\,\Gamma\left[n_3+2\alpha' n_3 k_3\cdot p_1\right]}{\Gamma\left[-n_2 -\frac{n_2}{2\alpha' k_1\cdot p_2}\right]\Gamma\left[1-n_1-2\alpha' n_1 k_1\cdot p_2\right]\Gamma\left[1 +2\alpha' n_3 k_3\cdot p_1\right]}\nonumber\\
					&&-(-1)^{n_1+n_2} \,\frac{\varepsilon^{(1)i_1}_{n_1}\cdot \varepsilon^{(3)i_3}_{n_3}\,\varepsilon^{(2)i_2}_{n_2}\cdot p_1}{(n_1+2\alpha' n_3 k_3\cdot p_1)}\,\frac{ \Gamma\left[-\frac{n_1}{2\alpha' k_3\cdot p_1}\right]\Gamma \left[1-2\alpha' n_3 k_3 \cdot p_1\right]\,\Gamma\left[n_2+2\alpha' n_2 k_2\cdot p_3\right]}{\Gamma\left[-n_1 -\frac{n_1}{2\alpha' k_3\cdot p_1}\right]\Gamma\left[1-n_3-2\alpha' n_3 k_3\cdot p_1\right]\Gamma\left[1 +2\alpha' n_2 k_2\cdot p_3\right]}\nonumber\\
					&&+(-1)^{n_1+n_3}\frac{ \varepsilon^{(2)i_2}_{n_2}\cdot \varepsilon^{(2)i_3}_{n_3}\,\varepsilon^{(1)i_1}_{n_1}\cdot p_2}{(n_3+2\alpha' n_2 k_2\cdot p_3)}\,\frac{ \Gamma\left[-\frac{n_3}{2\alpha' k_2\cdot p_3}\right]\Gamma \left[1-2\alpha' n_2 k_2 \cdot p_3\right]\,\Gamma\left[n_1+2\alpha' n_1 k_1\cdot p_2\right]}{\Gamma\left[-n_3 -\frac{n_3}{2\alpha' k_2\cdot p_3}\right]\Gamma\left[1-n_2-2\alpha' n_2 k_2\cdot p_3\right]\Gamma\left[1 +2\alpha' n_1 k_1\cdot p_2\right]}\nonumber\\
					&&-2\alpha'\,(-1)^{n_1+n_2+n_3} \varepsilon^{(1)i_1}_{n_1}\cdot p_2\,\varepsilon^{(2)i_2}_{n_2}\cdot p_1\,\varepsilon^{(3)i_3}_{n_3}\cdot p_2\,\prod_{a=1}^3\frac{\Gamma \left[ n_a+2\alpha' n_ak_a\cdot p_{a+1}\right]}{\Gamma\left[1+2\alpha' n_ak_a\cdot p_{a+1}\right]}\Bigg]\,.\label{3coherent}
				\end{eqnarray}
				The previous correlator, when specified for the case of three photons, reproduces the three-photon amplitude computed within the covariant formulation of string quantization. Covariant polarizations are introduced through the identification:  $\lambda_\mu=\lambda_i\,\varepsilon_\mu^i$.  Starting from it, it is straightforward to compute the amplitude with one-photon and two DDF-states of the generic level $n$. This amplitude is the sum of two contributions coming from the sum of the two non-inequivalent permutations of the external states. 
				One of these contributions is readily obtained by setting in Eq.\eqref{3coherent} $n_1=1$ for the photon's external leg and $n_2=n_3=n$ for the DDF-legs, resulting in:    
				\begin{eqnarray}
					&&C^{123}_{pDD}= g\,\sqrt{2\alpha'} \,\Bigg\{\frac{\lambda^{(1)}_1\cdot \lambda^{(2)}_n \,\lambda^{(3)}_n\cdot p_2}{\Gamma[n]^2}\,\frac{- \Gamma[-1-2\alpha' n k_2\cdot p_1]\,\Gamma[n+2\alpha' nk_3\cdot p_1]}{\Gamma[-n -2\alpha' n k_2\cdot p_1]\Gamma[1+2\alpha' n k_3\cdot p_1]}\nonumber\\
					&&+\frac{\lambda^{(1)}_1 \cdot \lambda^{(3)}_n \,\varepsilon^{(2)}_n\cdot p_1}{\Gamma[n]^2}\,\frac{ (-1)^n \Gamma[n +2\alpha' k_2\cdot p_3]\Gamma[-1-2\alpha' n k_3\cdot p_1]}{\Gamma[1+2\alpha' n k_2\cdot p_3]\Gamma[-n-2\alpha' n k_3\cdot p_1]}\nonumber\\
					&&-(-1)^n\frac{\lambda_n^{(2)}\cdot \lambda_n^{(3)}\,\lambda^{(1)}\cdot p_2}{n\,\Gamma[n]^2}\frac{\Gamma[1-2\alpha' n k_3\cdot p_2]\,\Gamma[1-2\alpha' n\,k_2\cdot p_3 ]}{\Gamma[1-n -2\alpha'n k_3\cdot p_2 ]\Gamma[1-n -2\alpha' n k_2\cdot p_3]}\nonumber\\
					&&+ 2\alpha'\frac{ \lambda_1^{(1)}\cdot p_2\, \lambda_n^{(2)}\cdot p_1 \,\lambda_n^{(3)}\cdot p_2}{\Gamma[n]^2} \,\frac{ \Gamma[n+2\alpha' n k_2\cdot p_3]\Gamma[n+2\alpha' n k_3\cdot p_1]}{\Gamma[1+2 \alpha' n k_2\cdot p_3]\Gamma[1+2\alpha' n k_3\cdot p_1]}\Bigg\}\,,\label{8.16}
				\end{eqnarray} 
				where we have introduced the gauge coupling constant $g$ and shortly denoted $\lambda= \lambda_i\,\varepsilon^i$. The amplitude can be simplified through the identities\footnote{In the case $n=1$ the ratios of  Gamma-functions are equal to one.}
				\begin{eqnarray}
					\frac{ \Gamma[-1 - 2\alpha' n k_a\cdot p_b]}{\Gamma[-n - 2\alpha' n k_a\cdot p_b]}=(-1)^{n-1}\prod_{l=2}^n(l+2\alpha' n k_a\cdot p_b) ~~;~~\frac{\Gamma[n+2\alpha' n k_a\cdot p_b]}{\Gamma[1+2\alpha' n k_a\cdot p_b]}=\prod_{l=1}^{n-1}(l+2\alpha' n k_a\cdot p_b)\,.\nonumber
				\end{eqnarray} 
				Moreover,  by using the conditions
				$2\alpha' n k_a\cdot  p_b= -n -2\alpha' nk_a \cdot p_1$, with $a,b\neq 1$, 	Eq. \eqref{8.16} can be written only in terms of the photon momentum $p_1$:
				\begin{eqnarray}
					&&C^{123}_{pDD}=\frac{g\sqrt{2\alpha'}}{\Gamma[n]^2} (-1)^n\Bigg\{\lambda^{(1)}_1\cdot \lambda_n^{(2)} \,\lambda_n^{(3)}\cdot p_2\,\prod_{l=2}^n (l+2\alpha' nk_2\cdot p_1)\prod_{l=1}^{n-1}(l+2\alpha' nk_3\cdot p_1)\nonumber\\
					&&+ \,\lambda_1^{(1)}\cdot \lambda_n^{(3)} \,\lambda_n^{(2)}\cdot p_1 \prod_{l=1}^{n-1}(l+2\alpha' nk_2\cdot p_1)\prod_{l=2}^{n}(l+2\alpha' n k_3\cdot p_1)\nonumber\\
					&&-\frac{1}{n} \lambda_n^{(2)}\cdot \lambda_n^{(3)} \lambda_1^{(1)}\cdot p_2\,\prod_{l=0}^{n-1}(l+2\alpha' nk_3\cdot p_1) \prod_{l=0}^{n-1}(l+ 2\alpha' n k_2\cdot p_1)\nonumber\\
					&&-2\alpha'  \lambda^{(1)}_1\cdot p_2 \, \lambda^{(2)}_n \cdot p_1 \lambda^{(3)}_n\cdot p_2 \,\prod_{l=1}^{n-1}(l+2\alpha' n k_2\cdot p_1)\prod_{l=1}^{n-1}(l+2\alpha' n k_3\cdot p_1)\Bigg\}\,.
				\end{eqnarray}  
				The previous on-shell three-point correlator is zero due to the kinematic. To avoid such a limitation we consider the momenta complex.
				By adding to this expression the contribution obtained from swapping the external legs $2$ with $3$, we obtain zero. This leads to the conclusion that the interaction between a photon and two DDF-states of the same level vanishes, mirroring the behavior observed in the three-photon amplitude.
				
				The functional generator of the correlation functions of DDF-states derived  in the previous section allows us to easily obtain the three-point with  two  DDF-states of the generic level $s$ of the  form:
				\begin{eqnarray}
					\lambda_{i_1}\dots \lambda_{i_s} :A_{-1}^{i_i}:\dots :A_{-1}^{i_s}: |  0,p_T\rangle
				\end{eqnarray}
				and a photon. Here,  we have written the symmetric polarization of the generic level $s$ of the DDF-state, in the factorized product of $s$-polarization tensors,  $\lambda_{i_1,\dots \lambda_{i_s}}= \lambda_{i_1}\,\dots \lambda_{i_s}$. The result  can be written as the sum of four contributions that collect terms with different dependence on  the trace of the polarizations of the external higher spin-states $\lambda^{(a)}_{1\,i}\,\lambda^{(a)i}_{1}$ with $a=1,2$ :
				\begin{eqnarray}
					&&C^{123}_{3,s}(\lambda^{(1)},  k_1,p_1; \lambda^{(2) }p_2; \lambda^{(3)}, k_3,\,p_3)= \lambda ^{(1)}_{i_1}\dots \lambda^{(1)}_{i_s} \,\lambda^{(2)}_k\,\lambda^{(3)}_{j_1}\dots \lambda^{(3)}_{j_s}\nonumber\\
					&&\frac{\partial ^s}{\partial_{\lambda_{-1\,i_1}^{(1)}}\dots \partial_{\lambda_{-1\,i_s}^{(1)}}}\frac{\partial}{\partial \lambda_{-1 \,k}^{(2)}}\,\frac{\partial ^s}{\partial_{\lambda_{-1\,j_1}^{(3)}}\dots \partial_{\lambda_{-1\,j_s}^{(3)}}} {\cal W}(\lambda^1,x_1,k_1,p_T^1;\dots)\Bigg|_{\lambda_{-1}^{(a)}=0} \nonumber\\
					&&=\frac{1}{(x_1-x_2)(x_1-x_3)(x_2-x_3)} %-\sqrt{2\alpha'}
					\sum_{a=1}^4{\cal A}_a(\lambda^{(1)},k_1,\,p_1;\lambda^2,\,p_2,\lambda^{(3)},\,k_3,\,p_3)\,. \nonumber\\
				\end{eqnarray}
				The first term collects all the contributions to the correlator without any trace on the external polarizations
				\begin{eqnarray}
					&&{\cal A}_1= -\sqrt{2\alpha'}\,s \,[(s-1)!]^2\sum_{n=0}^s \frac{(\lambda^{(1)}_{1\,i}\varepsilon^{(1)\,i}_{1}\cdot \varepsilon^{(3)\,j}_{1}\lambda^{(3)}_{3\,j})^n}{n![(s-n)!]^2}\nonumber\\
					&&\times \Bigg[s(s-n) \Big(  \lambda^{(1)}_{1\,i}\varepsilon^{(1)\,i}_{1}\cdot \varepsilon^{(2)\,k}_{1}\lambda^{(2)}_{2\,k}\,\lambda^{(3)}_{3\,j}\varepsilon^{(3)\,j}_1\cdot p_2- \lambda^{(3)}_{3\,i}\varepsilon^{(3)\,i}_{1}\cdot \varepsilon^{(2)\,k}_{1}\lambda^{(2)}_{2\,k}\,\lambda^{(1)}_{1\,j}\varepsilon^{(1)\,j}_1\cdot p_2\Big)\nonumber\\
					&&-2\alpha' \, \lambda^{(2)}_k\varepsilon^{(2)\,k}_1 \cdot p_1\,  \lambda^{(1)}_{1\,j}\varepsilon^{(1)\,j}_1\cdot p_2\,\lambda^{(3)}_{3\,j}\varepsilon^{(3)\,j}_1\cdot p_2\Bigg]( -2\alpha' \, \lambda^{(1)}_{1\,j}\varepsilon^{(1)\,j}_1\cdot p_2\,\lambda^{(3)}_{3\,j}\varepsilon^{(3)\,j}_1\cdot p_2)^{s-n-1}\,.\nonumber\\
				\end{eqnarray}
				The second contribution collects terms depending only on the trace of the first polarization tensor $\lambda_{1\,i}^{(1)}\,\lambda_{1}^{(1)\,i}$ 
				\begin{eqnarray}
					&&{\cal A}_1=\frac{(-1)^s}{2}\sqrt{2\alpha'} \sum_{l=1}^{\frac{s_{even}}{2}}\sum_{n=0}^{s-2l}\frac{ s!}{n!(s-n)!}\frac{s!}{l!(s-2l-n)!} \frac{(-1)^n}{2^{l-1}} \left( \lambda^{(1)}_{1\,j}\varepsilon^{(1)\,j}_1\cdot\varepsilon^{(1)\,j}_1\lambda^{(1)}_{1\,j}\,\alpha' k_1\cdot p_3 (1+2\alpha' k_1\cdot p_3)\right)^l\nonumber\\
					&&\times \left(\lambda^{(1)}_{1\,i}\varepsilon^{(1)\,i}_{1}\cdot \varepsilon^{(3)\,j}_{1}\lambda^{(3)}_{3\,j}\right)^n\Big[(s-n-2l) \lambda^{(1)}_{1\,i}\varepsilon^{(1)\,i}_{1}\cdot \varepsilon^{(2)\,k}_{1}\lambda^{(2)}_{2\,k}\lambda^{(3)}_{3\,j}\varepsilon^{(3)\,j}_1\cdot p_2\nonumber\\
					&&+(s-n)\lambda^{(3)}_{3\,i}\varepsilon^{(3)\,i}_{1}\cdot \varepsilon^{(2)\,k}_{1}\lambda^{(2)}_{2\,k}\,\lambda^{(1)}_{1\,j}\varepsilon^{(1)\,j}_1\cdot p_2-2\alpha'  \lambda^{(2)}_k\varepsilon^{(2)\,k}_1 \cdot p_1\,  \lambda^{(1)}_{1\,j}\varepsilon^{(1)\,j}_1\cdot p_2\,\lambda^{(3)}_{3\,j}\varepsilon^{(3)\,j}_1\cdot p_2\Bigg]\nonumber\\
					&&\times ( \sqrt{2\alpha'} \,  \lambda^{(1)}_{1\,j}\varepsilon^{(1)\,j}_1\cdot p_2)^{s-2l-n-1}(\sqrt{2\alpha'}\lambda^{(3)}_{3\,j}\varepsilon^{(3)\,j}_1\cdot p_2)^{s-n-1}\,.\nonumber\\
				\end{eqnarray}
				Here, $s_{even} =s$ for $s$ even, while $s_{even}=s-1$ for $s$-odd. The contribution to the correlator collecting the terms with the trace of the third polarization tensor reads:   
				\begin{eqnarray}
					&&{\cal A}_3=\frac{(-1)^s}{2}\sqrt{2\alpha'} \sum_{q=1}^{\frac{s_{even}}{2}}\sum_{n=0}^{s-2q}\frac{ s!}{n!(s-n)!}\frac{s!}{q!(s-2q-n)!} \frac{(-1)^n}{2^{q-1}} \left( \lambda^{(3)}_{1\,j}\varepsilon^{(3)\,j}_1\cdot\varepsilon^{(3)\,j}_1\lambda^{(3)}_{1\,j}\,\alpha' k_3\cdot p_2 (1+2\alpha' k_3\cdot p_2)\right)^l\nonumber\\
					&&\times \left(\lambda^{(1)}_{1\,i}\varepsilon^{(1)\,i}_{1}\cdot \varepsilon^{(3)\,j}_{1}\lambda^{(3)}_{3\,j}\right)^n \Big[(s-n) \lambda^{(1)}_{1\,i}\varepsilon^{(1)\,i}_{1}\cdot \varepsilon^{(2)\,k}_{1}\lambda^{(2)}_{2\,k}\lambda^{(3)}_{3\,j}\varepsilon^{(3)\,j}_1\cdot p_2\nonumber\\
					&&+(s-n-2q)\lambda^{(3)}_{3\,i}\varepsilon^{(3)\,i}_{1}\cdot \varepsilon^{(2)\,k}_{1}\lambda^{(2)}_{2\,k}\,\lambda^{(1)}_{1\,j}\varepsilon^{(1)\,j}_1\cdot p_2-2\alpha'  \lambda^{(2)}_k\varepsilon^{(2)\,k}_1 \cdot p_1\,  \lambda^{(1)}_{1\,j}\varepsilon^{(1)\,j}_1\cdot p_2\,\lambda^{(3)}_{3\,j}\varepsilon^{(3)\,j}_1\cdot p_2\Big]\nonumber\\
					&&\times( \sqrt{2\alpha'} \,  \lambda^{(1)}_{1\,j}\varepsilon^{(1)\,j}_1\cdot p_2)^{s-n-1}(\sqrt{2\alpha'}\lambda^{(3)}_{3\,j}\varepsilon^{(3)\,j}_1\cdot p_2)^{s-2q-n-1}\,.\nonumber\\
				\end{eqnarray}
				The last contribution collects terms with both traces in the external polarizations:
				\begin{eqnarray}
					&&{\cal A}_4=(-1)^s\sum_{l=1}^{\frac{s_{even}}{2}}\sum_{q=1}^{\frac{s_{even}}{2}}\sum_{n=0}^{s-2min.\{l,\,q\}}\frac{1}{2l!\,2q!\,n!} \frac{s!}{(s-2q -n)!}\frac{s!}{(s-2l-n)!} \frac{(-1)^n}{2^{l-1} 2^{q-1} }\nonumber\\
					&&\times\left( \lambda^{(1)}_{1\,j}\varepsilon^{(1)\,j}_1\cdot\varepsilon^{(1)\,j}_1\lambda^{(1)}_{1\,j}\,\alpha' k_1\cdot p_3 (1+2\alpha' k_1\cdot p_3)\right)^l\left( \lambda^{(3)}_{1\,j}\varepsilon^{(3)\,j}_1\cdot\varepsilon^{(3)\,j}_1\lambda^{(3)}_{1\,j}\,\alpha' k_3\cdot p_2 (1+2\alpha' k_3\cdot p_2)\right)^l\nonumber\\
					&&\times \left(\lambda^{(1)}_{1\,i}\varepsilon^{(1)\,i}_{1}\cdot \varepsilon^{(3)\,j}_{1}\lambda^{(3)}_{3\,j}\right)^n\nonumber\\
					&&\times \big[(s-n-2l) \lambda^{(1)}_{1\,i}\varepsilon^{(1)\,i}_{1}\cdot \varepsilon^{(2)\,k}_{1}\lambda^{(2)}_{2\,k}\lambda^{(3)}_{3\,j}\varepsilon^{(3)\,j}_1\cdot p_2+(s-n-2q)\lambda^{(3)}_{3\,i}\varepsilon^{(3)\,i}_{1}\cdot \varepsilon^{(2)\,k}_{1}\lambda^{(2)}_{2\,k}\,\lambda^{(1)}_{1\,j}\varepsilon^{(1)\,j}_1\cdot p_2\nonumber\\
					&&-2\alpha'  \lambda^{(2)}_k\varepsilon^{(2)\,k}_1 \cdot p_1\,  \lambda^{(1)}_{1\,j}\varepsilon^{(1)\,j}_1\cdot p_2\,\lambda^{(3)}_{3\,j}\varepsilon^{(3)\,j}_1\cdot p_2\Big]( \sqrt{2\alpha'} \,  \lambda^{(1)}_{1\,j}\varepsilon^{(1)\,j}_1\cdot p_2)^{s-n-2l-1}(\sqrt{2\alpha'}\lambda^{(3)}_{3\,j}\varepsilon^{(3)\,j}_1\cdot p_2)^{s-2q-n-1}\,.\nonumber\\
					\label{43.L}
				\end{eqnarray}
				We observe that by choosing the external polarizations squaring to zero $(\lambda_{-1}^{(a)})^2=0$ with $a=1,3$, as the circular polarization introduced in \cite{Gross:2021gsj}, only the first term of the correlator ${\cal A}_1$ survives. In this case, the three points becomes formally identical to the three-point vertex giving the interaction between two spin-$s$ states lying on the leading Regge trajectory and a photon in the bosonic string theory.

		\subsection{From DDF to Covariant Amplitudes} 
		
		In this subsection, having in mind the relationship between DDF and covariant states given in Eq. \eqref{10L}, we explore the connection between DDF and covariant amplitudes for the first massive string levels.

		We start by considering the first excited level $n=2$ in detail. This level can be represented with the DDF operator as:
		
		\begin{eqnarray}
			\left(\lambda_i\,A_{-2}^i+\lambda_{ij} :A^i_{-1}:\,:A_{-1}^j:\right)|0,\,p_T\rangle~.\label{N=2DDF}
		\end{eqnarray}
		It contains $d-2+\frac{(d-2)(d-1)}{2}$ components that coincide with
		the degrees of freedom of a traceless symmetric $d$-dimensional
		tensor. With the use of eq. \eqref{10L} and after some algebra, it can
		be equivalently written in the form:
		\begin{eqnarray}
			\left(T_\mu \alpha^\mu_{-2} +S_{\mu\nu}\,\alpha_{-1}^\mu\,\alpha_{-1}^\nu\right)|0, p\rangle ~~;~~p\equiv p_T-2k~~,~~p^2=-\frac{1}{\alpha'},\label{45L}
		\end{eqnarray}
		where we have introduced the covariant polarizations emerging from the
		DDF representation of the physical states\footnote{The relation
			between the DDF and covariant polarization tensors for the level $N=2$
			is also given in \cite{Skliros:2011si}, this coincides with
			Eq. \eqref{1.12} for traceless polarization tensor $\lambda^i_i=0$.}:
		\begin{eqnarray}
			T_\mu&& \equiv \lambda_i\, \varepsilon_{(2)\mu}^i-\sqrt{\frac{\alpha'}{2}}\lambda_{ij}\,\varepsilon^i_{(1)}\cdot \varepsilon_{(1)}^j\,k_\mu\nonumber\\
			S_{\mu\nu}&&\equiv\lambda_{ij}\,\varepsilon^i_{(1)\rho } \,\varepsilon_{(1)\sigma}^j\left( \delta^\rho_\mu\,\delta^\sigma_\nu  +\alpha'\,\eta^{\rho\sigma}\,k_\mu\,k_\nu\right)-\sqrt{2\alpha'} \lambda_i\left( \varepsilon^i_{(2)\mu} \,k_\nu+\varepsilon^i_{(2)\nu} \,k_\mu\right)\label{1.12}
		\end{eqnarray}
		Here,  we have introduced the  polarization defined in Eq. \eqref{eq:covariant projector}
		which has the property of being  transverse both to $k_\mu$ and the tachyon momentum  $p_T$    \cite{Gross:2021gsj}. 
		It can be readily  verified that the DDF covariant polarizations satisfy Virasoro's conditions as given in Eq.s \eqref{1.5}:
		\begin{eqnarray}
			\sqrt{2\alpha'}p^\nu\,S_{\mu\nu}+T_\mu=0~~;~~S^\mu_{~\mu}+2\sqrt{2\alpha'}p^\mu\,T_\mu=0~,\label{Virasoro2}
		\end{eqnarray}
		and we can associate these polarization tensors with their covariant counterparts.  Consequently, Eq.s \eqref{1.12} provides a  parametrization of the covariant polarizations that arise from the DDF representation of the physical states.
		
		The vertex of the massive state of the level $N=2$ is then  given by:
		\begin{eqnarray}
			V_{N=2}(z) = (\lambda_i :A^{i}_2: +\lambda_{ij}\,:A_{-1}^i:\, :A_{-1}^j:)V_T(p_T,z) ~~;~~S_{ij}=S_{ji}~.\label{48L}
		\end{eqnarray}
		
		By using this vertex, we now evaluate, as example, the correlator of this vertex with two tachyons. Such correlator is easily evaluated by using  the functional generator  \eqref{eq:corr_func_coher_states} specified for case $M=3$ and with the expressions    \eqref{26} and \eqref{27} for the three-point residues appearing in its definition.  In detail:
		\begin{eqnarray}
			C^{123}_3(n=2,\,n=0,\,n=0)=\left(\lambda^i\frac{\partial}{\partial \lambda_{2}^{(1)i}}+\lambda^{ij}\frac{\partial^2}{\partial\lambda_{1}^{(1)i}\partial \lambda_{1}^{(1)j}}\right)A_3(\lambda^{(1)},\lambda^{(2)}\,\lambda^{(3)})\Bigg|_{\lambda_i=0}
		\end{eqnarray}
		The results of  such a calculation turns out to be:
		\begin{eqnarray}
			C^{123}_3(n=2,\,n=0,\,n=0)&&=\alpha' \,\lambda_{ij} \left[ 2\varepsilon_1^i\cdot p_2\, \varepsilon_1^j\cdot p_2+k_1\cdot p_2\,\varepsilon_1^i \cdot\varepsilon_1^j (1+2 \alpha' k_1\cdot p_2) \right]\nonumber\\
			&&- \sqrt{2\alpha'}\,\lambda_i\varepsilon_2^i\cdot p_2 (1+4 \alpha'\,k_1\cdot p_2)~.
		\end{eqnarray}
		By using the relation between the covariant and DDF-polarization as determined in \eqref{1.12} we have:
		\begin{eqnarray}
			C^{123}_3(n=2,\,n=0,\,n=0)= 2\alpha' p_{2\mu}\,p_{2\nu} S^{\mu\nu}- \sqrt{2\alpha'} \,p_{2\mu}\,T^\mu~. \label{9.6}
		\end{eqnarray}
		The scattering amplitude is obtained by summing the cyclically inequivalent correlators, which are derived by exchanging the two tachyon vertices. This corresponds to exchanging $p_2$ and $p_3$ in Eq. \eqref{9.6} getting:
		\begin{eqnarray}
			C^{132}_3(n=2,\,n=0,\,n=0)= 2\alpha' p_{3\mu}\,p_{3\nu} S^{\mu\nu}- \sqrt{2\alpha'} \,p_{3\mu}\,T^\mu~. \label{9.7}
		\end{eqnarray}
		By using the momentum conservation and the Virasoro's conditions, the amplitude turns out to be:
		\begin{eqnarray}
			&&A_3(n=2,\,n=0,\,n=0)=C^{132}_3+C^{123}_3= 4\alpha' p_{2\mu}\,p_{2\nu} S^{\mu\nu}-2\, \sqrt{2\alpha'} \,p_{2\mu}\,T^\mu~. 
		\end{eqnarray}
		The gauge transformation defined in Eq. \eqref{gcov}, transforms to zero, $T_\mu\rightarrow T_\mu'=0$, and one can introduce the symmetric traceless polarization as given in \eqref{A.89} here rewritten:
		\begin{eqnarray}
			S'_{\mu\nu}=S_{\mu\nu}-\sqrt{\frac{\alpha'}{2}}\left(p_\mu\,T_\nu+p_\nu\,T_\mu\right) -\frac{p\cdot T}{\sqrt{2} (d-1)}\Bigg( -5\eta_{\mu\nu} +\alpha' (d-6) p_\mu\,p_\nu\Bigg) ~.\label{A.89}
		\end{eqnarray}
		This polarization tensor is symmetric and traceless as a consequence of the Virasoro's conditions, and   satisfies:
		\begin{eqnarray}
			2\alpha'p_{2}^\mu\,p_{2}^\nu\, S'_{\mu\nu}= -\sqrt{2\alpha'} p_1\cdot T \,\frac{d-26}{d-1} + 2\alpha'p_{2}^\mu\,p_{2}^\nu\, S_{\mu\nu}-\sqrt{2\alpha'} p_2\cdot T~.
		\end{eqnarray}
		where we have used the on-shell condition $p_1\cdot p_2=\frac{1}{2\alpha'}$. In the critical dimensions, the first term is vanishing and, when this identity is used in Eq. \eqref{9.6}, we recover the standard expression for the amplitude which coincides with the first term in Eq. \eqref{9.6} being, in the new gauge, the vector polarization tensor vanishing:
		\begin{eqnarray}
			A_3(n=2,\,=0,\,n=0)= 4\alpha' p_{2\mu}\,p_{2\nu} S'^{\mu\nu}\qquad;\qquad p_1^\mu\,S'_{\mu\nu}=S^{'\mu}_{~\mu}=0~.\label{53L}
		\end{eqnarray}
		
		The DDF operators exhibit an intriguing feature, they transform under the transverse $SO(d-2)$ Lorentz group but yield covariant amplitudes. However, it's worth noting that these amplitudes are expressed in a rather unconventional gauge for the polarization of the external particles.
		A further remark is that the circular polarization of the DDF-states, $\lambda_i=0$  and $\lambda_i^{~i}=0$, provides the following parametrization of the covariant polarization:
		\begin{eqnarray}
			S_{\mu\nu} = \lambda_{ij} \,\varepsilon_{(1)\mu}^i\, \varepsilon_{(1)\nu}^j, \quad T_\mu = 0, \quad S_\mu^{~\mu} = p^\mu S_{\mu\nu} = 0.
		\end{eqnarray}
		This parametrization satisfies the Virasoro conditions, although not all components of the rank-2 symmetric and traceless polarization tensor are activated. It can be considered a consistent truncation of the transverse and symmetric polarization, compatible  with the Virasoro conditions. With this truncated polarization, the amplitude with one DDF-state and two tachyons contains only the first term in Eq. \eqref{9.6}, but with the rank-2 tensor being transverse and traceless. By promoting the truncated polarization to the full polarization tensor $S'_{\mu\nu}$,  one gets the correct result as given in Eq. \eqref{53L}. The same observation applies to the amplitude with two DDF-states and one photon, where the corresponding covariant amplitude coincides with the interaction of two massive spin-2 states lying on the first Regge trajectory, as observed after Eq. \eqref{43.L}.

		Eq. \eqref{9.6} and Eq. \eqref{53L} are two equivalent expressions of the same amplitude. In the former, the external physical state is given in Eq. \eqref{45L}, while in computing the amplitude in Eq. \eqref{53L}, the external on-shell string states lying on the leading Regge trajectory are represented as:
		\begin{eqnarray}
			S'_{\mu\nu} \alpha^\mu_{-1}\,\alpha_{-1}^\nu|0,p\rangle~.\label{54L}
		\end{eqnarray}
		The two different representations of the spin-2 massive state must have  the same second Casimir eigenvalue defined in terms of the square of the Pauli-Lubraski $d-3$ form \cite{Brink:2002zx}:
		\begin{eqnarray}
			W_{\mu_0 \dots \mu_{d-4}} = \frac{1}{2\sqrt{(d-3)!}} \epsilon_{\mu_0\dots \mu_{d-4}\nu_1\nu_2\nu_3}\,\hat{p}^{\nu_1}\,S^{\nu_2\nu_3} ~.
		\end{eqnarray} 
		with $S^{\mu\nu}$ is the string spin operator \cite{Green:1987sp} %defined in appendix Eq. \eqref{StrSpin}
		and $\hat{p}^\mu$ is the momentum operator. The string expression for the second Casimir operator turns out to be \cite{Bekaert:2006py}\footnote{Sometimes, in the literature, this operator is referred to as the quartic Casimir operator.}:
		\begin{eqnarray}
			\hat{C}_2(\mathfrak{iso}(D-1,1)) =\frac{1}{2} \left[- \hat{p}^2\hat{S}_{\mu_1\mu_2}\,\hat{ S}^{\mu_1\mu_2} + \hat{S}_{\sigma_3\sigma_1}\,\hat{p}^{\sigma_1} \, \hat{S}^{\sigma_3\sigma_2}\,\hat{p}_{\sigma_2}\right]~.
		\end{eqnarray}
		The second Casimir of the massive spin-2 representation given in Eq. \eqref{54L} can be  obtained from the expression:
		\begin{eqnarray}
			\langle p,\,0|\alpha_{1}^{\rho_1}\alpha_{1}^{\rho_2}\,{\cal P}_{\rho_1 \rho_2}^{\sigma_1\sigma_2} \,\hat{C}_2(\mathfrak{iso}(D-1,1))\,{\cal P}_{\sigma_1\sigma_2}^{\mu_1\mu_2} \alpha_{-1\mu_1}\,\alpha_{-1\mu_2}|0,p\rangle= 2M^2(d-2)(d-1)(d+1)~,\label{60L}
		\end{eqnarray}
		where ${\cal P}$  is the transverse and traceless projector defined in Eq. \eqref{235L} of Appendix \ref{app:notations}. Eq. \eqref{60L} gives the trace of the second Casimir operator. By dividing this expression by the dimension of the transverse and traceless symmetric representation, i.e.,
		$d_{\tiny \ydiagram{2}}=\frac{1}{2} (d-2)(d+1) $, we obtain the corresponding eigenvalues which  turn to be $4M^2(d-1)$, twice the expected value of a spin-2 state \cite{Firrotta:2024qel}. %the correct value computed in Eq. \eqref{231L}. 
		This overall extra factor two is related to the level of string excitation and can be eliminated by normalizing the sting states with the factor $1/\sqrt{2}$.
		
		To compute the second Casimir of the same spin-2 representation, now given by the string states of the form in Eq. \eqref{45L}, it is convenient to use the Virasoro conditions to rewrite these states in the form:
		\begin{eqnarray}
			S_{\mu\nu} \left[\alpha_{-1}^\mu\,\alpha_{-1}^\nu -\sqrt{\frac{\alpha'}{2}} \left(\hat{p}^\mu \,\alpha_{-2}^\nu +\hat{p}^\nu\,\alpha_{-2}^\mu \right)\right]|0,p\rangle\equiv S_{\mu\nu}\,{\cal \hat A}^{\mu\nu}   |0,p\rangle~.
		\end{eqnarray} 
		The Casimir eigenvalues of this representation are obtained from the following trace:
		\begin{eqnarray}
			&&\langle 0,p| {\cal \hat A}^{\dagger}_{\mu\nu}   \,\hat{C}_2(\mathfrak{iso}(D-1,1)){\cal \hat A}^{\mu\nu}   |0,p\rangle\nonumber\\
			&& =2M^2 (d-2)(d-1) (d+2) -2M^2 (d-2) (d-1)=2M^2 (d-2)(d-1)(d+1)~.\label{58L}
		\end{eqnarray}
		Eq. \eqref{58L} coincides with Eq. \eqref{60L}, showing that they represent  two different realizations of the same massive spin-2 particle.
		
		At the massive level $N=3$, the spectrum of bosonic string theory, as reviewed in Appendix \ref{Covpol}, contains a symmetric three index tensor and a two index antisymmetric tensor. These are described by the following DDF vertex operators:
		\begin{eqnarray}
			V_{N=3}(z)=\left( \lambda_i\,A^i_{-3}(z) + \lambda_{ij} \,:A_{-2}^1:\,:A_{-1}^j:(z) +\lambda_{ijk}\,:A_{-1}^i:\,:A_{-1}^j:\,:A_{-1}^k(z)\right)\,V_T(p_T,\,z)~,
		\end{eqnarray}
		where the DDF operators are defined in Eq. \eqref{8}. The connection with the covariant formulation is carried out similarly to the level $N=2$. Details are given in Appendix \ref{Covpol}; here, we provide only the relation between the covariant and DDF polarizations:
		\begin{eqnarray}
			T_\mu&&= \lambda_i\varepsilon_{(3)\mu}^i
			-\frac{2}{3} \lambda_{ij} \varepsilon_{(2)}^i\cdot \varepsilon^j_{(1)}\sqrt{2\alpha'} k_\mu\nonumber\\
			T_{\rho\nu}&&=-3 \sqrt{2\alpha'}\lambda_i\varepsilon_{(3)\rho}^i\,k_\nu -\frac{3}{2} \sqrt{2\alpha'}\lambda_i\varepsilon_{(3)\nu}^i\,k_\rho+\lambda_{ij}\varepsilon_{(2)\rho}^i\,\varepsilon_{(1)\nu}^j+\lambda_{ij}  \varepsilon_{(2)}^i\cdot \varepsilon_{(1)}^j \,2\, (2\alpha')k_\rho\,k_\nu\nonumber\\
			&&-\sqrt{\frac{\alpha'}{2}}S_{ijk}( \varepsilon_{(1)}^i\cdot\varepsilon_{(1)}^j \varepsilon_{(1)\nu}^k + \varepsilon_{(1)}^i\cdot\varepsilon_{(1)}^k \varepsilon_{(1)\nu}^j+\varepsilon_{(1)}^j\cdot\varepsilon_{(1)}^k \varepsilon_{(1)\nu}^i)k_\rho \nonumber\\
			S_{\mu\nu\sigma}&&=\lambda_{ijk} \varepsilon^i_{(1)\sigma}\varepsilon^j_{(1)\nu}\varepsilon^k_{(1)\mu}+(2\alpha')\frac{3}{2} \left( \lambda_i\varepsilon_{(3)\sigma}^ik_\nu k_\mu+\lambda_i\varepsilon_{(3)\nu}^ik_\sigma k_\mu+\lambda_i\varepsilon_{(3)\mu}^ik_\nu k_\sigma \right)\nonumber\\
			&&-2\sqrt{2\alpha'}\frac{1}{6}\left(\lambda_{ij} \varepsilon^i_{(2)\sigma} \, 
			k_\nu\,\varepsilon^j_{(1)\mu}+\lambda_{ij} \varepsilon^i_{(2)\sigma} \, 
			k_\mu\,\varepsilon^j_{(1)\nu}+\lambda_{ij} \varepsilon^i_{(2)\nu} \, 
			k_\sigma\,\varepsilon^j_{(1)\mu}+\lambda_{ij} \varepsilon^i_{(2)\nu} \, 
			k_\mu\,\varepsilon^j_{(1)\sigma} \right.\nonumber\\ 
			&&\left.+\lambda_{ij} \varepsilon^i_{(2)\mu}\,k_\nu\,\varepsilon^j_{(1)\sigma}+\lambda_{ij} \varepsilon^i_{(2)\mu} \, 
			k_\sigma\,\varepsilon^j_{(1)\nu}\right)-\frac{4}{3} \lambda_{ij} \varepsilon_{(2)}^i \cdot \varepsilon_{(1)}^j(2\alpha')^\frac{3}{2}k_\sigma\,k_\nu\,k_\mu\nonumber\\
			&& +\alpha'\frac{1}{3}\lambda_{ijk}\left[ (\varepsilon_{(1)}^i\cdot \varepsilon_{(1)}^j\varepsilon_{(1)\sigma}^k+\varepsilon_{(1)}^i\cdot \varepsilon_{(1)}^k\varepsilon_{(1)\sigma}^j+\varepsilon_{(1)}^j\cdot \varepsilon_{(1)}^k\varepsilon_{(1)\sigma}^i)k_\nu k_\mu\right.\nonumber\\
			&&+(\varepsilon_{(1)}^i\cdot \varepsilon_{(1)}^j\varepsilon_{(1)\nu}^k+\varepsilon_{(1)}^i\cdot \varepsilon_{(1)}^k\varepsilon_{(1)\nu}^j+\varepsilon_{(1)}^j\cdot \varepsilon_{(1)}^k\varepsilon_{(1)\nu}^i)k_\sigma k_\mu\nonumber\\
			&&\left. +(\varepsilon_{(1)}^i\cdot \varepsilon_{(1)}^j\varepsilon_{(1)\mu}^k+\varepsilon_{(1)}^i\cdot \varepsilon_{(1)}^k\varepsilon_{(1)\mu}^j+\varepsilon_{(1)}^j\cdot \varepsilon_{(1)}^k\varepsilon_{(1)\mu}^i)k_\nu k_\sigma\right]~,\label{63L}
		\end{eqnarray}  
		where in the covariant formulation of the string, we have parametrized  the generic state of the level $N=3$ as follows:
		\begin{eqnarray}
			\left(T_\mu \alpha_{-3}^\mu+ T_{\mu\nu} \,\alpha_{-2}^\mu\,\alpha_{-1}^\nu +S_{\mu\nu\rho}\,\alpha_{-1}^\mu\,\alpha_{-1}^\nu\,\alpha_{-1}^\rho\right)|0,\,p\rangle~.
		\end{eqnarray} 
		The DDF-correlator between the states of this level and two tachyons is\footnote{The amplitude is obtained by adding to this expression the cyclically inequivalent ordering, where the two tachyon vertices are exchanged in the correlator.} :
		\begin{eqnarray}
			&&C^{123}_3(p_1,\lambda_i^{(1)},\lambda_{ij}^{(1)},\,\lambda_{ijk}^{(1)},p_2,\,p_3)=\int \frac{\prod_{i=1}^3\,dz_i}{dV_{abc}}\,\langle V_{N=3}(z_1)\,V_T(z_2,\,p_2)\,V_T(z_3,\,p_3)\rangle\nonumber\\
			&&= \sqrt{2\alpha'}\,\lambda_i^{(1)} \varepsilon_{(3)}^i\cdot p_2\, (1+3\alpha' k_1\cdot p_2)(1+6\alpha' k_1\cdot p_2)-2\alpha'\,\lambda_{ij}^{(1)}\,\varepsilon_{(1)}^i\,\cdot p_2 \varepsilon_{(2)}^j\cdot p_2 (1+4\alpha' k_1\cdot p_2)\nonumber\\
			&&-\frac{2}{3} \lambda_{ij}^{(1)}\varepsilon_{(1)} ^i\cdot \varepsilon_{(2)}^j (2\alpha' k_1\cdot p_2)(1+2\alpha' k_1\cdot p_2)(1+4 \alpha' k_1\cdot p_2)+(2\alpha' )^{\frac{3}{2}} \,\lambda_{ijk}^{(1)} \varepsilon_{(1)}^1\cdot p_2\varepsilon_{(1)}^j\cdot p_2 \varepsilon_{(1)}^k\cdot p_2\nonumber\\
			&&+(2\alpha')^{\frac{3}{2}}\,\frac{3}{2} \lambda_{ijk}^{(1)} \,\varepsilon_{(1)}^i\cdot p_2 \, \varepsilon_{(1)}^i\cdot \varepsilon_{(1)}^j\, (1+2\alpha' k_1\cdot p_2) k_1\cdot p_2 ~. 
		\end{eqnarray}
		By using the identification between DDF and covariant polarization the previous correlator can be written as:
		\begin{eqnarray}
			&&C^{123}_3(p_1,\,p_2,\,p_3)= \sqrt{2\alpha'} p_2^\mu \, T_\mu - 2\alpha' p_2^\mu\,p_2^\nu \, T_{\mu\nu} + (2\alpha')^{\frac{3}{2}}\,p_2^\mu \,p_2^\nu \,p_2^\rho \,S_{\mu\nu\rho}~.\label{65L}
		\end{eqnarray}
		The scattering amplitude is obtained by adding to the previous expression 
		the cyclically not equivalent correlator:
		\begin{eqnarray}
			&&C^{132}_3(p_1,\,p_2,\,p_3)= \sqrt{2\alpha'} p_3^\mu \, T_\mu - 2\alpha' p_3^\mu\,p_3^\nu \, T_{\mu\nu} + (2\alpha')^{\frac{3}{2}}\,p_3^\mu \,p_3^\nu \,p_3^\rho \,S_{\mu\nu\rho}~.\label{65L2}
		\end{eqnarray}   
		The Virasoro conditions impose the following constraints on the covariant polarizations
		\begin{eqnarray}
			&&3T_\mu+\sqrt{2\alpha'} p^\nu T_{\mu\nu}=0~~;~~T_{\mu\nu}+T_{\nu\mu}+ 3\sqrt{2\alpha'} S_{\mu\nu\rho}p^\rho=0\nonumber\\
			&&2\sqrt{2\alpha'} p^\mu T_{\mu\nu} + 3 T_\nu + 3 \eta^{\mu\rho} S_{\mu\rho\nu}=0~~;~~3\sqrt{2\alpha'}p^\mu T_\mu+ 2\eta^{\mu\nu}T_{\mu\nu}=0 ~.
		\end{eqnarray}
		By using the momentum conservation and the previous constraints, we can rewrite Eq. \eqref{65L2} in the form:
		\begin{eqnarray}
			C_3^{132}=-C_3^{123}%\sqrt{2\alpha'} \,p_2^\mu\,T_\mu+ 2\alpha'\,p_2^\mu\,p_2^\nu \,T_{\mu\nu}-(2\alpha')^{\frac{3}{2}} \,p_2^\mu\,p_2^\nu\,p_2^\rho\,S_{\mu\nu\rho}
		\end{eqnarray}
		The scattering amplitude turns out to be:
		\begin{eqnarray}
			A_3(p_1,\lambda_i^{(1)},\lambda_{ij}^{(1)},\,\lambda_{ijk}^{(1)},p_2,\,p_3)=C_3^{123}+C_3^{132}=0%2\sqrt{2\alpha'} \,p_2^\mu\,T_\mu  %+4 \alpha'\,p_2^\mu\,p_2^\nu \,T_{\mu\nu}
		\end{eqnarray}
	We finally extend the analyses to the level $N=4$ where, as  recalled in appendix  \ref{Covpol}, the spectrum contains a spin-2 particle, a scalar, a symmetric, transverse and  traceless rank-2 tensor and a rank-2,1 transverse tensor. These states, in the covariant formulation of the string theory, are collected by the following state:
			\begin{eqnarray}
				\left(\xi_\mu\,\alpha_{-4}^\mu+ \xi_{\mu\nu}\,\alpha_{-3}^\mu\,\alpha_{-1}^\nu\, +S_{\mu\nu}\,\alpha_{-2}^\mu\,\alpha_{-2}^\nu + \xi_{\mu\nu\rho}\,\alpha_{-2}^\mu\,\alpha_{-1}^\nu\,\alpha_{-1}^\rho +S_{\mu\nu\rho\sigma}\,\alpha_{-1}^\mu\,\alpha_{-1}^\nu\,\alpha_{-1}^\rho\,\alpha_{-1}^\sigma\right)|0,\,p\rangle~.\label{69L}
			\end{eqnarray} 
			where the polarizations satisfy the Virasoro's constraints:
			\begin{eqnarray}
				&&4\eta^{\mu\nu}\,S_{\mu\nu} + 3 \eta^{\mu\nu} \,\xi_{\mu\nu} + 4\sqrt{2\alpha'}\,p^\mu\xi_\mu=0~~ ;~~2\eta^{\mu\nu}(\xi_{\mu\nu\rho} +\xi_{\mu\rho\nu}) + 3 \sqrt{2\alpha'}p^\mu\xi_{\mu\rho}+ 4\,\xi_\rho=0 \nonumber\\
				&&\xi_{\mu\nu\rho}\,\eta^{\nu\rho} + 4 \sqrt{2\alpha'}p^\nu \,S_{\nu\mu}+ 4 \xi_\mu=0~~;~~6 S_{\mu\nu\rho\gamma}\,\eta^{\rho\gamma} +\frac{3}{2} (\xi_{\mu\nu} +\xi_{\nu\mu}) + 2\sqrt{2\alpha'} p^\rho\,\xi_{\rho\mu\nu}=0\nonumber\\
				&&\sqrt{2\alpha'} \,p^\nu\xi_{\mu\nu} + 4\xi_\nu=0~~;~~2\sqrt{2\alpha'} p^\rho \,\xi_{\mu\nu\rho} + 4 S_{\mu\nu} + 3 \xi_{\mu\nu}=0 \nonumber\\
				&&4\sqrt{2\alpha'} \,p^\gamma  S_{\mu\nu\rho\gamma}+\frac{2}{3!}\left( \xi_{\mu\nu\rho} + {\rm perm.}\right)=0~.
			\end{eqnarray}
			Th DDF-vertex associated with those states is
			\begin{eqnarray}
				V_{N=4}(z)=&&\left[\lambda_i :A^i_{-4}(z) +\lambda_{ij}  :A_{-3}^i:\,:A_{-1}^j:(z)+S_{ij} :A_{-2}^i:
				:A_{-2}^j:(z)\right.\nonumber\\
				&&\left.+\lambda_{ijk} \,:A_{-2}^i:\,:A_{-1}^j:\,:A_{-1}^k:(z)+S_{ijkl}\,:A_{-1}^i:\,:A_{-1}^j:\,:A_{-1}^k:\,::A_{-1}^l:\right]\,V_T(p,z)~.\nonumber\\
			\end{eqnarray}
			As for the lower string levels we compute the interaction between this vertex operator and two tachyons:
			\begin{eqnarray}
				C^{123}_{4,\,T\,T}=\int \frac{\prod_{i=1}^3dz_i}{dV_{abc}}\,\langle V_{N=4}(z_1) \,V_T(p_2,z_2)\,V_T(p_3,z_3)\rangle  =\sum_{i=1}^5 A_i^{(3)}~.
			\end{eqnarray}
			It is the sum of five terms one for each DDF-state in the following evaluated:
			\begin{eqnarray}
				A_1^{(3)}&&=-\sqrt{2\alpha'} \lambda_i^{(1)}\varepsilon_{(4)}^i\cdot p_2\,\frac{\Gamma[4+8 \alpha'\,k_1\cdot p_2]}{3!\,(8\alpha' k_1\cdot p_2)!}\nonumber\\
				A_2^{(3)}&&= 2\alpha' \,( 1+3\alpha' k_1\cdot p_2) (1+6\alpha' k_1\cdot p_2) \lambda_{ij}^{(1)}\left[ \varepsilon_{(1)}^i\cdot p_2 \varepsilon_{(3)}^j\cdot p_2 + \frac{3}{4} \,k_1\cdot p_2\varepsilon_{(1)}^i\cdot \varepsilon_{(3)}^j (1+2 \alpha' k_1\cdot p_2)\right]\nonumber\\
				A_3^{(3)}&&= 2\alpha'  \,(1+4\alpha' k_1\cdot p_2)^2\,S_{ij}^{(1)}\left[\varepsilon_{(2)}^i\cdot p_2\varepsilon_{(2)}^j\cdot p_2+k_1\cdot p_2 \varepsilon_{(2)}^i\cdot \varepsilon_{(2)}^j (1+2\alpha' k_1\cdot p_2)\right]\nonumber\\
				A_4^{(3)}&&=-(2\alpha')^{\frac{3}{2}}\frac{(1+4 \alpha' k_1\cdot p_2)}{3!}\lambda_{ijk}^{(1)}\left[ 6\,\varepsilon_{(2)}^i\cdot p_2 \,\varepsilon_{(1)}^j\cdot p_2\,\varepsilon_{(1)}^k \cdot p_2+3 \,\varepsilon_{(2)}^i\cdot p_2 \, \varepsilon_{(1)}^j\cdot \varepsilon_{(1)}^k\,k_1\cdot p_2\, (1+ 2\alpha' k_1\cdot p_2)\right. \nonumber\\
				&&\left.+ 8\, \varepsilon_{(2)}^i\cdot \varepsilon_{(1)}^j\,\varepsilon_{(1)}^k\cdot p_2\,k_1\cdot p_2 (1+2\alpha' k_1\cdot p_2)\right]  \nonumber\\
				A_5^{(3)}&&= (2\alpha')^2S_{ijkl}^{(1)}\left[ \varepsilon_{(1)}^i\cdot p_2\,\varepsilon_{(1)}^j\cdot p_2\,\varepsilon_{(1)}^k\cdot p_2\, \varepsilon_{(1)}^l\cdot p_2+3\varepsilon_{(1)}^i\cdot p_2\,\varepsilon_{(1)}^j\cdot p_2 \varepsilon_{(1)}^k\cdot \varepsilon_{(1)}^l\,k_1\cdot p_2(1+2\alpha' k_1\cdot p_2)\right.\nonumber\\
				&&\left.+ \frac{3}{4} \varepsilon_{(1)}^i\cdot \varepsilon_{(1)}^j\,\varepsilon_{(1)}^l\cdot \varepsilon_{(1)}^k\,(k_1\cdot p_2)^2(1+2\alpha' k_1\cdot p_2)^2\right]~.
				\label{87l}
			\end{eqnarray}
			The expression of the amplitude is involved but when expressed in terms of the covariant polarizations becomes very compact,  being:
			\begin{eqnarray}
				C_{4,TT}^{123}
				=
				-\sqrt{2\alpha'} \,\xi_\mu p_2^\mu +2\alpha'
				\,(\xi_{\mu\nu}+S_{\mu\nu}) \, p_2^\mu\,p_2^\nu
				-
				(2\alpha')^{\frac{3}{2}}\xi_{\mu\nu\rho}\,p_2^\mu\,p_2^\nu\,p_2^\rho
				+
				(2\alpha')^2
				S_{\mu\nu\rho\sigma} p_2^\mu\,p_2^\nu\,p_2^\rho\,p_2^\sigma~.
				\label{88l}
			\end{eqnarray}
			Here,  we have used the relation between the DDF and covariant polarizations derived in app. \ref{Covpol}. The full amplitude is obtained by adding to Eq. \eqref{88l} the contribution from exchanging $p_2$ with $p_3$. However, using the identity $2\alpha' k_1 \cdot p_1 = 1$, it is straightforward to verify that Eqs. \eqref{87l} remain invariant under this exchange. We thus conclude that $C_{4,TT}^{132} = C_{4,TT}^{123}$ and:
			\begin{eqnarray}
				A_{4TT}(p_1,\,p_2,\,p_3)= 2\,C_{4,TT}^{123}~.
			\end{eqnarray}

		\newcommand{\dperp}{{d_\perp}}
		
		%%%%%%%%%%%%%%%%%%%%%%%%%%%%%%%%
		%%%Framed DDF examples 3 p.t%%%
		\section{Three point Framed DDF Scattering Amplitudes with massive scalars and spin 2}
		
		\label{sec:3pt_scattering}
		In this section we use the correlator of the coherent framed DDF
		states discussed in the previous sections to compute some three point
		bosonic string amplitudes in the framed DDF formalism
		\cite{Firrotta:2024qel} involving %DDF scalars of the excited string
		levels $N=4$ and $N=6$ and spin 2 symmetric tensors
		\cite{Pesando:2024lqa}.
		In the next section, we compute some four point amplitudes.
		We use the notation we denote by $\ket{S_N}$ to denote the scalar at
		level $N$ (i.e., having mass $M_N^2 = (N-1)/\ap$) and massive DDF
		state. The tachyon state is denoted by $|T\rangle$.
		%%%%%%%%%%%%
		\subsection{Three point Scattering Amplitude for  $\ket{S_4} \rightarrow 2\ket{T}$}
		
		The most general expression for a physical state of the level $N=4$ describing a $SO(d-2)$ scalar particle, is 
		\eq{
			\ket{S_4} 
			=
			\left(
			c_1 \sum_{i=1}^{d-2} \uA^i_{-1}\uA^i_{-3} + 
			c_2 \sum_{j=1}^{d-2} \uA^j_{-2}\uA^j_{-2} 
			+ 
			c_3 \sum_{l=1}^{d-2}\uA^l_{-1}\uA^l_{-1} \sum_{k=1}^{d-2}\uA^k_{-1}\uA^k_{-1}
			\right)\ket{k_T},
			\label{eq:S4expr}
		}
		where the summations run over the transverse directions and $c_1$,
		$c_2$ and $c_3$ are arbitrary constants.
		For generic values of these constants the state in
		Eq. \eqref{eq:S4expr} describes a scalar for $SO(d-2)$ rotations but
		it is actually generically a tensor for the $d$-dimensional Lorentz
		group $SO(d-1,1)$.
		In the following we will fix the values of the constants $c_i/c_1$, with
		$i=\,2,\,3$ to describe a $SO(d-1,1)$-scalar by computing three point
		amplitudes involving $|S_4\rangle$ and two tachyons.
		In order to do so we require that such an on shell amplitude becomes a
		number.
		This requirement follows from the observation that the only Lorentz
		invariant amplitude one can write for a three scalar scattering in the
		momentum space is of the form $f(\alpha' \up_a \cdot \up_b)$, where
		$a,b \in \{1,2,3\}$.
		From the three point kinematic along with the on shell conditions we
		have $\alpha' \,\up_a \cdot \up_b$ are just a number being for a level
		$N=4$ massive state and two tachyons:
		\begin{eqnarray}
			\alpha' \,\up_1\cdot \up_2
			=
			\alpha'\,\up_1\cdot \up_3 =\frac {3}{2}
			~~;~~
			\alpha' \,\up_2\cdot \up_3= -\frac{5}{2}
			.
		\end{eqnarray}
		
		Recalling the expression of the $M$ point correlator for the coherent
		SDS vertex operators $\cS(x; \{\lambda_{n\,i}\},\up)$ in the framed
		DDF formalism
		\eq{ \mathcal{W}_M
			&\equiv \left\langle
			\prod_{a=1}^M\cS(x_a; \lambda^a_{n},\up_{T}^a)\right
			\rangle
			=
			\exp\Bigg\{
			\sum_{i=1}^{d-2}
			\Bigg(\sum_a \sum_{n \in \Z}\lambda^a_{n\, i}
			{\tilde {\oldI}}^{i}_{a, n}(\{x\};\, \{\up^i \};\, \{\up^+ \})
			\nonumber \\
			+&~ \frac{1}{2}
			\sum_{a,b}\sum_{n,m\in \mathbb{Z}}\lambda^a_{n\, i}\lambda^b_{m\, i}
			\oldJ_{a,n;\, b, m}\left(\{x\},\, \{\up^+\};\, x_b,\, \up^b \right)
			\Bigg)\Bigg\}
			\times
			\prod_{b=1}^{M-1} \prod_{a=b+1}^M x_{b a}^{\dap \up_{T}^b \cdot  \up_{T}^a }
			\times
			\delta^d\left(\sum_{a=1}^M \up^{a}\right),
			\label{eq:cohcorrsumm}
		}
		the ordered $1 2 3$ correlator for the process $\ket{S_4} \rightarrow
		2 \ket{T}$ is obtained by acting on \eqref{eq:cohcorrsumm} the
		differential operator (with respect to $\{\lambda^i\}$) corresponding
		to
		\eqref{eq:S4expr} \eq{ \cD \left(\{\lambda^a_{n\, i}\};
			\ket{S_4}\right) :=
			\left( c_1 \sum_i\del_{\lambda^a_{1\, i}}\del_{\lambda^a_{3\, i}}
			+ c_2 \sum_j\del_{\lambda^a_{2\, j}}\del_{\lambda^a_{2\, j}}
			+ c_3 \sum_l\del_{\lambda^a_{1\, l}}\del_{\lambda^a_{1\,
					l}}\sum_k\del_{\lambda^a_{1\, k}}\del_{\lambda^a_{1\, k}}\right)
		}
		Here in the expression of the ${\underline{\cal  A}}$ integrals
		we have extracted the exponential factor, as in Eq. \eqref{25}, which
		has been used to obtain the delta function on the momenta.
		Therefore the correlator, stripped from the momentum conserving delta
		function, for the $1 2 3$ ordering $\cC^{\ket{S_4}}_{123}$ of the
		(Chan-Paton) orderings is given by
		\eq{
			\cD \left(\{\lambda^{a=1}_{n\, i}\}; \ket{S_4}\right)\cW_3
			&=
			\sum_i c_1 \left({{\oldI}}^i_{1,1} {{\oldI}}^i_{1,3} +
			\oldJ_{1,1;1,3}\right)
			+ \sum_i c_2\left({{\oldI}}^i_{1,2}{{\oldI}}^i_{1,2}
			+ \oldJ_{1,2;1,2}\right)
			\nonumber \\
			+ ~c_3 \Bigg(\sum_{l, k}
			{{\oldI}}^l_{1,1}{{\oldI}}^l_{1,1}{
				{\oldI}}^k_{1,1}{{\oldI}}^k_{1,1}
			+ \sum_l&
			{{\oldI}}^l_{1,1}{{\oldI}}^l_{1,1} \oldJ_{1,1;1,1}
			+ \sum_i\,
			(\dperp+2)
			\oldJ_{1,1;1,1}{{\oldI}}^i_{1,1}{
				{\oldI}}^i_{1,1}
			+ \dperp (\dperp+2)
			\oldJ_{1,1;1,1}^2\Bigg),
			\label{eq:corrS4TT}
		}
		where
		$\dperp= d-2= 24$ is the number of transverse dimensions
		and
		${{\oldI}}^i_{a,n}
		\equiv {  {\oldI}}^i_{a, n}(\{x\}; \{\up^i\}; \{\up^+\})$,
		$\oldJ_{a,n; b,m} \equiv \oldJ_{a, n;\, b, m}(\{x\};\, \{\up^+ \})$.

		For the on shell three point correlator
		in critical dimension
		the $x_{ab}$ dependence (from the Koba-Nielsen terms and the $\oldI,\oldJ$ integrals) cancels out on-shell and is not included in the formula above for clarity.
		Now for $M=3$ the integrals are given in Eq.\eqref{26} and \eqref{27}, here we quote again their 
		explicit expressions
		\eq{
			\oldI^i_{a=1,n} &= \left(\ualpha^i_{a+1} - \rho_{a+1\,a}\ualpha^i_{a}\right)\holdI_{a,n}, 
			\nonumber \\
			\holdI_{a=1,n} = - &\frac{1}{(n-1)!} \prod_{k=1}^{n-1}( n \rho_{a+1\, a} +k)\times
			\left( \frac{ - x_{2 3} }{ x_{1 2} x_{1 3} } \right)^n
			\,
			\prod_{b=a+1}^{3} (x_{a b} )^{ -n \rho_{b a} } % \frac{\up^+_{T \sN s} }{\up^+_{T \sN r}} }
		,
	}
and
\eq{
\oldJ_{a,n;a,m} = \frac{mn}{m+n}\rho_{a+1~a}(\rho_{a+1~a}+1)\holdI_{a,n}\holdI_{a,m}
.
}
For further simplification we choose the rest frame of the massive particle.

In particular for a three particles scattering process one can always rotate to a two-dimensional spatial plane (the case we shall consider below) characterized by 
\eq{
	&\up_{1} = (m,\textbf{0}_{d-1})~~;~
	&\up_{2} = -\left(\frac{m}{2},P\cos\theta,\textbf{0}_{d-3},P\sin\theta\right)~~;~
	&\up_{3} = -\left(\frac{m}{2},-P\cos\theta,\textbf{0}_{d-3},-P\sin\theta\right),
	\label{eq:3ptkin}
}
where $\textbf{0}_n \equiv (0,0,..)$ has 
$n$ components and
\eq{
	&P = |\vec \up_{2}| 
	= |\vec \up_{3}| = \sqrt{m^2/4 - m_T^2}
	;~~
	m^2 = M_4^2 = 3/\ap 
	\nonumber \\
	\implies 
	\up^+_2 = -\frac{1}{\sqrt{2}}&\left(\frac{m}{2} + P\sin\theta\right),~\up^+_3 = -\frac{1}{\sqrt{2}}\left(\frac{m}{2} - P\sin\theta\right),~\up^2_2 = -P \cos\theta = - \up^2_3
	.
}
with $m_T^2= -1/\alpha'$ the mass of the tachyon.

Finally we add to the amplitude \eqref{eq:corrS4TT} above the other Chan-Paton ordering $\cC^{\ket{S_4}}_{132}$ 
%\RA{
	obtained by exchanging the
	%} 
particles 2 and 3 to get the following total amplitude\footnote{ We notice that 
	however $\cC^{\ket{S_4}}_{132}=\cC^{\ket{S_4}}_{123}$.},
\eq{
	\cM_{\ket{S_4}\rightarrow 2\ket{T}}
	=& \cC^{\ket{S_4}}_{123}+\cC^{\ket{S_4}}_{132}
	\nonumber\\  
	=&
	\frac{1}{4}\Bigg[8 c_2(1+2 \rho_{21})^2[2\ap (\up^\perp)^2 + \dperp
	\rho_{21}(1+\rho_{21})]
	\nonumber \\
	&+
	c_1[2+9\rho_{21}(1+\rho_{21})]
	[8\ap (\up^\perp)^2 + 3 \dperp \rho_{21}(1+\rho_{21})]
	\nonumber \\
	&+
	2c_3[16(\ap)^2 (\up^\perp)^4 + 8\ap(\dperp+2)
	(\up^\perp)^2\rho_{21}(1+\rho_{21}) +
	\dperp(\dperp+2)\rho_{21}^2(1+\rho_{21})^2]
	\Bigg]
	\label{eq:ampS4TT}
	,
}
where, $ (\up^\perp_{2})^2 = \sum_{i=1}^{d-2}(\up^i_{2})^2$ is the modulus of the transverse momentum.
Substituting the kinematics we obtain the simplified expression
\eq{
	\cM_{\ket{S_4}\rightarrow 2\ket{T}} &= \frac{1}{4608} 
	\Big[
	\cos 4 \theta \left(-10584 c_1 + 1323 \dperp c_1 - 
	9408 c_2 + 1568 \dperp c_2 + 
	9408 c_3 - 2156 \dperp c_3 + 
	98 \dperp^2 c_3\right) 
	\nonumber \\
	+ &     
	\cos 2 \theta \left(-2016 c_1 - 2772 \dperp c_1 - 
	3584 \dperp c_2 + 48384 c_3 - 
	4144 \dperp c_3 - 56 \dperp^2 c_3\right) 
	\nonumber \\
	+& (8568 c_1 + 1665 \dperp c_1 + 
	9408 c_2 + 2016 \dperp c_2 + 
	38976 c_3 - 1476 \dperp c_3 + 
	102 \dperp^2 c_3)\Big].
	\nonumber \\
	%  &= -4
	\label{eq:A4TT}
}
The requirement that the amplitude is a number fixes, in the critical dimensions, the values:
\begin{eqnarray}
	\frac{c_2}{c_1}=-\frac{7}{10} ~~;~~\frac{c_3}{c_1}=-\frac{1}{10}
	,
\end{eqnarray}
so that we get
\begin{eqnarray}
	\cM_{\ket{S_4}\rightarrow 2\ket{T}}= \frac{2 \,c_1}{5}
	.
\end{eqnarray}
This choice of coefficients is consistent with the values $c_1 = -10,\, c_2 = 7, \,c_3 = 1$ 
	necessary to describe the scalar at the level $N=4$ in
	Eq. \eqref{eq:S4expr} (upto rescaling) \cite{Pesando:2024lqa,Markou:2023ffh,Basile:2024uxn}.

	%%%%%%%%%%%%%%%%%%%%%%%%%%%%%%%%%%%%%%%%%%%%%%
	
	%%%%%%%%%%%%%%%%%%%%%%%%%%%%%%%%%%%%%%%%%%%%%%
	\subsection{Three-point Scattering Amplitude for $\ket{S_6} \rightarrow 2\ket{T}$}
	The general transverse scalar at level $N=6$ is given by,
	\eq{
		\ket{S_6} =
		\Big(c_{1131}&\, \left(1 , 1\right)\,\left(3 , 1\right) +c_{1122}\, \left(1 , 1\right)\,\left(2 , 2\right) +c_{2121}\, \left(2 , 1\right)^2 + b_{1,1,1,1}\, \left(1 , 1\right)^3 
		\nonumber\\
		+c_{51}&\, \left(5 , 1\right) + c_{42}\, \left(4 , 2\right) + c_{33} \left(3 , 3\right)\Big)\ket{\up_{T}},
	}
	where we have used the compact notation $(m,n) := \sum_{i=1}^{d-2} \uA^i_{-m}\uA^i_{-n}$.
	Upto rescaling, $\ket{S_6}$ is a Lorentz scalar only for 
	$c_{1131}=1,
	c_{1122}=-\frac{9}{14},
	c_{2121}=-\frac{2}{7},
	b_{1111}=-\frac{5}{84},
	c_{33}=-\frac{10}{3},
	c_{42}=4,
	c_{51}=-\frac{2}{7}$, 
	and in critical dimension $\dperp=24$.
	Following the exact procedure as the previous case the corresponding differential operator
	\eq{
		\cD \left(\{\lambda^{a=1}_{n\,i}\}; \ket{S_6}\right)
		:= \
		\Bigg(c_{1131} \sum_i
		\del_{\lambda^1_{1\,i}}\del_{\lambda^1_{1\,i}}\sum_j
		\del_{\lambda^1_{3\,j}}\del_{\lambda^1_{1\,j}}
		+ c_{1122}\sum_i
		\del_{\lambda^1_{1\,i}}\del_{\lambda^1_{1\,i}}\sum_j
		\del_{\lambda^1_{2\,j}}\del_{\lambda^2_{2\,j}}
		\nonumber\\
		+~c_{2121}\sum_i \del_{\lambda^1_{2\,i}}\del_{\lambda^1_{1\,i}}\sum_j
		\del_{\lambda^1_{2\,j}}\del_{\lambda^1_{1\,j}}
		+ b_{1111}\sum_i \del_{\lambda^1_{1\,i}}\del_{\lambda^1_{1\,i}}\sum_j \del_{\lambda^1_{1\,j}}\del_{\lambda^1_{1\,j}}\sum_k \del_{\lambda^1_{1\,k}}\del_{\lambda^1_{1\,k}}
		\nonumber\\
		+ c_{51} \sum_i \del_{\lambda^1_{5\,i}}\del_{\lambda^1_{1\,i}}
		+ c_{42} \sum_i \del_{\lambda^1_{4\,i}}\del_{\lambda^1_{2\,i}}
		+ c_{33}  \sum_i \del_{\lambda^1_{3\,i}}\del_{\lambda^1_{3\,i}}\Bigg),
	}
	acting on \eqref{eq:cohcorrsumm} gives
	\eq{
		\cC^{\ket{S_6}}_{123}
		=
		&c_{1131} \Bigg[\dperp(\dperp+2)\oldJ_{1,1;1,3}
		+ (\dperp+2)\oldJ_{1,1;1,1} \sum_k \oldI^k_{1,1}\oldI^k_{1,3}
		\nonumber\\
		&
		\phantom{c_{1131}}
		+ \sum_j \oldI^j_{1,1}\oldI^j_{1,1}
		\left((\dperp+2)\oldJ_{1,1;1,3} + \sum_l
		\oldI^l_{1,1}\oldI^l_{1,3}\right)\Bigg]
		\nonumber \\
		&+ c_{1122} \Bigg[ 2 \dperp \oldJ^2_{1,1;1,2}
		+ 2\oldJ_{1,1;1,2}\sum_k \oldI^k_{1,1}\oldI^k_{1,2}
		\nonumber\\
		&
		\phantom{c_{1122}}
		+\left(\dperp \oldJ_{1,1;1,1} + \sum_l
		\oldI^l_{1,1}\oldI^l_{1,1}\right)
		\left(\dperp \oldJ_{1,2;1,2} + \sum_l
		\oldI^l_{1,2}\oldI^l_{1,2}\right)
		\Bigg]
		\nonumber\\
		+& c_{2121} \Bigg[
		\dperp(\dperp+1)\oldJ^2_{1,2;1,1} + \oldJ_{1,2;1,2}\left(\dperp
		\oldJ_{1,1;1,1}
		+ \sum_k \oldI^k_{1,1}\oldI^k_{1,1}\right)
		+ (2\dperp+2)\oldJ_{1,2;1,1} \sum_j \oldI^j_{1,1}\oldI^j_{1,2}
		\nonumber\\ 
		&
		\phantom{c_{2121}}
		+ \left(\sum_l \oldI^l_{1,1}\oldI^l_{1,2}\right)^2
		+ \oldJ_{1,1;1,1}\sum_j \oldI^j_{1,2}\oldI^j_{1,2}\Bigg]
		\nonumber\\
		&
		+ b_{1111}\Bigg[\dperp(\dperp+2)(\dperp+4)\oldJ^3_{1,1;1,1}
		+ 3(\dperp+2&)(\dperp+4)\oldJ^2_{1,1;1,1} \sum_k
		\oldI^k_{1,1}\oldI^k_{1,1}
		\nonumber\\ 
		&
		\phantom{b_{1111}}
		+ 3(\dperp+4)\oldJ_{1,1;1,1}
		\left(\sum_j \oldI^j_{1,1}\oldI^j_{1,1}\right)^2 + \left(\sum_l
		\oldI^l_{1,1}\oldI^l_{1,1}\right)^3
		\Bigg]
		\nonumber \\
		+c_{51}& \Bigg[\dperp \oldJ_{1,1;1,5}
		+ \sum_k \oldI^k_{1,1}\oldI^k_{1,5}\Bigg]
		+ c_{42} \Bigg[\dperp \oldJ_{1,2;1,4}
		+ \sum_k \oldI^k_{1,2}\oldI^k_{1,4}\Bigg]
		+ c_{33} \Bigg[\dperp \oldJ_{1,1;1,5} + \sum_k \oldI^k_{1,1}\oldI^k_{1,5}\Bigg],
		\label{eq:CS6123}
	}
	
	%% \textbf{Kinematics:}
	Once again we choose the reference frame where the massive level $N=6$
	particle is at rest as before in \eqref{eq:3ptkin}, with the
	difference being only in the mass, so we have
	\eq{ &P = |\vec \up_2| = |\vec \up_3| = \sqrt{m^2/4 - m_T^2};
		~~m^2 = M_6^2= 5/\ap
		\nonumber \\ \
		\implies \up^+_2 =
		-\frac{1}{\sqrt{2}}&\left(\frac{m}{2} + P\sin\theta\right),~\up^+_3
		= -\frac{1}{\sqrt{2}}\left(\frac{m}{2} - P\sin\theta\right),~\up^2_2
		= -P \cos\theta = - \up^2_3.  }
	Adding to \eqref{eq:CS6123} the other Chan-Paton ordering we get the
	full amplitude as% \RA{(computed in Mathematica notebook)},
	\begin{align}
		\cM_{\ket{S_6}\rightarrow 2\ket{T}} 
		=
		\frac{1}{2048000} 
		\Bigg\{
		&
		\Bigg[
		85726080 b_{1111}
		+360 (10282 c_{33}+9984 c_{42}+9975 c_{51}+25152 c_{1122}
		\nonumber \\
		&+18650 c_{1131}
		+25104 c_{2121})
		\nonumber\\
		&
		+\dperp \big(-6863680 b_{1111}+1042500 c_{33}+962560 c_{42}+666750 c_{51}-88960 c_{1122}
		\nonumber \\
		&+105420 c_{1131} -67520 c_{2121}\big)
		\nonumber \\
		&
		+ \dperp^2 \big(212280 b_{1111}+37440 c_{1122}+26610 c_{1131}
		+33280 c_{2121}\big) 
		\nonumber \\
		&
		- 980 b_{1111} \dperp^3 
		\Bigg]
	\end{align}
	\begin{align}
		+& 
		5 \cos(2 \theta) 
		\Bigg[24501312 b_{1111}-443592 c_{33} 
		-497664 c_{42}-564300 c_{51}+684288 c_{1122}
		\nonumber \\
		&-105480 c_{1131}+668736 c_{2121} \nonumber \\ 
		&+ \dperp \big(-2354400 b_{1111}-375138 c_{33}-339968 c_{42}-232975 c_{51}-157888 c_{1122}
		\nonumber \\
		& -77094 c_{1131}-143456 c_{2121}\big) \nonumber \\
		&+ \dperp^2 
		\big(63396 b_{1111}-10080 c_{1122}-8937 c_{1131}-8960 c_{2121}\big) \nonumber \\
		&-918 b_{1111} \dperp^3 \Bigg]
	\end{align}
	\begin{align}
		+& 
		18 \cos(4\theta) 
		\Bigg[
		2341440 b_{1111}-197640 c_{33}-199680 c_{42}-60 (3125 c_{51}
		+8384 c_{1122}
		\nonumber \\
		&+10350 c_{1131}+8368 c_{2121}) \nonumber \\
		&+\dperp \big(-339552 b_{1111}+64638 c_{33}+59392 c_{42}+40625 c_{51}+12224 c_{1122}
		\nonumber \\
		&+19962 c_{1131}+11488 c_{2121}\big)
		\nonumber \\
		&+\dperp^2 \big(11556 b_{1111}
		+2016 c_{1122} +1431 c_{1131}+1792 c_{2121}\big) \nonumber 
		\\
		&-54 b_{1111} \dperp^3
		\Bigg] 
	\end{align}
	\begin{align}
		+&
		81 \cos(6\theta)
		\Bigg[66240 b_{1111}+29160 c_{33}+30720 c_{42}+60 (625 c_{51}-704 c_{1122}
		\nonumber \\
		&-810 c_{1131}-688 c_{2121})
		\nonumber \\
		&+ \dperp \big(-15264 b_{1111}-4374 c_{33}-4096 c_{42}-3125 c_{51}+8128 c_{1122}
		\nonumber \\
		&+7614 c_{1131}+7136 c_{2121}\big) \nonumber \\
		&+\dperp^2 \big(972 b_{1111}-288 c_{1122}-243 c_{1131} 
		-256 c_{2121}\big) 
		\nonumber\\
		&-18 b_{1111} \dperp^3
		\Bigg]
		\Bigg\} \nonumber \\
		&= \frac{8}{7}.
		\label{eq:A6TT}
	\end{align}
	This amplitude reduces to a number (as in the last line of \eqref{eq:A6TT}) when the previously quoted $c$s and $b$s are used.
	% \IP{Which is the value?} \DB{Mentioned the value in the last line of the eq. above.}
	
	%%%%%%%%%%%%%%%%%%%%%%%%%%%%
	\subsection{Three point Scattering Amplitude for $\ket{S^{4}_{\mu\nu}} \rightarrow 2\ket{T}$}
	To obtain a non trivial angular dependence for on-shell three point amplitudes in critical dimension we look at the richer spin 2 level $N=4$ massive DDF state scattering with two tachyons.
	
	The relevant 2-tensor at level $N=4$ is
	\eq{
		\ket{{S}^{4}_{\mu\nu}}
		=
		\left(c_{13}\, {1}^{i}\,{3}^{j} + c_{13}\, {1}^{j}\,{3}^{i} - c_{22}\, {2}^{i}\,{2}^{j} - c_{1111}\, \left(1 , 1\right)\,{1}^{i}\,{1}^{j}\right)\ket{\up_{T}} ,
		~~~~
		i \ne j
		\nonumber\\
	}
	where as before $n^i := \uA^i_{-n}$. 
	$\ket{S^4_{\mu\nu}}$ is a spin 2 under $SO(1,d-1)$ only for values proportional to
	$c_{1111}=1,
	~c_{22} = 7,
	~c_{13} = 4$
	and $i\ne j$ because of the traceless condition.
	We shall confirm the same values after computing the amplitude.
	
	The corresponding differential operator to act on the coherent correlator \eqref{eq:cohcorrsumm} is given by
	\eq{
		\cD \left(\{\lambda^1_{n\,i}\}; \ket{S^4_{\mu\nu}}\right) := c_{13}\left(\del_{\lambda^1_{1\,i}}\del_{\lambda^1_{3\,j}} +\del_{\lambda^1_{3\,i}}\del_{\lambda^1_{1\,j}} \right) - c_{22} \del_{\lambda^1_{2\,i}}\del_{\lambda^1_{2\,j}} - c_{1111}\left(\sum_{k=1}^{d-2} \del_{\lambda^1_{1\,k}}\del_{\lambda^1_{1\,k}}\right)\del_{\lambda^1_{1\,i}}\del_{\lambda^1_{1\,j}}.
		\label{eq:diffopSmn4}
	}
	The second order derivatives are easy to compute. 
	Focusing only on the fourth order term we get
	\eq{
		&\left(\sum_{k=1}^{d-2} \del_{\lambda^1_{1\,k}}\del_{\lambda^1_{1\,k}}\right)\del_{\lambda^1_{1\,i}}\del_{\lambda^1_{1\,j}}
		\exp\left\{\sum_i \sum_{n \in \mathbb{Z}} \lambda^1_{n\,i} \oldI_{1,n} + \sum_i\sum_{n,m \in \mathbb{Z}}\frac{1}{2}\lambda^1_{n\,i}\lambda^1_{m\,j}\oldJ_{1,n;1,m}\right\} 
		\nonumber \\
		=& \left(\sum_{k=1}^{d-2} \del_{\lambda^1_{1\,k}}\del_{\lambda^1_{1\,k}}\right) \Bigg[\oldI^i_{1,1}\oldI^j_{1,1} + \oldI^i_{1,1} \left( \sum_n \lambda^1_{n\,j}\oldJ_{1,1;1,n}\right) + \oldI^j_{1,1}\left( \sum_n \lambda^1_{n\,i}\oldJ_{1,1;1,n}\right)
		\nonumber \\
		&+ \left(\sum_n \lambda^1_{n\,i}\oldJ_{1,1;1,n}\right)\left( \sum_m \lambda^1_{n\,j}\oldJ_{1,1;1,m}\right) \Bigg]\exp{S} \Big\lvert_{\{\lambda^1_{n}\}\rightarrow 0} 
		\nonumber \\
		=&~ \oldI^i_{1,1}\oldI^j_{1,1} \left(\sum_{k=1}^{d-2} (\oldI^k_{1,1})^2 + \dperp \oldJ_{1,1;1,1}\right) + 2 \oldI^i_{1,1}(\oldJ_{1,1;1,1})\oldI^j_{1,1} + 2 \oldI^j_{1,1}(\oldJ_{1,1;1,1})\oldI^1_{1,1} 
		\nonumber \\
		&= \oldI^i_{1,1}\oldI^j_{1,1} \left[(\dperp+4)  \oldJ_{1,1;1,1} + \sum_{k=1}^{d-2} (\oldI^k_{1,1})^2 \right].
	}
	This gives all the terms dependent on $\dperp$ in the correlator. 
	Then the correlator 
	$\cC^{\ket{S^4_{\mu\nu}}}_{123}$ is given by
	\eq{
		\cC^{\ket{S^4_{\mu\nu}}}_{123} = c_{13}\left[\oldI^i_{1,3}\oldI^j_{1,1}+\oldI^j_{1,3}\oldI^i_{1,1}\right] - c_{22} \oldI^i_{1,2}\oldI^j_{1,2} - c_{1111}\oldI^i_{1,1}\oldI^j_{1,1} \left[(\dperp+4)  \oldJ_{1,1;1,1} + \sum_{k=1}^{d-2} (\oldI^k_{1,1})^2 \right] 
		.
		\nonumber \\
	}
	As before we want the general amplitude consistent with Lorentz invariance
	for a single spin 2 %DDF 
	state scattering with two tachyons in order to compare with the explicit result.
	The strategy is as follows.
	\begin{enumerate}
		\item 
		We write down the most general possible covariant three point amplitude involving a massive spin 2 field and two scalars (labeled '1', '2' and '3' respectively) as
		\eq{
			\mathcal{M} = \eps_{\mu\nu} \left[
			\sum_{a,b \in \{1,2,3\}} A^{(a,b)}(p_c \cdot p_d)
			~p_a^\mu p_b^\nu
			\right],
		}
		where, $A^{(a,b)}(p_c \cdot p_d)$ with $c,d \in \{1,2,3\}$ are some functions of the momentum scalar products.
		Again these reduce to numbers for on shell three point amplitudes.
		$\eps_{\mu\nu}$ is the general 2 index polarization tensor for the spin 2 particle. 
		\item 
		We choose the rest frame of the massive state and write down the kinematics.
		\item
		Having chosen the frame we can solve in an explicit way the transversality condition for the polarization tensor.
		We specialize to $\eps_{\mu\nu} \simeq \eps^{(ij)}_{\mu\nu} = \eps^{i}_{\mu}\eps^j_{\nu} 
		\equiv E^{\underline{i}}_\mu E^{\underline{j}}_\nu$ with $i \ne j$ because of the trace condition. 
		Therefore
		\eq{
			\mathcal{M}  
			= 
			\eps^{(ij)}_{\mu\nu} \left[
			\sum_{a,b \in \{1,2,3\}} 
			A^{(a,b)}(p_c\cdot p_d)
			~p_a^\mu p_b^\nu\right] =  
			\left[\sum_{a,b \in \{1,2,3\}}
			A^{(a,b)}(\uk_c \cdot \uk_d)
			~\up^i_{a}\up^j_{b}
			\right].
			\label{eq:Msp2TT}
		}
		Moreover using momentum conservation 
		$\sum_a \up_a = 0$ 
		and using transversality 
		$\uk_1^i = 0$ 
		we can simplify the amplitude in \eqref{eq:Msp2TT} to
		\eq{
			\mathcal{M} 
			=
			\tilde{A}\,\up_{2}^i \up_{2}^j 
			= 
			\tilde{A}~\eps^{(ij)}_{\mu\nu} p_2^\mu p_2^\nu
			,
			\label{eq:sp2_lorinvamp}
		}
		where $\tilde{A} = \, \left(A^{(2,2)}-A^{(2,3)}-A^{(3,2)}\right)$ is once again a number on-shell strings.
		\item 
		We choose $i,j$ along two arbitrary but different, fixed transverse directions.
		For the purpose of being explicit, we choose $i=1,~j=2$.
		This is not restrictive since any traceless two tensor in $SO(25)$ can be rotated to this configuration.
		Given the choice of the rest frame and polarization
		the original Lorentz groups breaks as
		$SO(1,d-1) \rightarrow SO(d-3)$ 
		then we can use the surviving symmetry to rotate the $\up_2, \up_3$ vectors. 
		We use this symmetry along with momentum conservation (which implies 
		$\up_2^i = -\up_3^i$
		in directions $i=3,\dots d-1$)
		to orient these components  along the $1^{st}$, $2^{nd}$ and $(d-1)^{th}$ directions.
	\end{enumerate} 
	
	Thereby one can write down the kinematics 
	for the three-point scattering (in the rest frame of $\up_1$):
	\eq{
		&\up_1 = (m,\textbf{0}_{d-1}); \nonumber \\
		&\up_2 = -\left(\frac{m}{2}, P\cos\theta \sin\phi, P \cos\theta \cos \phi,\textbf{0}_{d-4}, P \sin\theta\right) \nonumber \\
		&\up_3 = -\left(\frac{m}{2},-\vec{\up}_{2}\right),
		\label{eq:Smn4TTkin}
	}
	where $P = \sqrt{\frac{m^2}{4}-m_T^2}$, $m$ is the mass of the DDF state (in this example, $m = M_4 = 3/\ap$) and $m_T$ is the mass of the tachyon. 
	The Lorentz invariant amplitude \eqref{eq:sp2_lorinvamp} in this case is of the form,
	\eq{
		\mathcal{M} & 
		= 
		\tilde A~ \up^{i=1}_{2}\up^{j=2}_{2} 
		\propto \cos^2\theta \sin(2\phi),
		\label{eq:MS4TTg}
	}
	where in the last line, we have used kinematics \eqref{eq:Smn4TTkin}.
	Substituting the kinematics and adding the other ordering we get the full amplitude as
	\eq{
		\cM_{\ket{S^4_{\mu\nu}}\rightarrow 2\ket{T}} = \frac{7}{96} \cos ^2\theta \sin (2 \phi ) &\Big[7 \cos (2 \theta ) \left(c_{1111} (\dperp-8)+8 c_{22} -18 c_{13}\right) \nonumber \\
		&-c_{1111} (\dperp+88)-56 c_{12}+114 c_{13}\Big].
		\label{eq:MS4TTc}
	}
	Clearly, the angular dependence matches with \eqref{eq:MS4TTg} in $\dperp=24$ for $c_{1111}=1;~c_{22} = 7;~c_{13} = 4$.
	For $M=3$ amplitudes all spin 2 tensors scattering into two tachyons with the massive state in the rest frame as \eqref{eq:Smn4TTkin} have the same angular dependence in the amplitude independent of the excitation level $N$.
	%%%%%%%%%%%%%%%%%%%%%%%%%%%%%%%%%%%%
	%%%%%4 pt. framed DDF examples%%%%
	\section{Four point Framed DDF Scattering Amplitudes}
	\label{sec:4pt_scattering}
	In this section we compute some explicit four point amplitudes.
	They involve one spin 2 and three tachyons.
	The spin 2 are chosen at level $N=2$, $N=4$
	and $N=6$.

	\subsection{Four point scattering amplitude for $\ket{S^{2}_{\mu\nu}} + \ket{T} \rightarrow 2\ket{T}$}
	We start with the simple case of the scattering of $\ket{S^{N=2}_{\mu\nu}}$ with three tachyons. At the level $N=2$, the corresponding differential operator is
	\eq{
		\cD \left(\{\lambda^{a=1}_{n\,i}\}; \ket{S^2_{\mu\nu}}\right) :=\left(\del_{\lambda^1_{1\,i}}\del_{\lambda^1_{1\,j}}\right)
		,~~~~
		i \ne j
		\label{eq:diffopSmn2}
	}
	acting on $\cW_4$.
	For $i \neq j$ the correlator $\cC^{\ket{S^2_{\mu\nu}}}_{1234}$ is simply,
	\eq{
		\cC^{\ket{S^2_{\mu\nu}}}_{1234} = \oldI^i_{1,1}\oldI^j_{1,1}
		\label{eq:CSmn21234}
	}
	For simplicity we shall examine here only the partial amplitude obtained by integrating \eqref{eq:CSmn21234} over the anharmonic ratio.
	
	Following the same line of reasoning provided in the previous section for the three point spin-2 amplitude, for the four-point scattering we choose again $i=1,~j=2$ to be the polarization directions. 
	Given this choice of polarization we can use the surviving rotational symmetry to set the massive state spacial momentum in the $(d-1)^{th}$ direction.
	Explicitly,
	the kinematics in the CM frame of the incoming strings ($1,4$) is then given by,
	
	\eq{
		&\up_1 = (E_1, \textbf{0}_{d-2},p_{in}), \nonumber \\
		&\up_4 = (E_4, \textbf{0}_{d-2}, - p_{in}), \nonumber \\
		&\up_2 = -(E_2, p_{out}\sin\theta\cos\phi,p_{out}\sin\theta\sin\phi,\textbf{0}_{d-4}, p_{out}\cos\theta), \nonumber \\
		&\up_3 = -(E_3, -\vec{\up}_{2}),
		\label{eq:4ptkin}
	}
	where,
	\eq{&E_1 = \frac{s+(M_2)^2-m_T^2}{2\sqrt{s}} = \frac{\sqrt{s}}{2} + \frac{1}{\ap \sqrt{s}} \nonumber \\
		&E_4 = \frac{s+m_T^2-(M_2)^2}{2\sqrt{s}}=\frac{\sqrt{s}}{2} - \frac{1}{\ap \sqrt{s}}  \nonumber \\
		&E_2 = \frac{s+m_T^2-m_T^2}{2\sqrt{s}}=\frac{\sqrt{s}}{2} =E_3
	}
	and,
	\eq{
		s = (\up_1& + \up_4)^2 = E_1^2 + E_4^2 + 2 E_1. E_4 \\
		=&~ 2 p_{in}^2 + (M_2)^2 + m_T^2 + 2 E_1.E_4 \\
		&\implies p_{in}^2 = \frac{s}{4}+ \frac{1}{(\ap)^2 s}, \\
		&\implies p_{out}^2 = \frac{s}{4}+\frac{1}{\ap}.
	}
	
	Given the previous kinematics we have to handle the $M=4$ integrals.
	We further choose the $SL(2,\R)$ symmetry by $x_1 \rightarrow +\infty, x_2 \rightarrow 1, x_4 \rightarrow 0$ to simply identify the anharmonic ratio $\omega = \frac{x_{14}x_{23}}{x_{13}x_{24}}$. 
	The correlator (for just one Chan-Paton ordering) is,
	\eq{
		\cC^{\ket{S^2_{\mu\nu}}}_{1234} = \omega^2 \up^i_{2} \up^j_{2} &\times \left[\frac{\omega^{2 \ap  \up_2\cdot \up_3+1} (1-\omega)^{2 \ap \up_3\cdot\up_4}}{x_1^2} (x_1-1)^{2 \ap  \up_1\cdot \up_2} x_1^{2 \ap  (\up_1\cdot \up_3+\up_1\cdot\up_4)}\right] \Bigg\lvert_{x_1 \rightarrow +\infty} \\
		&=  \omega ^2 \up^i_{2} \up^j_{2} \left[\omega^{2 \ap  \up_2\cdot \up_3+1} (1-\omega)^{2 \ap \up_3\cdot\up_4}\right] \\
		&=  \omega ^2 \up^i_{2} \up^j_{2} \left[\omega^{\ap s - 1} (1-\omega)^{\ap t - 2}\right], 
	}
	where, in getting to the last line, we have used,
	\eq{
		s &= (\up_1+\up_4)^2 = -\frac{1}{\ap} + \frac{1}{\ap} + 2 \up_1\cdot\up_4 \nonumber \\
		&\implies \ap s = 2 \ap \up_1\cdot \up_4 = 2 \ap \up_2\cdot \up_3+2, \\
		t &= (\up_1+\up_2)^2 = 2 \up_1\cdot \up_2 \nonumber \\
		&\implies \ap t - 2 = 2 \ap \up_3\cdot \up_4
	}
	Finally, integrating $\omega \in (0,1)$ we get the (partial) amplitude as,
	\eq{
		\cM^{(1234)}_{\ket{S^{2}_{\mu\nu}} \rightarrow 3\ket{T}} = \frac{\Gamma \left(\frac{s}{2}+2\right) \Gamma \left(\frac{t}{2}-1\right) \up^i_2 \up^j_2}{\Gamma \left(\frac{1}{2} (s+t+2)\right)} \\
		= 
		\frac{1}{2}\frac{\Gamma \left(\frac{s}{2}+2\right) \Gamma \left(\frac{t}{2}-1\right) \sin(2\phi)\sin^2 \theta}{\Gamma \left(\frac{1}{2} (s+t+2)\right)}
		\label{eq:ampstN2}
	}
	where we have used the kinematics described above to get the last expression.
	As a final crucial simplification, we note that,
	\eq{
		t = (\up_1+\up_2)^2 = -(E_1 - E_2)^2 + p_{in}^2 + p_{out}^2 - 2 p_{in} p_{out} \cos\theta \nonumber \\
		= \frac{s}{2}+\frac{1}{\ap} - 2 \sqrt{\left(\frac{s}{4}+ \frac{1}{(\ap)^2 s}\right)\left(\frac{s}{4}+\frac{1}{\ap}\right)}\cos\theta.
		\label{eq:stcosN2_rel}
	}
	Using \eqref{eq:stcosN2_rel} in \eqref{eq:ampstN2}, we can remove
	either the $\theta$-dependence or Mandelstam $t$.
	In terms of $(s,t)$,
	\eq{
		\cM^{(1234)}_{\ket{S^{2}_{\mu\nu}} \rightarrow 3\ket{T}}(s,t) = \frac{1}{2}\frac{\left(s^2 t-s t^2+4 s t+32\right) \Gamma \left[\frac{s}{2}+2\right] \Gamma \left[\frac{t}{2}-1\right] \sin (2\phi )}{\left(s^2+16\right) \Gamma \left[\frac{1}{2} (s+t+2)\right]} \nonumber \\
		= \frac{1}{2}\frac{\left(s^2 t-s t^2+4 s t+32\right) \sin (2\phi )}{\left(s^2+16\right)} \beta\left(\frac{s}{2}+2,\frac{t}{2}-1\right) ,
	}
	where $\beta(a,b)$ is the Beta function
	and in terms of $(s,\theta)$ it is,
	
	\eq{
		\cM^{(1234)}_{\ket{S^{2}_{\mu\nu}} \rightarrow 3\ket{T}}(s,\theta) = \frac{(s+8) \Gamma \left[\frac{s}{2}+2\right] \Gamma \left[\frac{1}{4} \left(s-\sqrt{(s+8) \left(s+\frac{16}{s}\right)} \cos (\theta )\right)\right]}{8 \Gamma \left[\frac{1}{4} \left(3 s-\sqrt{(s+8) \left(s+\frac{16}{s}\right)} \cos (\theta )+8\right)\right]} \sin ^2(\theta ) \sin (2 \phi).
	}
	%%%%%%%%%%%%%%%%%%%%%%%%%%%%
	\subsection{Four-point Scattering Amplitude for $\ket{S^{4}_{\mu\nu}} + \ket{T}\rightarrow 2\ket{T}$}
	Following the exact same calculation pattern as the $N=2$ level we perform the analysis for $\ket{S_{\mu\nu}^4}$ scattering with 3 tachyons. The formal expression for the correlator is given by (same as the three-point case),
	\eq{
		\cC^{\ket{S_{\mu\nu}^4}}_{1234} =  -\oldI^i_{1,1}\oldI^j_{1,1} \left[(\dperp+4)  \oldJ_{1,1;1,1} + \sum_{k=1}^{d-2} (\oldI^k_{1,1})^2 \right] + 4\left(\oldI^i_{1,3}\oldI^j_{1,1}+\oldI^j_{1,3}\oldI^i_{1,1}\right) - 7 \oldI^i_{1,2}\oldI^j_{1,2}.
	}
	
	%\textbf{Kinematics:} 
	For the 4 point scattering, the kinematics is given in the CM frame of the incoming strings by
	\eqref{eq:4ptkin}
	with the updated quantities,
	\eq{&E_1 = \frac{s+M^2-m_T^2}{2\sqrt{s}} = \frac{\sqrt{s}}{2} + \frac{2}{\ap \sqrt{s}} \nonumber \\
		&E_4 = \frac{s+m_T^2-M^2}{2\sqrt{s}}=\frac{\sqrt{s}}{2} - \frac{2}{\ap \sqrt{s}}  \nonumber \\
		&E_2 = \frac{s+m_T^2-m_T^2}{2\sqrt{s}}=\frac{\sqrt{s}}{2} =E_3
		\label{eq:E1E2E3S4T3}
	}
	and,
	\eq{
		s = (\up_1& + \up_4)^2 = E_1^2 + E_4^2 + 2 E_1. E_4 \nonumber \\
		=&~ 2 p_{in}^2 + M^2 + m_T^2 + 2 E_1.E_4 \nonumber \\
		&\implies p_{in}^2 = \frac{s}{4} - \frac{1}{\ap}+ \frac{4}{(\ap)^2 s}, \\
		&\implies p_{out}^2 = \frac{s}{4}+\frac{1}{\ap}
		\label{eq:sN4S4T3}
	}
	
	%\textbf{Computation and Results:} 
	We again choose the gauge $x_1 \rightarrow +\infty, x_2 \rightarrow 1, x_4 \rightarrow 0$ to simply express the calculations in terms of the anharmonic ratio $\omega$. 
	The correlator is,
	\eq{
		\cC^{\ket{S_{\mu\nu}^4}}_{1234} &= \frac{1}{2} \omega ^2 \up^i_{2} \up^j_{2} \Big[-\omega ^2 \left((\dperp-12) \rho_{31}^2+(\dperp-12) \rho_{31} + 2 (\up^i_2)^2+2 (\up^j_2)^2-2\right) \nonumber \\
		-&2 \omega  \rho_{41} ((\dperp-12) \rho_{31}+4)-(\dperp-12) \rho_{41}^2 + (20-\dperp) \rho_{41}\Big] \nonumber \\
		&\times \left[\frac{\omega^{2 \ap  \up_2\cdot \up_3+1} (1-\omega)^{2 \ap \up_3\cdot\up_4}}{x_1^5} (x_1-1)^{2 \ap  \up_1\cdot \up_2-1} x_1^{2 \ap  (\up_1\cdot \up_3+\up_1\cdot\up_4)}\right] \Bigg\lvert_{x_1 \rightarrow +\infty} \\
		&= \frac{1}{2} \omega ^2 \up^i_2 \up^j_2 \Big[-\omega ^2 \left((\dperp-12) \rho_{31}^2+(\dperp-12) \rho_{31} + 2 (\up^i_2)^2+2 (\up^j_2)^2-2\right) \nonumber \\
		-&2 \omega  \rho_{41} ((\dperp-12) \rho_{31}+4)-(\dperp-12) \rho_{41}^2 + (20-\dperp) \rho_{41}\Big]  \left[\omega^{2 \ap  \up_2\cdot \up_3+1} (1-\omega)^{2 \ap \up_3\cdot\up_4}\right]  \nonumber \\
		&= \frac{1}{2} \omega ^2 \up^i_2 \up^j_2 \Big[-\omega ^2 \left((\dperp-12) \rho_{31}^2+(\dperp-12) \rho_{31} + 2 (\up^i_2)^2+2 (\up^j_2)^2-2\right) \nonumber \\
		&-2 \omega  \rho_{41} ((\dperp-12) \rho_{31}+4)-(\dperp-12) \rho_{41}^2 + (20-\dperp) \rho_{41}\Big]  \left[\omega^{\ap s - 1} (1-\omega)^{\ap t - 2}\right], 
	}
	where, in getting to the last line, we have used,
	\eq{
		s &= (\up_1+\up_4)^2 = -\frac{3}{\ap} + \frac{1}{\ap} + 2 \up_1\cdot\up_4 \nonumber \\
		&\implies \ap s + 2 = 2 \ap \up_1\cdot \up_4 = 2 \ap \up_2\cdot \up_3 + 4, \\
		t &= (\up_1+\up_2)^2 = 2 \up_1\cdot \up_2 \nonumber \\
		&\implies \ap t - 2 = 2 \ap \up_3\cdot \up_4
	}
	Integrating over $\omega \in (0,1)$ we get the (partial) amplitude as,
	\eq{
		\cM^{(1234)}_{\ket{S^{4}_{\mu\nu}} \rightarrow 3\ket{T}} = &\frac{\up^i_2 \up^j_2 \Gamma \left[\frac{s}{2}+2\right] \Gamma \left[\frac{t}{2}-1\right] }{2 \Gamma \left[\frac{1}{2} (s+t+2)\right]}\Bigg[-\frac{(s+4) (s+6) \left((\dperp-12) \rho_{31} (\rho_{31}+1)+2 (\up^i_2)^2+2 (\up^j_2)^2-2\right)}{(s+t+2) (s+t+4)} \nonumber \\
		-&\left((\dperp-12) \rho_{41}^2\right)-\dperp \rho_{41}-\frac{2 (s+4) \rho_{41} ((\dperp-12) \rho_{31}+4)}{s+t+2}+20 \rho_{41}\Bigg] \\
		= &
		\frac{1}{2}\frac{\Gamma \left[\frac{s}{2}+2\right] \Gamma \left[\frac{t}{2}-1\right] \sin(2\phi)\sin^2 \theta}{\Gamma \left[\frac{1}{2} (s+t+2)\right]} F_4(s,t) = \frac{1}{2} \beta\left(\frac{s}{2}+2,\frac{t}{2}-1\right)F_4(s,t) \sin(2\phi)\sin^2 \theta,
		\label{eq:ampstN4}
	}
	where, we have collected the terms in the square bracket in $F_4(s,t)$ and used the kinematics described above to get the last expression
	%\footnote{\RA{Maybe we should give the explicit expression of $F_4$}}. 
	As a final crucial simplification, we note that,
	\eq{
		t = (\up_1+\up_2)^2 = -(E_1 - E_2)^2 + p_{in}^2 + p_{out}^2 - 2 p_{in} p_{out} \cos\theta \nonumber \\
		= \frac{s}{2} - 2 \sqrt{\left(\frac{s}{4} - \frac{1}{\ap}+ \frac{4}{(\ap)^2 s}\right)\left(\frac{s}{4}+\frac{1}{\ap}\right)}\cos\theta.
		\label{eq:stcosN4_rel}
	}
	As before, using \eqref{eq:stcosN4_rel} in \eqref{eq:ampstN4}, we can remove the $\theta$-dependence or Mandelstam $t$. 
	%The form of the full amplitude is omitted for clarity (check the Mathematica notebooks).
	The expression for $F_4(s,t)$ after simplification using the full kinematics \eqref{eq:4ptkin} and \eqref{eq:E1E2E3S4T3}-\eqref{eq:sN4S4T3}, in critical dimension $\dperp=24$ is given by,
	\eq{
		F_4(s,t) = \frac{2 \left[-s^2t^2+3 s^2 t-4 s t^2 -5 s^2-18 t^2-2 s t-10 s-20 t+40+\frac{(s+1) (t-2) \left(s^2t-4s^2+2 s t+48 t+64\right)}{\sqrt{s^2-8 s+64}}\right]}{3 (s+t+2) (s+t+4)}
	}

	\subsection{Four-point Scattering Amplitude for $\ket{S^{6}_{\mu\nu}} +\ket{T}\rightarrow 2\ket{T}$}
	Performing the same procedure as the preceding sections, we also explicitly computed the amplitude for the spin-2 massive DDF state at $N=6$ scattering with three tachyons. The intermediate calculations are computationally extensive, but do not feature any new techniques.% For the interested reader, the expressions are listed in the Mathematica notebook.
	
	For the massive spin-2 DDF scattering with three tachyons, we observe that the on-shell (partial) amplitude in the chosen kinematics is of the form,
	\eq{
		\cM^{(1234)}_{\ket{S_{\mu\nu}^N}\rightarrow 3 \ket{T}} = \frac{1}{2} \beta\left(\frac{s}{2}+2,\frac{t}{2}-1\right)F_N(s,t) \up^i_2\up^j_2,
	}
	where, the form of $F_N(s,t)$ gets increasingly complicated with increasing $N$. 
	It would be interesting, even if computationally extensive to study $F_N(s,t)$ in different regimes (Regge and high energy) for large $N$. This is important for understanding the overall peak structure of the full amplitude. 
	
	%We look at some plots of the spin-2 massive DDF and three tachyon scattering amplitudes for the levels $N=2,4,6$. 
	%in Fig. \ref{fig:ampN246SmnTTT}. 
	For fixed $s$, the number of peaks and zeros increases with the level $N$ of the DDF state (for the large $s$ regime, this is expected from the asymptotic forms listed in Table. \ref{tab:Fn_asymps}). 
	\begin{table}[h]
		\centering
		\begin{tabular}{|c|c|c|c|}
			\hline
			Large $s$ limit & $N=2$ & $N=4$ & $N=6$ \\
			\hline
			$F_N(s,t)$ & 1   &  $(2 - 2 t) + \mathcal{O}(\frac{t^2}{s}$)  & $\frac{1}{8} (-2368 + 384 t + 1386 t^2) s + \mathcal{O}(t^3)$    \\
			\hline
		\end{tabular}
		\caption{Asymptotic behavior of $F_N(s,t)$ for large $s$.}
		\label{tab:Fn_asymps}
	\end{table}
	Furthermore, as expected, the amplitudes become soft in the large $s$ limit. Once again, the exact asymptotic behavior would be interesting to study in comparison to the well-known Veneziano amplitude and other relevant results.
	
	%%%%%%%%%%%%%%%%%%%%%%%%%%%%%%%%
	
		\section{Summary and conclusion}
		\label{section7}
		In this paper, we propose a natural extension of the Sciuto-Della Selva-Saito  vertex operator to describe DDF coherent states in bosonic string theory. This extension is formulated for both the standard and framed definitions of the DDF operators. We have computed the correlator for an arbitrary number of SDS vertices and, based on this, introduced a generating functional for correlators involving any number of DDF states.
		
		This approach offers an alternative method for computing DDF scattering amplitudes, where the interaction details are encoded in integrals centered around the Koba-Nielsen variables, corresponding to the insertion points of vertex operators on the complex plane of the string worldsheet. These complex integrals only need to be evaluated once for a given number of external states and yield polynomials in the anharmonic ratios of the Koba-Nielsen variables. We have computed various three- and four-point amplitudes to verify the general formula of the generating functional.
		
		The remaining challenge lies in connecting these integrals to the geometry of the Riemann surface that describes the worldsheet of interacting strings. It is, in fact, possible to show that DDF amplitudes computed on the upper half-plane correspond to lightcone amplitudes on Mandelstam surfaces \cite{biswas2024LCDDF}. A deeper understanding of this connection could establish a clear parallel between the generating functional for DDF amplitudes and the M-Reggeon vertex, which was constructed in the 1980s to derive standard perturbative string amplitudes

		Another aspect we began exploring in this paper is the correspondence between string amplitudes involving covariant and DDF string states. We have investigated this one-to-one correspondence for some of the lowest massive string states, relating DDF amplitudes to conventional string amplitudes. This involves calculating a superposition of DDF states, and the analysis is simplified using the functional generators. In the case of scalar amplitudes, where the external state's polarization is absent, we obtained the fully covariant amplitude. For the spin-2 state at the $N=4$ string level, we focused our analysis on a specific choice of external polarization, $\varepsilon_{\mu\nu}^{ij}=\varepsilon_\mu^i\otimes \varepsilon_{\nu}^j$ with $i\neq j$  which is transverse and traceless by construction. 
		This is not a limitation, and we expect that analyzing amplitudes with the most general choice of external polarization would necessarily involve the physical null states. These states provide residual gauge transformations on the polarization of external particles, which should be utilized to simplify the expressions of the amplitudes. 
		At the second excited level, we observed that the covariant polarizations emerging from the one-to-one correspondence with DDF states become transverse and traceless only after applying the residual gauge transformations permitted by the Virasoro algebra. 
		We expect this property will hold at all string levels within the bijective correspondence.

		A further application aimed at gaining a deeper understanding of possible chaotic phenomena in string theory involves computing amplitudes with only external scalar particles. These amplitudes are simpler as they do not require polarization tensors, yet they reveal the presence of chaos in higher-spin three-point amplitudes. Despite being conceptually straightforward, the computations are technically and computationally demanding. For instance, a four-point amplitude involving four level-8 scalars can contain approximately $10^{20}$ terms.
		%%%%%%%%%%%%%%%%%%%%%%%%%%%%%%%%%%%%%%%%%%%%%%%%%%%%%%%%%%%%%%%%%%%%%%
		%%%%%%%%%%%%%%%%%%%%%%%%%%%%%%%%%%%%%%%%%%%%%%%%%%%%%%%%%%%%%%%%%%%%%%
		%%%%%%%%%%%%%%%%%%%%%%%%%%%%%%%%%%%%%%%%%%%%%%%%%%%%%%%%%%%%%%%%%%%%%%

\bigskip	
	
	{\bf Acknowledgments:} 
{R.M. thanks Paolo Di Vecchia and Mritunjay Verma for their stimulating discussions.} 	
		\appendix
		
		\section{Notations}
		\label{app:notations}
		We expand the string coordinates as
		\begin{equation}
			X^\mu(u, \bar u)
			= L^\mu(u) + R^\mu(\bar u)
			,~~~~
			u\in C
			,~~
			Im(u)\ge 0
			,
		\end{equation}
		with 
		\begin{equation}
			R^\mu(\bar u) = L^\mu(\bar u)
			,
		\end{equation}
		and
		$L^\mu(z)= L^{(-)\,\mu}(z) +L^{(+)\,\mu}(z)$
		($z\in C$)
		with:
		\begin{eqnarray}
			L^{(-)}(z)
			=
			\hat{q}
			-i \sum_{n=1}^\infty \frac{\alpha_{-n}}{n} \,z^n
			,
			~~~~
			L^{(+)}(z)
			=
			-i\sqrt{2\ap}\,\hat{p}\ln z
			+ i\sum_{n=1}^\infty \frac{\alpha_n}{n} z^{-n}
			,
		\end{eqnarray}
		where the operators satisfy the algebra
		\begin{equation}
			[\hat q, \hat p]= \frac{i}{\sqrt{2\ap}}
			,~~~
			[\alpha_m, \alpha_n] = m\, \delta_{m+n,0}
			.
		\end{equation}
		In the ordering, we use the commutation relations:
		\begin{eqnarray}
			&&
			[L^{(+)}(z_1),\,L^{(-)}(z_2)]= -\log(z_1-z_2)
			,
			~~~~
			|z_1|>|z_2|
			.
		\end{eqnarray}
		The operator  ${\cal P}$, which projects a generic rank-two tensor onto a transverse and traceless two-index tensor, is given by the following expression: 
		\begin{eqnarray}
			{\cal P}^{\mu_1\mu_2}_{\nu_1\nu_2}= \pi^{\mu_1}_{ \nu_1}\,\pi^{\mu_2}_{\nu_2} -\frac{\pi^{\mu_1\mu_2}\,\pi_{\nu_1\nu_2}}{d-1}~~;~~\pi_\rho^\tau=  \delta_\rho^\tau + \frac{p_\rho\,p^\tau}{M^2}~~;~p \cdot \pi=0\label{215L}\label{235L}
		\end{eqnarray}

		\section{SDS vertex and  its expression for framed DDF}
		\label{app:SDS_vertex}
		Let us remind how this works.
		Given any state, physical or unphysical, $|\phi\rangle$ in an Hilbert
		space $\cH$
		we can explicitly compute the
		corresponding operator in the Hilbert space $\widetilde \cH$
		$  V[\tilde X](x;\, |\phi\rangle) $
		using the Sciuto-Della Selva-Saito vertex as
		\begin{equation}
			\widetilde \cS(x) |\phi\rangle
			=
			V[\tilde X](x;\, |\phi\rangle)
			.
		\end{equation}
		Explicitly the Sciuto-Della Selva-Saito vertex is given by
		\begin{align}
			\widetilde \cS(x)
			&=
			\langle x=0, 0_{a} |
			:
			e^{
				\frac{i}{2}
				g_{\mu \nu}
				\sum_{n=0}^\infty \frac{\alpha^\mu_{n}}{n!}
				\partial^n_x \tilde X^\nu(x, \bar x)
			}
			:
			\nonumber\\
			&=
			\langle x=0, 0_{ a} |
			:
			e^{
				-
				g_{\mu \nu}
				\oint_{z=0,\, |z|<|x|} \frac{d z}{ 2\pi i}
				\partial_z L^\mu(z)
				\tilde L^\nu(x+z)
			}
			:
			\nonumber\\
			=&
			\langle x=0, 0_{a} |
			:
			e^{
				i
				g_{\mu \nu}
				\left(
				\alpha^\mu_0 \tilde L^\nu(x)
				+
				\sum_{n=1}^\infty \frac{\alpha^\mu_{n}}{n!}
				\partial^n_x \tilde L^\nu(x)
				\right)
			}
			:
			~~~~\mbox{when }x>0
			,
			\label{eq:SDS}
		\end{align}
		where the normal ordering is acting with respect to $\tilde \alpha$ operators
		since only for them there are creators and annihilators,
		the vacuum $\langle 0_a|$ is the one for the $\alpha$ operators
		acting in $\cH$ and similarly
		$\langle q=0|$ is the null eigenstate of $\hat q$.
		
		Notice that the second and third expressions involve only chiral operators
		and is the one used in computing the Reggeon vertex which therefore
		uses only the chiral Green function.
		
		An explicit example reads
		\begin{align}
			\widetilde \cS(x) \alpha^\mu_N |k \rangle
			=
			i \frac{N}{N!} :\partial^N \tilde L^\mu(x)
			e^{i \sdap k_\nu  \tilde L^\nu(x)} :
			.
		\end{align}
				This means that
				the SDS vertex realizes in an explicit way the state to operator
				correspondence.

		A slightly different point of view is to assert that
		$  \widetilde \cS(x)$ is a coherent state.
		In facts if we define the coherent state
		\begin{equation}
			\Big(\Big( z_{0 \mu},\, z_{n \mu}\Big|
			=\langle q=0; 0_a|
			exp\left[
			z_{0 \mu} \hat p^\mu + \sum_{n=1}^\infty \frac{1}{n }z_{n \mu}\,\alpha^\mu_n
			\right]
			,
		\end{equation}
		we can write
		\begin{align}
			\widetilde \cS(x)
			=
			: \Big(\Big( \sdap k_\mu =  \tilde L_\mu(x);
			z_{\mu n} =   i \frac{1}{n!} :\partial^n \tilde L_\mu(x)
			\Big|
			:
			,
		\end{align}
		where the normal ordering is such that we can treat $\tilde L$ as
		c-numbers.
		Now remember that for coherent states we have
		\begin{align}
			(z| & a^\dagger
			=
			\langle 0 | e^{z a} a^\dagger
			=
			\langle 0 | e^{z a} z
			,
			\nonumber\\
			\langle q=0 | & e^{i \sdap \delta\, \hat  p} f( \hat q ) | p=0 \rangle 
			=
			\langle q=0 |  f( \hat q + \delta)  e^{i \sdap \delta\, \hat p} | p=0 \rangle
			=
			f( \delta )
			.
		\end{align}
		Looking to the SDS vertex as a coherent state allows to immediately
		formulate the substitution rule
		\begin{align}
			\widetilde \cS(x) 
			f( \hat q^\mu, \alpha^\mu_{-n} )
			|p=0; 0_a\rangle
			=
			:
			f( \hat q^\mu \Rightarrow \tilde L^\mu(x),\,
			\alpha^\mu_{-n}
			\Rightarrow i  \frac{ \sqrt{n} }{n!}
			:\partial^n \tilde L^\mu(x)
			) 
			:
			.
		\end{align}
		
		This implies the useful rule for any functional
		\begin{align}
			\widetilde \cS(x)
			&
			f[ L^{(-)}(z) ]
			|0\rangle
			=
			\widetilde \cS(x)
			f\left[ \hat q
			-
			i \sum_{n=1}^\infty \frac{1}{ n} \alpha_{-n} z^n
			\right]
			\,
			|p=0; 0_a\rangle
			%
			%% %
			%% \nonumber\\
			%% %
			%% =&
			%% :
			%% f\left[
			%%   \tilde L(x)
			%%   -
			%%   i \sum_{n=1}^\infty \frac{1}{\sqrt{n}} z^n
			%%   \times
			%%   i \ishap \frac{ \sqrt{n} }{n!}
			%%   :\partial^n \tilde L^\mu(x)
			%%   \right]
			%% %
			%%  \nonumber\\
			%
			%%  =&
			=
			:
			f\left[ \tilde L(z+x) \right]
			:
			,
			\label{eq: SDS substitution rule}
		\end{align}
		which is the sought after mapping.

		\subsection{DDF coherent state}
		We define the framed DDF coherent state without normal ordering  as
		\begin{align}
			\ket{\lambda^i_n, \uk_T}
			\equiv
			e^{
				\sum_{i=1}^{d-1}
				\sum_{n=1}^\infty
				\lambda^i_n \uA^i_{-n}
			}
			| \uk_T \rangle
			,
			\label{eq:coherent DDF state}
		\end{align}
		since the $\uA_{-n}$ with $n\ge 1$ commute.
		
		Now we normal order the $\ualpha^i$ in $\uA^i$.
		Luckily they enter in a linear way in the exponent
		while all $\ualpha^+$ behave as c-number.
		We have however to be careful since the operatorial Green function is
		only defined for the modulus of the argument of the annihilator part
		bigger than the modulus of the argument of the creator part so we
		write
		\begin{align}
			| \{\lambda^i_n\}, \uk_T )
			=&
			e^{
				\sum_{i=1}^{d-1}
				\sum_{n=1}^\infty
				\lambda^i_n
				%    \ishap
				i\,
				\left[
				\oint_{z=0, |z|<|w|} \frac{d z}{ 2\pi i}
				\partial_z \uL^{i (-)} (z)
				e^{- i n \frac{ \uL^+(z) }{\oldap \upzero^+} } 
				+
				\oint_{w=0} \frac{d w}{ 2\pi i}
				\partial_w \uL^{i (+)} (w)
				e^{- i n \frac{ \uL^+(w) }{\oldap \upzero^+} } 
				\right]
			}
			| \uk_T \rangle
			\nonumber\\
			=
			:\dots:
			&
			e^{
				+\oh
				\sum_{i,j=1}^{d-1}
				\sum_{n.m=1}^\infty
				\lambda^i_n\,    \lambda^j_m
				\oint_{z=0, |z|<|w|} \frac{d z}{ 2\pi i}
				\oint_{w=0} \frac{d w}{ 2\pi i}
				\left[
				\partial_z \uL^{i (-)} (z)
				e^{- i n \frac{ \uL^+(z) }{\oldap \upzero^+} } 
				,\,
				\partial_w \uL^{j (+)} (w)
				e^{- i j \frac{ \uL^+(w) }{\oldap \upzero^+} } 
				\right]
			}
			| \uk_T \rangle
			.
		\end{align}
		Paying attention to the fact we have $[L^{(-)},\, L^{(+)}]$ which
		implies a $-1$ in front of the correlator, e get therefore  
		\begin{align}
			| \lambda^i_n,
			&
			\uk_T )
			=
			: 
			\exp \left(
			\sum_{i=1}^{d-1}
			\sum_{n=1}^\infty
			\lambda^i_n
			i
			%%    \ishap
			\oint_{z=0} \frac{d z}{ 2\pi i}
			\partial_z \uL^i(z)
			e^{- i n \frac{ \uL^+(z) }{\oldap \upzero^+} } 
			\right)
			:
			\nonumber\\
			\exp
			&
			\left(
			+\oh
			\sum_{i=1}^{d-1}
			\sum_{n_1, n_2=1}^\infty
			\lambda^i_{n_1}
			\lambda^i_{n_2}
			\oint_{z_1=0, |z_1| > |z_2|} \frac{d z_1}{ 2\pi i}
			\oint_{z_2=0} \frac{d z_2}{ 2\pi i}
			\frac{1}{(z_1 -z_2)^2}
			e^{
				-i
				\frac{
					n_1  \uL^+(z_1)
					+
					n_2\uL^+(z_2)
				}{\oldap \upzero^+}
			}  
			\right)
			| \uk_T \rangle
			.
		\end{align}
		After normal ordering the coherent state we can apply the annihilators
		on $|\uk_T\rangle$.
		In particular in the exponent
		$- i n \frac{ \uL^+(z) }{\oldap \upzero^+}$
		we have the contribution
		$\frac{-n \sdap \upzero^+ \log(z)}{ (\sdap  \upzero^+) }$
		which gives $\frac{1}{z^n}$ and
		$\upzero^+$ is replaced with $\uk_T^+$
		so we get
		\begin{align}
			| \lambda^i_n, &\uk_T )
			=
			\exp \left(
			\sum_{i=1}^{d-1}
			\sum_{n=1}^\infty
			\lambda^i_n
			\oint_{z=0} \frac{d z}{ 2\pi i}
			\frac{1}{z^n}
			\left(
			i
			%%    \ishap
			\partial_z \uL^{i (-)}(z)
			+
			\sdap \uk^i_T
			\frac{1}{z}
			\right)
			e^{-i n \frac{ \uL^{+ (-)} (z) }{\oldap \uk^+_T} } 
			\right)
			\nonumber\\
			\exp
			&
			\left(
			\oh
			\sum_{i=2}^{D-2}
			\sum_{n_1, n_2=1}^\infty
			\lambda^i_{n_1}
			\lambda^i_{n_2}
			\oint_{z_1=0, |z_1| > |z_2|} \frac{d z_1}{ 2\pi i}
			\oint_{z_2=0} \frac{d z_2}{ 2\pi i}
			\frac{
				e^{-i
					\frac{
						n_1  \uL^{+ (-)}(z_1)
						+
						n_2\uL^{+ (-)}(z_2)
					}{\oldap \uk^+_T}
				}
			}{
				z_1^{n_1} z_2^{n_2} (z_1 -z_2)^2
			}
			\right)
			| \uk_T \rangle
			.
		\end{align}

		The big advantage is now that {\sl all} fields are only creators,
		i.e. we have got rid of annihilators.

		\subsection{The vertex associated with the coherent state}
		Now we can act with it on the SDS and use the substitution rule
		\eqref{eq: SDS substitution rule} and get the SDS vertex for the
		coherent state as (we replace immediately $\tilde L \rightarrow L$)
		\begin{align}
			\widetilde \cS(x)
			&
			| \lambda^i_n, \uk_T )
			\equiv
			\cS( x; \lambda^i_n, \uk_T)
			\nonumber\\
			=&
			:
			\exp \left(
			\sum_{i=1}^{d-1}
			\sum_{n=1}^\infty
			\lambda^i_n
			\oint_{z=0,\, |z|\rightarrow 0} \frac{d z}{ 2\pi i}
			\frac{1}{z^n}
			\left(
			i
			%%    \ishap
			\partial_z \uL^{i}(z+x)
			+
			\sdap \uk^i_T
			\frac{1}{z}
			\right)
			e^{-i n \frac{ \uL^{+} (z+x) }{\oldap \uk^+_T} } 
			\right)
			\nonumber\\
			\exp
			&
			\left(
			\oh
			\sum_{i=1}^{d-1}
			\sum_{n_1, n_2=1}^\infty
			\lambda^i_{n_1}
			\lambda^i_{n_2}
			\oint_{z_1=0, |z_1| > |z_2|} \frac{d z_1}{ 2\pi i}
			\oint_{z_2=0} \frac{d z_2}{ 2\pi i}
			\frac{
				e^{
					-i
					\frac{
						n_1  \uL^+(z_1+x)
						+
						n_2\uL^+(z_2+x)
					}{\oldap \uk^+_T}
				}
			}{
				z_1^{n_1} z_2^{n_2} (z_1 -z_2)^2
			}
			\right)
			\,
			e^{i \sdap \uk_{T \mu} \uL^\mu(x)}
			:
			,
			\label{app:DDF S vertex}
		\end{align}
		where $|z|\rightarrow 0$ means that the contour must not
		encircle other vertices.
		
		\subsection{Comparing with the expression of the vertex
			in the usual formalism}
		\label{sect:compaing single vertex}
		We can now compare with the result obtained using the usual formalism
		in Eq.\eqref{19}.
		The first step is to integrate by part the single integral as
		\begin{align}
			\oint_{z=0}
			&
			\frac{d z}{ 2\pi i}
			\frac{1}{z^n}
			\left(
			i
			\ishap
			\partial_z \uL^{i}(z+x)
			+
			\sdap \uk^i_T
			\frac{1}{z}
			\right)
			e^{-i n \frac{ \uL^{+} (z+x) }{\oldap \uk^+_T} } 
			\nonumber\\
			=&
			\frac{1}{(n-1)!}
			\frac{d^{n-1}}{d x^{n-1} }
			\left(
			i
			\ishap
			\partial_x \uL^{i}(x)
			e^{-i n \frac{ \uL^{+} (x) }{\oldap \uk^+_T} } 
			\right)
			+
			\sdap \uk^i_T
			\frac{1}{n!}
			\frac{d^{n}}{d x^{n} }
			\left(
			e^{-i n \frac{ \uL^{+} (x) }{\oldap \uk^+_T} } 
			\right)
			\nonumber\\
			=&
			\frac{1}{(n-1)!}
			\frac{d^{n-1}}{d x^{n-1} }
			\left[
			i
			\ishap
			\left(
			\partial_x \uL^{i}(x)
			-
			\frac{\uk_T^i}{ \uk_T^+}
			\partial_x \uL^{+}(x)
			\right)
			e^{-i n \frac{ \uL^{+} (x) }{\oldap \uk^+_T} }
			\right]
			\nonumber\\
			=&
			\oint_{z=0}
			\frac{d z}{ 2\pi i}
			\frac{1}{z^n}
			i
			\ishap
			\left(
			\delta^{\underline{i}}_{\underline{\mu}}
			-
			\frac{\uk_T^i}{ \uk_T^+}
			\delta^{\underline{+}}_{\underline{\mu}}
			\right)
			\partial_z \uL^{ \underline{\mu} }(z+x)
			e^{-i n \frac{ \uL^{+} (z+x) }{\oldap \uk^+_T} }     
			.
			\label{eq:app integration by part single integral}
		\end{align}
		This form reveals the covariant projector Eq.\eqref{eq:covariant projector}
		\begin{equation}
			\underline{\Pi}^i_{\underline{\mu}}
			=
			%%      \left(
			\delta^{\underline{i}}_{\underline{\mu}}
			-
			\frac{\uk_T^i}{ \uk_T^+}
			\delta^{\underline{+}}_{\underline{\mu}}
			%%    \right)
			,
		\end{equation}
		written in flat coordinates.
		The identification between the two forms of the projector requires
		the following mapping which is also needed
		to match the $Q$ integrals with the ones in
		Eq.\eqref{eq:integrals in usual DDF vertex} used in
		\eqref{app:DDF S vertex} 
		we utilize the map 
		\begin{align}
			z_a
			&
			\rightarrow
			z + x_a
			,
			\nonumber\\
			\epsilon^i_\mu
			&
			\rightarrow E^i_\mu
			,~~
			\varepsilon^i_\mu \rightarrow
			\underline{\Pi}^i_j\,E^j_\mu
			,
			\nonumber\\
			k_\mu
			&
			\rightarrow
			\frac{ E^+_\mu}{ \dap E^+_\nu\, p_T^\nu}
			,~~
			n_{a} \rightarrow n_{a}
			.
			\label{app:mapping framed usual}
		\end{align}
		In the framed DDF approach it is natural to take all $\epsilon_{n}$ be
		equal so that we get
		\begin{equation}
			\varepsilon_{(\na)}^i  \,\cdot\varepsilon_{(\nb)}^j
			\rightarrow
			\delta^{i j}
			.
		\end{equation}

		%%%%%%%%%%%%%%%%%%%%%%%%%%%%%%%%%%%%%%%%%%%%%%%%%%%%%%%%%%%%%%%%%%%%%
		%%%%%%%%%%%%%%%%%%%%%%%%%%%%%%%%%%%%%%%%%%%%%%%%%%%%%%%%%%%%%%%%%%%%%
		%%%%%%%%%%%%%%%%%%%%%%%%%%%%%%%%%%%%%%%%%%%%%%%%%%%%%%%%%%%%%%%%%%%%%
		%%%%%%%%%%%%%%%%%%%%%%%%%%%%%%%%%%%%%%%%%%%%%%%%%%%%%%%%%%%%%%%%%%%%%
		
		\section{Correlator of coherent states} 
		\label{app:correlator}
		In this section we give details of the calculation of the correlator
		of an arbitrary number of coherent states.
		
		Correlators are symbolically like 
		\begin{eqnarray}
			{\cal W}_M
			=&
			\langle 0_a,\, p=0|\,  
			{\cal N}_1\, :\, e^{\sum_{Z_1} \hat A_{1 Z_1}\, e^{ \hat B_{1 Z_1}}}\, e^{C_1} :\,
			\dots
			{\cal N}_M\, :\, e^{\sum_{z_M} \hat A_{M Z_M}\, e^{\hat B_{M Z_M}}}\, e^{C_M} :\,
			| 0_a,\, p=0 \rangle
			,
		\end{eqnarray}
		with
		\begin{align}
			{\cal N}_a
			=&
			e^{\sum_{n, m=0}^\infty\, \lambda^a_{n\,i}\, \lambda^a_{m\,j}\, Q^{ij}_{n\,m}(x_a)}\,
			,
			\nonumber\\
			e^{\sum_{Z_a} \hat A_{1 Z_a}\, e^{ \hat B_{1 Z_a}}}
			=&
			e^{\sum_{n=1}^\infty \lambda^a_{n\,i}\,  Q^i_{-n}(x_a) }
			,
			\nonumber\\
			e^{C_a}
			=&
			e^{i\sqrt{2\alpha'} p^a_{T\, \mu}\, L^\mu(x_a)}
			,
		\end{align}
		where $Q$ is defined in Eq. \ref{eq:integrals in usual DDF vertex}.
		We used $\sum_{Z_a}$ to symbolically mean
		$\sum_{n=1}^\infty\, \oint_{x_a}$.
		
		The correlator is computed in two steps.
		First, we move on the left and right respectively, the creation and
		annihilation operators contained in the definition of the tachyon
		vertices entering in Eq. \eqref{19} and
		denoted $e^{C_a}$ in the previous equations
		\begin{eqnarray}
			e^{i\sqrt{2\alpha'} \ppTa \cdot L(x_a)}
			=
			e^{i\sqrt{2\alpha'} \ppTa \cdot L^{(-)}(x_a)}\,
			e^{i\sqrt{2\alpha'} \ppTa \cdot L^{(+)}(x_a)}
			.
		\end{eqnarray}

		Schematically for any operators ${\cal O}$ acting on the string
		coordinate we have (notice the exponential of an exponential):
		\begin{eqnarray}
			&&
			e^{e^{i {\cal O}\cdot L^{(+)}(w)}}~
			e^{i\sqrt{2\alpha'}  p_T\cdot L^{(-)}(z)}
			=
			\sum_{n=0}^\infty  \frac{1}{n!} e^{i n{\cal O}\cdot L^{(+)}(w)}~
			e^{i\sqrt{2\alpha'} p_T\cdot L^{(-)}(z)}
			\nonumber\\
			&&
			=
			e^{i\sqrt{2\alpha'}  p_T\cdot L^{(-)}(z)}\sum_{n=0}^\infty 
			\frac{1}{n!} e^{i n{\cal O}\cdot L^{(+)}(z)}\,
			e^{-n[{\cal O}\cdot L^{(+)}(w),\,  p_T\cdot L^{(-)}(z)]}
			\nonumber\\
			&&
			= e^{i\sqrt{2\alpha'}  p_T\cdot L^{(-)}(z)}\,
			e^{e^{i {\cal O}\cdot L^{(+)}(w)
					+ \sqrt{2\alpha'} p_T\cdot {\cal O}_{w} \log(w-z)}}
			,~~~~|w|>|z|
			,
		\end{eqnarray}
		and similarly
		\begin{eqnarray}
			&&
			e^{i\sqrt{2\alpha'}  p_T\cdot L^{(+)}(z)}\,
			e^{e^{i {\cal O}\cdot L^{(-)}(w)}}
			=
			e^{
				e^{i {\cal O}\cdot L^{(-)}(w) + \sqrt{2\alpha'} p_T\cdot {\cal O}_{w} \log(z-w)}
			}\,
			e^{i\sqrt{2\alpha'}  p_T\cdot L^{(+)}(z)}
			,~~~~|
			z|>|w|
			.
		\end{eqnarray}

		When we consider a path with $|z-x_a|> |x_b|$ in the definition of $Q$s
		these relations imply: 
		\begin{eqnarray}
			\exp
			&&
			{\mkern-30mu}
			\left\{
			\oint_{x_a} \frac{dz}{2\pi i\, (z-x_a)^n}
			\left.
			:\,
			e^{
				i (\lambda^a_{n\,i} \varepsilon_{(n)}^{a\,i}\partial_z
				-n\sqrt{2\alpha'}\, k) \cdot L(z)
			}
			\,:
			\right|_{linear\,in\,\lambda}
			\right\}\,
			e^{i\sqrt{2\alpha'} \ppTb\cdot L^{(-)}(x_b)}
			\nonumber\\
			=
			e^{i\sqrt{2\alpha'} \ppTb\cdot L^{(-)}(x_b)}
			&&
			\exp\left\{
			\oint_{x_a} \frac{dz}{2\pi i\, (z-x_a)^n}
			(z-x_b)^{-2\alpha' n k\cdot p_T^b}\,
			\left.
			:\,
			e^{i
				(\lambda^a_{n\,i} \varepsilon_{(n)}^{a\,i} \partial_z
				-n\sqrt{2\alpha'} k) L(z)
				+\frac{ \lambda^a_{n\,i}
					\sqrt{2\alpha'}\,\varepsilon_{(n)}^{a\,i} \cdot p_T^b}{(z-x_b)}}
			\,:
			\right|_{linear\,in\,\lambda}
			\right\}
			\nonumber\\
			=
			e^{i\sqrt{2\alpha'} \ppTb\cdot L^{(-)}(x_b)}
			&&
			\exp\left\{
			\lambda^a_{n\,i}\,
			\oint_{x_a} \frac{dz}{2\pi i\, (z-x_a)^n}
			(z-x_b)^{-2\alpha' n k\cdot p_T^b}\,
			e^{-i\,n \sqrt{2\alpha'}\, k \cdot L(z)}\,
			\varepsilon_{(n)}^{a\,i} \cdot
			\left(
			i \partial_z L
			+
			\frac{\sqrt{2\alpha'}\, p_T^b}{(z-x_b)}
			\right)
			\right\}
			,
		\end{eqnarray}
		and similarly  
		\begin{eqnarray}  
			e^{i\sqrt{2\alpha'} p^b_T\cdot L^{(+)}(x_b)}\,
			&&
			{\mkern-30mu}
			e^{
				\oint_{x_a} \frac{dz}{2\pi i\, (z-x_a)^n}
				\left.
				:\,
				e^{i(\lambda^a_{n\,i}\varepsilon_{(n)}^{a\,i}
					-n\sqrt{2\alpha'} k) L(z)}
				\,:
				\right|_{lin.\lambda}
			}\,\,
			\nonumber\\
			=
			&&
			e^{
				\oint_{x_a} \frac{dz}{2\pi i\, (z-x_a)^n}
				(x_b-z)^{-2\alpha' n k\cdot p_T^b}
				\left.
				e^{i( \lambda^a_{n\,i} \varepsilon_{(n)}^{a\,i} \partial_z
					-n\sqrt{2\alpha'} k) L(z)
					+
					\frac{
						\lambda^a_{n\,i} \sqrt{2\alpha'} \varepsilon_{(n)}^{a\,i} \cdot  p_T^b
					}{(z-x_b)}
				}\right|_{lin.\lambda}}\,
			e^{i\sqrt{2\alpha'} p^a_T\cdot L^{(+)}(x_b)}  .
		\end{eqnarray}
		We then need also the usual relation
		\begin{eqnarray}
			&&
			e^{i\sqrt{2\alpha'} p^a_T\cdot L^{(+)}(x_b)}\,
			e^{i\sqrt{2\alpha'} p^b_T\cdot L^{(-)}(x_b)}
			=
			(x_a-x_b)^{2\alpha'p_T^a\cdot p_T^b}\,
			e^{i\sqrt{2\alpha'} p^b_T\cdot L^{(-)}(x_b)}\,
			e^{i\sqrt{2\alpha'} p^a_T\cdot L^{(+)}(x_b)}
			.
		\end{eqnarray}  
		After the first step we are left with operators, schematically,  of the form:
		\begin{eqnarray}
			{\cal W}_M
			=&
			\prod_{a=1}^M{\cal N}_a\,
			\prod_{a=1}^{M-1}\, \prod_{b=a+1}^{M}\,  (x_a-x_b)^{2\alpha'p_T^a\cdot p_T^b}\,
			\nonumber\\
			&
			\langle p= -\sum_{a=1}^M \ppTa |
			:\, e^{\sum_{Z_1}\, A_{1 Z_1}\,e^{B_{1 Z_1}}}:\,
			%%  :\,e^{A_2\,e^{B_2}}:
			\dots
			:\, e^{\sum_{Z_M}\, A_{M Z_M}\,e^{B_{M Z_M}}}:\,
			|p=0\rangle
			,
		\end{eqnarray}
		where the double product
		$  \prod_{a=1}^{M-1}\, \prod_{b=a+1}^{M}\,$
		comes from reordering the operators $e^{C_a}$
		and
		\begin{equation}
			\langle p= -\sum_{a=1}^M \ppTa | = | p= -\sum_{a=1}^M \ppTa
			\rangle^\dagger
			,
		\end{equation}
		from the action of $e^{\sum_a C_a^{(-)}}$.
		In the previous expression we have the new operators
		\begin{align}
			&:\, e^{\sum_{Z_a}\, A_{a Z_a}\,e^{B_{a Z_a}}}:\,
			\nonumber\\
			&=
			:\,
			\exp\left\{
			\sum_{n_a=1}^\infty\,
			\lambda^a_{n_a\, i}\,
			\oint_{x_a} \frac{dz}{2\pi i\, (z-x_a)^{n_a}}\,
			\varepsilon_{(n_a)}^{a\,i} \cdot
			\left(
			i \partial_z L
			+
			\sum_{b=1,\, b\ne a}^M
			\frac{\sqrt{2\alpha'}\, p_T^b}{(z-x_b)}
			\right)
			\,
			(z-x_b)^{-2\alpha' n_a k\cdot p_T^b}\,
			e^{-i\,n_a \sqrt{2\alpha'}\, k \cdot L(z)}\,
			\right\}
			:  
		\end{align}
		so that
		$A_{a Z_a}
		=
		i \varepsilon_{(n_a)}^{a\, i} \cdot
		\left( \partial_z L
		-   \sum_{b=1,\, b\ne a}^M
		\frac{i \sqrt{2\alpha'}\, p_T^b}{(z-x_b)} \right)
		$
		and most importantly the rest  $e^{B_{a Z_a}}$ behave as c-numbers because
		$k^2=k\cdot \varepsilon=0$.

		We have then (up to a point on zero modes in $e^{B_{a Z_a}}$ discussed below)
		\begin{eqnarray}
			\langle A_{a Z_a}\,A_{b Z_b}\rangle \neq 0
			~~,~~
			\langle A_{a Z_a}\,B_{b Z_b}\rangle = 0
			~~,~~
			\langle B_{a Z_a} \rangle \neq 0
			.
			\label{A.34}
		\end{eqnarray}

		The previous expression can then be evaluated to
		\begin{align}
			{\cal W}_M
			=&
			\prod_{a=1}^M{\cal N}_a\,
			\prod_{a=1}^{M-1}\, \prod_{b=a+1}^{M}\,  (x_a-x_b)^{2\alpha'p_T^a\cdot p_T^b}\,
			e^{
				\sum_{a<b}\, \sum_{Z_a} \,\sum_{Z_b}\,
				\langle A_{a Z_a}\,A_{b Z_b}\rangle\,
				e^{\langle B_{a Z_a} \rangle}\, e^{ \langle B_{b Z_b} \rangle}
			}
			.
		\end{align}

		The only subtleties are the zero mode of the exponential operator
		$e^{-i n_a \sqrt{2\ap} k \cdot L(z)}$. These zero modes commute with
		all the operators in the correlator but when acting on the vacuum
		change the momentum and impose the correct delta on the momenta. This
		is the reason of the exponential factors $e^{-in \sqrt{2\alpha'}k\cdot
			q}$ that appear in eq.s \eqref{25} and \eqref{26}.

		\section{Correlators of framed DDF operators}
		\label{app:framed_DDF_correlator}

		In this appendix we would like to give the same derivation of Appendix 
		\ref{app:correlator} but from the framed DDF point of view.
		In particular we are interested in computing the correlator
		\begin{align}
			\cW_M
			=&
			\langle p=0, 0_a |
			\cS( x_1; \lambda^{1 }_{(n) i}, \uk^1_{T} )
			\dots
			\cS( x_M; \lambda^M_{(n) i} , \uk^M_{T} )
			|p=0, 0_a \rangle
			,
			~~~~
			|x_1|>|x_2|> \dots >|x_M|
			.
		\end{align}

		The first idea is to normal order
		the $\partial \uL^i$ operators
		and then to let them act on the bra and ket vacua.
		This does not work since $\partial \uL^i$ are bound to $e^{\dots \uL^+}$
		which do not commute with the tachyon vertexes.
		
		Therefore the first step is to move the $\uL^-$ in the tachyonic
		vertexes.
		Explicitly we move
		$  e^{ i \sdap\uk^b_{-} \uL^{- (+)}(x_b) }$ operators to the right
		and let them act on the vacuum.
		At the same time the $  e^{ i \sdap\uk^b_{-} \uL^{- (-)}(x_b) }$
		can be moved to the right.
		
		In order to  move $  e^{ i \sdap\uk_{\snr -} \uL^{- (+)}(\xr) }$ to
		the right
		and have a well defined operatorial result
		we need $|x_b| > |x_c|$ in the following expression
		\begin{align}
			e^{ i \sdap\uk^b_{-} \uL^{- (+)}(x_b) }
			&
			F\left[
			e^{ -i  \frac{m}{\oldap \uk^+_{c} }
				( \uL^{+ (-)}(z+x_c) +
				\uL^{+ (+)}(z+x_c) ) }
			\right]
			\nonumber\\
			=&
			F\left[
			e^{ -i  \frac{m}{\oldap \uk^+_{c} }( \uL^{+ (-)}(z+x_c) +
				\uL^{+ (-)}(z+x_c) ) }
			(x_b -x_c- z)^{ -m \frac{\uk^+_{b} }{\uk^+_{c}} }
			\right]
			e^{ i \sdap\uk^b_{-} \uL^{- (+)}(x_b) }
			.
		\end{align}
		This is valid for any functional $F[]$.
		It can be proved easily for all functional entering the vertex.
		This happens because they are exponentials which can be expanded, then
		the previous equation is valid for each factor of each addend of the expansion.
		
		We also need the corresponding relation for
		$  e^{ i \sdap \uk^b_{ -} \uL^{- (-)}(x_b) }$
		when $|\xr| > |\xs|$
		\begin{align}
			&
			F\left[
			e^{ -i  \frac{m}{\oldap \uk^+_{c} }(
				\uL^{+ (-)}(z+x_c) +
				\uL^{+ (+)}(z+x_c) ) }
			\right]
			e^{ i \sdap\uk^b_{-} \uL^{- (-)}(x_b) }
			\nonumber\\
			=&
			e^{ i \sdap\uk^b_{-} \uL^{- (-)}(x_b) }
			F\left[
			e^{ -i  \frac{m}{\oldap \uk^+_{c} }
				( \uL^{+ (-)}(z+x_c) +
				\uL^{+ (+)}(z+x_c) ) }
			(z +x_c  -x_b )^{ -m \frac{\uk^+_{b} }{\uk^+_{c}} }
			\right]
			.
		\end{align}

		In performing the previous reordering 
		and explicitly evaluating  $\langle \up^+=0| \dots | \up^+=0\rangle$
		we find
		\begin{align}
			=&
			\langle \up^-=\up^i=0, 0_{a} |
			\nonumber\\
			%
			% exp 1 linear
			%
			:
			\exp
			&
			\Bigg(
			\sum_{i=1}^{d-2}
			\sum_{n=1}^\infty
			\lambda^{a=1}_{n\, i}
			\oint_{z=0} \frac{d z}{ 2\pi i}
			\frac{1}{z^n}
			\left(
			i
			\ishap
			\partial_z \uL^{i}(z+x_1)
			+
			\sdap \uk^i_{T a=1}
			\frac{1}{z}
			\right)
			e^{-i n \frac{ \uL^{+} (z+x_1) }{\oldap \uk^+_{T a=1}} } 
			\nonumber\\
			&
			\phantom{\exp}
			\prod_{\rrr 2}^{M}
			(x_1 -x_b + z)^{ -n \frac{\uk^+_{T b} }{\uk^+_{T  a=1}} }
			\Bigg)
			\,
			e^{i \sdap\uk_{T a=1}^i \uL^i(x_1) - i \sdap\uk_{T a=1}^- \uL^+(x_1)}
			:
			\,
			\prod_{b=2}^M
			(x_1-x_b)^{\dap \uk_{T a=1}^- \uk_{T b}^+}
			\nonumber\\
			%
			% exp 1 quadratric
			%
			\exp
			&
			\Bigg(
			%  +
			\oh
			\sum_{i=1}^{d-2}
			\sum_{n_1, n_2=1}^\infty
			\lambda^{a=1}_{n_1\, i}
			\lambda^{a=1}_{n_2\, i}
			\oint_{z_1=0, |z_1| > |z_2|} \frac{d z_1}{ 2\pi i}
			\oint_{z_2=0} \frac{d z_2}{ 2\pi i}
			\frac{
				e^{
					-i
					\frac{
						n_{ 1}\uL^+(z_1+x_1)
						+
						n_{ 2}\uL^+(z_2+x_1)
					}{\oldap \uk^+_{T a=1} }
				}
			}{
				z_1^{n_1} z_2^{n_2} (z_1 -z_2)^2
			}
			\nonumber\\
			&
			\phantom{\exp}
			\prod_{b=2}^{M}
			(x_1 - x_b + z_1)^{ -n_1 \frac{\uk^+_{T b} }{\uk^+_{T a=1}} }
			(x_1 - x_b + z_2)^{ -n_2 \frac{\uk^+_{T b} }{\uk^+_{T a=1}} }
			\Bigg)
			\nonumber
		\end{align}
		%
		%
		% factor 2
		%
		%
		\begin{align}
			%
			% exp 2 linear
			%
			\times
			:
			\exp
			&
			\Bigg(
			\sum_{i=1}^{d-2}
			\sum_{n=1}^\infty
			\lambda^{a=2}_{n\,i}
			\oint_{z=0} \frac{d z}{ 2\pi i}
			\frac{1}{z^n}
			\left(
			i
			\ishap
			\partial_z \uL^{i}(z+x_2)
			+
			\sdap \uk^i_{T a=2}
			\frac{1}{z}
			\right)
			e^{-i n \frac{ \uL^{+} (z+x_2) }{\oldap \uk^+_{T a=2}} } 
			\nonumber\\
			&
			\phantom{\exp}
			(x_1 - x_2 - z)^{ -n \frac{\uk^+_{T a=1} }{\uk^+_{T a=22}} }
			\nonumber\\
			&
			\phantom{\exp}
			\prod_{b=3}^{M}
			(x_2 -x_b + z)^{ -n \frac{\uk^+_{T b} }{\uk^+_{T a=2}} }
			\Bigg)
			\,
			e^{i \sdap\uk_{T a=2}^i \uL^i(x_2) - i \sdap\uk_{T a=2}^- \uL^+(x_2)}
			:
			\,
			(x_1-x_2)^{\dap \uk_{T a=2}^- \uk_{T b=1}^+}
			\prod_{b=3}^M
			(x_2-x_b)^{\dap \uk_{T a=2}^- \uk_{T b}^+}
			\nonumber\\
			%
			% exp 2 quadratic
			%  %
			\exp
			&
			\Bigg(
			%+
			\oh
			\sum_{i=1}^{d-2}
			\sum_{n_1, n_2=1}^\infty
			\lambda^{a=2}_{n_1\, i}
			\lambda^{a=2}_{n_2\, i}
			\oint_{z_1=0, |z_1| > |z_2|} \frac{d z_1}{ 2\pi i}
			\oint_{z_2=0} \frac{d z_2}{ 2\pi i}
			\frac{
				e^{
					-i
					\frac{
						n_{ 1}\uL^+(z_1+x_2)
						+
						n_{ 2}\uL^+(z_2+x_2)
					}{\oldap \uk^+_{T a=2} }
				}
			}{
				z_1^{n_1} z_2^{n_2} (z_1 -z_2)^2
			}
			\nonumber\\
			&
			\phantom{\exp}
			(x_1 - x_2 - z_1)^{ -n_1 \frac{\uk^+_{T b=1} }{\uk^+_{T a=2}} }
			(x_1 - x_2 - z_2)^{ -n_2 \frac{\uk^+_{T b=1} }{\uk^+_{T a=2}} }
			\nonumber\\
			&
			\phantom{\exp}
			\prod_{b=3}^{M}
			(x_2 - x_b + z_1)^{ -n_1 \frac{\uk^+_{T b} }{\uk^+_{T a=2}} }
			(x_2 - x_b + z_2)^{ -n_2 \frac{\uk^+_{T b} }{\uk^+_{T a=2}} }
			\Bigg)
			\nonumber
		\end{align}
		
		\begin{align}
			%
			% exp N linear
			%
			\times
			:
			\exp
			&
			\Bigg(
			\sum_{i=1}^{d-2}
			\sum_{n=1}^\infty
			\lambda^{a=M}_{n\, i}
			\oint_{z=0} \frac{d z}{ 2\pi i}
			\frac{1}{z^n}
			\left(
			i
			\ishap
			\partial_z \uL^{i}(z+x_M)
			+
			\sdap \uk^i_{T a=M}
			\frac{1}{z}
			\right)
			e^{-i n \frac{ \uL^{+} (z+x_M) }{\oldap \uk^+_{T a=M}} } 
			\nonumber\\
			&
			\phantom{\exp}
			\prod_{b=1}^{M-1}
			(x_b - x_M- z)^{ -n \frac{\uk^+_{T b} }{\uk^+_{T a=M}} }
			\Bigg)
			\,
			e^{i \sdap\uk_{T a=M}^i \uL^i(x_M) - i \sdap\uk_{T a=M}^- \uL^+(x_M)}
			:
			\,
			\prod_{b=1}^{M-1}
			(x_b-x_M)^{\dap \uk_{T a=M}^- \uk_{T b}^+}
			\nonumber\\
			%
			% exp N quadratic
			%
			\exp
			&
			\Bigg(
			%  -
			\oh
			\sum_{i=1}^{d-2}
			\sum_{n_1, n_2=1}^\infty
			\lambda^{a=M}_{n_1\, i}
			\lambda^{a=M}_{n_2\, i}
			\oint_{z_1=0, |z_1| > |z_2|} \frac{d z_1}{ 2\pi i}
			\oint_{z_2=0} \frac{d z_2}{ 2\pi i}
			\frac{
				e^{
					-i
					\frac{
						n_{ 1}\uL^+(z_1+x_M)
						+
						n_{ 2}\uL^+(z_2+x_M)
					}{\oldap \uk^+_{T a=M} }
				}
			}{
				z_1^{n_1} z_2^{n_2} (z_1 -z_2)^2
			}
			\nonumber\\
			&
			\phantom{\exp}
			\prod_{b=1}^{M-1}
			(x_b - x_M - z_1)^{ -n_1 \frac{\uk^+_{T b} }{\uk^+_{T a=M}} }
			(x_b - x_M - z_2)^{ -n_2 \frac{\uk^+_{T b} }{\uk^+_{T a=M}} }
			\Bigg)
			\nonumber\\
			&
			| \up^-=\up^i=0, 0_{a} \rangle
			\,
			\delta\left( \sum_{b=1}^M \uk^+_{b} \right)
			\nonumber\\
			.
		\end{align}

			Notice that $\sum_{b=1}^M \uk^+_{b}=0$ requires at least one
			$\uk^+<0$. This means a state created on $\langle 0|$ by $\uA_{+n}$.
			This state is the hermitian conjugate of the ones considered and can
			be obtained by taking $\uk^+<0$ in the computed vertex and sending
			$x\rightarrow\infty$.

		Now we can act with $\uL^+$ on the vacua and
		the only non trivial contribution is  from $\uxzero^+$, i.e.
		$  e^{-i n \frac{ \uL^{+} (z+\xr) }{\oldap \uk^+_{T b}} } 
		\rightarrow
		e^{ -i n \frac{\uxzero^+}{\dueoldap \uk^+_{T b}} }
		$.

			Actually
			the previous substitution in terms with $\partial \uL^i$ cannot be
			done in this way without a justification.
		The reason is that the $\partial \uL^i$s at different $\xr$  do not
		commute and this does not allow to act with $\uL^+$ on the vacuum.
		The result is however right. The reason is the following.
		We keep $\uL^+$ as they are and we normal order $\partial \uL^i$s.
		This yield terms with $\partial \uL^{i (+)}$ on the right,
		with $\partial \uL^{i (-)}$ on the left and terms with c-numbers
		coming from the normal ordering.
		The terms with $\partial \uL^{i (\pm)}$ vanish and we are left with
		terms with c-numbers
		coming from the normal ordering.
		In these latter term we can perform the stated substitution.
		Reversing the process we see that we can make the substitution at the
		beginning.

		The result of the previous computation can be written in a compact way as
		\begin{align}
			\cW_M
			=&
			\langle \up^-=\up^i=0, 0_{a} |
			\nonumber\\
			%
			% exp number a linear
			%
			\prod_{a=1}^M
			\Bigg\{
			:
			\exp
			&
			\Bigg(
			\sum_{i=1}^{d-2}
			\sum_{n=1}^\infty
			\lambda^a_{n\, i}
			\oint_{z=0} \frac{d z}{ 2\pi i}
			\frac{1}{z^n}
			\left(
			i
			\ishap
			\partial_z \uL^{i}(z+x_a)
			+
			\sdap \uk^i_{T a}
			\frac{1}{z}
			\right)
			e^{-i n \frac{ \uxzero^+ }{\dueoldap \uk^+_{T a}} } 
			\nonumber\\
			&
			\phantom{\exp}
			\prod_{b=1}^{a-1}
			(x_b - x_a - z)^{ -n \frac{\uk^+_{T b} }{\uk^+_{T a}} }
			\nonumber\\
			&
			\phantom{\exp}
			\prod_{b=a+1}^{M}
			(x_a - x_b + z)^{ -n \frac{\uk^+_{T b} }{\uk^+_{T a}} }
			\Bigg)
			\,
			e^{i \sdap\uk_{T a}^i \uL^i(x_a) 
				- i  \uk_{T a^-} \uxzero^+ }
			:
			\nonumber\\
			%
			% exp number a quadratic
			%  %
			\exp
			&
			\Bigg(
			%  +
			\oh
			\sum_{i=1}^{d-2}
			\sum_{n_1, n_2=1}^\infty
			\lambda^a_{n_1\, i}
			\lambda^a_{n_2\, i}
			\oint_{z_1=0, |z_1| > |z_2|} \frac{d z_1}{ 2\pi i}
			\oint_{z_2=0} \frac{d z_2}{ 2\pi i}
			\frac{
				e^{
					-i
					\frac{
						n_{ 1}  \uxzero^+
						+
						n_{ 2} \uxzero^+
					}{\dueoldap \uk^+_{T a} }
				}
			}{
				z_1^{n_1} z_2^{n_2} (z_1 -z_2)^2
			}
			\nonumber\\
			&
			\phantom{\exp}
			\prod_{b= 1}^{a-1}
			(x_b - x_a - z_1)^{ -n_1 \frac{\uk^+_{T a} }{\uk^+_{T b}} }
			(x-b - x_a - z_2)^{ -n_2 \frac{\uk^+_{T a} }{\uk^+_{T b}} }
			\nonumber\\
			&
			\phantom{\exp}
			\prod_{b=a+1}^{M}
			(x_a - x_b + z_1)^{ -n_1 \frac{\uk^+_{T b} }{\uk^+_{T a}} }
			(x_a - x_b + z_2)^{ -n_2 \frac{\uk^+_{T b} }{\uk^+_{T a}} }
			\Bigg)
			\Bigg\}
			\nonumber\\
			&
			| \up^-=\up^i=0, 0_{a} \rangle
			\,
			\prod_{a=1}^{M-1}
			\prod_{b=a+1}^M
			(x_a-x_b)^{\dap \uk_{T a}^- \uk_{T b}^+}  
			\,
			\delta\left( \sum_{b=1}^M k^+_{b} \right)
			.
		\end{align}
		
		We are finally left with normal order the $\uL^i$ and let them act on
		the vacuum.
		We use
		\begin{align}
			&
			e^{\oint  f_i(z) \partial \uL^{i (+)}(z+\xr)} \,
			e^{i \sdap\uk_{T b i} \uL^{i (+)}(x_b)}
			\,\times\,
			e^{\oint   g_i(z) \partial \uL^{i (-)}(z+x_a)} \,
			e^{i \sdap\uk_{T a j} \uL^{j (-)}(x_a)} 
			\nonumber\\
			=&
			e^{\oint   g_i(z)
				\left[
				\partial \uL^{i (-)}(z+x_a)
				+
				i
				\frac{\sdap \uk^i_{T a} }{ z +x_a -x_b}
				\right]
			} \,
			e^{i \sdap\uk_{\snt T j} \uL^{j (-)}(x_a)} 
			\,\times\,
			e^{\oint  f_i(z)
				\left[
				\partial \uL^{i (+)}(z+x_b)
				+
				i
				\frac{\sdap\uk^i_{T a} }{ z +x_b -x_a}
				\right]
			} \,
			e^{i \sdap\uk_{T b i} \uL^{i (+)}(x_b)}
			\nonumber\\
			&
			\times
			e^{
				-
				\oh\oldap
				\oint_{z_1=0} f_i(z_1)
				\oint_{z_2=0} g_i(z_2)
				\frac{1}{( z_1 -z_2 +\xr -\xt )^2}
			}
			\,
			(x_b - x_a)^{\samedap \uk^i_{T br} \uk^i_{T a} }
			.
		\end{align}

		If we define
		\begin{equation}
			x_{a b} = x_a - x_b
			,
		\end{equation}
		to get the compact expression
		\begin{align}
			\cW_M&(\{x\};\, \{\lambda\};\, \{p_T\}  
			)
			=
			\langle \up^i=\up^-=0|
			\nonumber\\
			%
			% exp number t linear
			%
			\prod_{a=1}^M
			\Bigg\{
			\exp
			&
			\Bigg(
			\sum_{i=1}^{d-2}
			\sum_{n=1}^\infty
			\lambda^a_{n\, i}
			\oint_{z=0} \frac{d z}{ 2\pi i}
			\frac{1}{z^n}
			\left(
			\sum_{b\ne a}
			\frac{\sdap \uk^i_{T a} }{z +x_{a b} }
			+
			\sdap \uk^i_{T a}
			\frac{1}{z}
			\right)
			e^{-i n \frac{ \uxzero^+ }{\dueoldap \uk^+_{T a}} } 
			\nonumber\\
			&
			\phantom{\exp}
			\prod_{b=1}^{a-1}
			(x_{b a}  - z)^{ -n \frac{\uk^+_{T b} }{\uk^+_{T a}} }
			\,
			\prod_{b=a+1}^{M}
			(x_{a b} + z)^{ -n \frac{\uk^+_{T b} }{\uk^+_{T a}} }
			\Bigg)
			\nonumber\\
			\nonumber
		\end{align}
		\begin{align}
			%
			% exp number t quadratic self interaction
			%  %
			\times
			\exp
			&
			\Bigg(
			% + 
			\oh
			\sum_{i=1}^{d-2}
			\sum_{n_1, n_2=1}^\infty
			\lambda^a_{n_1\, i}
			\lambda^a_{n_2\, i}
			\oint_{z_1=0, |z_1| > |z_2|} \frac{d z_1}{ 2\pi i}
			\oint_{z_2=0} \frac{d z_2}{ 2\pi i}
			\frac{
				e^{
					-i
					\frac{
						n_{ 1}  \uxzero^+
						+
						n_{ 2} \uxzero^+
					}{\dueoldap \uk^+_{T a} }
				}
			}{
				z_1^{n_1} z_2^{n_2} (z_1 -z_2)^2
			}
			\nonumber\\
			&
			\phantom{\exp}
			\prod_{b=1}^{a-1}
			(x_{b a} - z_1)^{ -n_1 \frac{\uk^+_{T b} }{\uk^+_{T a}} }
			\prod_{b=a+1}^{M}
			(x_{a b}     + z_1)^{ -n_1 \frac{\uk^+_{T b} }{\uk^+_{T a}} }
			\nonumber\\
			&
			\phantom{\exp}
			\prod_{b=1}^{a-1}
			(x_{b a}    - z_2)^{ -n_2 \frac{\uk^+_{T \sns} }{\uk^+_{T \snt}} }
			\,
			\prod_{b=a+1}^{M}
			(x_{a b}    + z_2)^{ -n_2 \frac{\uk^+_{T b} }{\uk^+_{T a}} }
			\Bigg)
			\Bigg\}
			\nonumber\\
			\nonumber
		\end{align}
		%%%%%%%
		\begin{align}
			%
			% exp number a  bquadratic 
			%  %
			\times
			\exp
			&
			\Bigg(
			%  +
			\sum_{i=1}^{d-2}
			\sum_{n_1, n_2=1}^\infty
			\sum_{a=1}^{M-1}
			\sum_{b=a+1}^M
			\lambda^a_{n_1\, i}
			\lambda^b_{n_2\, i}
			\oint_{z_1=0, |z_1| > |z_2|} \frac{d z_1}{ 2\pi i}
			\oint_{z_2=0} \frac{d z_2}{ 2\pi i}
			\frac{
				e^{
					-i
					\frac{
						n_{ 1}  \uxzero^+
					}{\dueoldap \uk^+_{T a} }
					-i
					\frac{
						n_{ 2} \uxzero^+
					}{\dueoldap \uk^+_{T b} }
				}
			}{
				z_1^{n_1} z_2^{n_2} (z_1 -z_2 +x_{a b}  )^2
			}
			\nonumber\\
			&
			\phantom{\exp}
			\prod_{c=1}^{a-1}
			(x_{c a}   - z_1)^{ -n_1 \frac{\uk^+_{T c} }{\uk^+_{T a}} }
			\prod_{c=a+1}^{M}
			(x_{a c}   + z_1)^{ -n_1 \frac{\uk^+_{T c} }{\uk^+_{T a}} }
			\nonumber\\
			&
			\phantom{\exp}
			\prod_{c=1}^{b-1}
			(x_{c b}    - z_2)^{ -n_2 \frac{\uk^+_{T c} }{\uk^+_{T b}} }
			\,
			\prod_{c=b+1}^{M}
			(x_{b c}    + z_2)^{ -n_2 \frac{\uk^+_{T c} }{\uk^+_{T b}} }
			\Bigg)
			\Bigg\}
			\nonumber
		\end{align}
		%%%%%%%%%
		\begin{align}
			&
			\prod_{\ttt 1}^M
			e^{i  \sdap \uk_{T \TTT}^i \uxzero^i 
				- i \sdap \uk_{T \TTT}^- \uxzero^+}
			\,
			| \up^i=\up^-=0 \rangle
			\,
			\nonumber\\
			&
			\prod_{\rrr 1}^{M-1} \prod_{\ttt \RRR+1}^M
			(\xr - \xt)^{\dap \uk_{T \RRR} \cdot \uk_{T \TTT} }
			\,
			\delta\left( \sum_{\rrr 1}^M \uk^+_{\RRR} \right)
			.
		\end{align}

		To make contact with the result given in the main text and the form of
		the vertex which uses the projector we notice that
		\begin{equation}
			\oint_{z=0} \frac{d z}{ 2\pi i}
			%  \frac{1}{z^n}
			\left(
			\sum_{b=1}^M
			\frac{ \uk^+_{T b} }{z + x_{a b}  }
			\right)
			\prod_{b=1}^{a-1}
			(x_{b a}  - z)^{ -n \frac{\uk^+_{T b} }{\uk^+_{T a}} }
			\,
			z^{ -n \frac{\uk^+_{T a} }{\uk^+_{T a}} } 
			\,
			\prod_{b=a+1}^{m}
			(x_{a b}    + z)^{ -n \frac{\uk^+_{T b} }{\uk^+_{T a}} }
			=0
			,
		\end{equation}
		since the integrand is a total derivative.
		
		Finally the quadratic terms can be cast in a unique form to get
		\begin{align}
			\cW_M
			=&
			\langle \up^i=\up^-=0|
			\nonumber\\
			%
			% exp number t linear
			%
			\prod_{a=1}^M
			\Bigg\{
			\exp
			&
			\Bigg(
			\sum_{i=1}^{d-2}
			\sum_{n=1}^\infty
			\lambda^a_{n\, i}
			\,
			\oldI^{i}_{a, n}(\{x\};\, \{\uk^i\};\, \{\uk^+\}
			%% ;\, \uxzero
			)\,
			e^{-i n \frac{ \uxzero^+ }{\dueoldap \uk^+_{T a}} } 
			\Bigg)
		\end{align}
		%%%%%%%
		\begin{align}
			%
			% exp number t quadratic  both self and inter
			%  %
			\exp
			&
			\Bigg(
			% +
			\oh
			\sum_{i=1}^{d-2}
			\sum_{n_1, n_2=1}^\infty
			\sum_{a,b= 1}^{M}
			\lambda^a_{n_1\, i}\,
			\lambda^b_{n_2\, i}\,
			\oldJ_{a, n_1;\, b, n_2}( \{x\};\, \{\uk^+\} )\,
			e^{
				-i
				\frac{
					n_{ 1}  \uxzero^+
				}{\dueoldap \uk^+_{T a} }
				-i
				\frac{
					n_{ 2} \uxzero^+
				}{\dueoldap \uk^+_{T b} }
			}
			%% \oint_{z_1=0, |z_1| > |z_2|} \frac{d z_1}{ 2\pi i}
			%% \oint_{z_2=0} \frac{d z_2}{ 2\pi i}
			%% \frac{
				%%   e^{
					%%     -i
					%%     \frac{
						%%       n_{ 1}  \uxzero^+
						%%     }{\dueoldap \uk^+_{T \snr} }
					%%     -i
					%%     \frac{
						%%       n_{ 2} \uxzero^+
						%%     }{\dueoldap \uk^+_{T \snt} }
					%%   }
				%% }{
				%%   z_1^{n_1} z_2^{n_2} (z_1 -z_2 +\xxrrtt  )^2
				%% }
			%% \nonumber\\
			%% &
			%% \phantom{\exp}
			%% \prod_{\sss 1}^{\RRR-1}
			%% (\xxssrr   - z_1)^{ -n_1 \frac{\uk^+_{T \sns} }{\uk^+_{T \snr}} }
			%% \prod_{\sss \RRR+1}^{N}
			%% (\xxrrss   + z_1)^{ -n_1 \frac{\uk^+_{T \sns} }{\uk^+_{T \snr}} }
			%% \nonumber\\
			%% &
			%% \phantom{\exp}
			%% \prod_{\sss 1}^{\TTT-1}
			%% (\xxsstt    - z_2)^{ -n_2 \frac{\uk^+_{T \sns} }{\uk^+_{T \snt}} }
			%% \,
			%% \prod_{\sss \TTT+1}^{N}
			%% (\xxttss    + z_2)^{ -n_2 \frac{\uk^+_{T \sns} }{\uk^+_{T \snt}} }
			\Bigg)
			\Bigg\}
			%
			%%  \nonumber\\
			%%  \nonumber
		\end{align}
		%%%%%%%%%
		\begin{align}
			\times
			&
			\prod_{a=1}^M
			e^{i \sdap \uk^{a i}_{T} \uxzero^i 
				- i \sdap  \uk^{a -}_{T} \uxzero^+}
			\,
			| \up^i=\up^-=0 \rangle
			\,
			%
			%% \nonumber\\
			%% \times
			%% &
			\prod_{a= 1}^{M-1} \prod_{b=a+1}^M
			\xxrrtt ^{\dap \uk^a_{T} \cdot \uk^b_{T} }
			\,
			\delta\left( \sum_{a=1}^M \uk^+_{a} \right)
			,
		\end{align}
		where we have introduced
		\begin{align}
			\oldI^{i}_{a, n}(\{x\};\, \{\uk^i\});\, \{\uk^+\}
			=&
			\oint_{z=0} \frac{d z}{ 2\pi i}
			\frac{1}{z^n}
			\sum_{b \ne a}
			\frac{\sdap  }{z + x_{a b}  }
			\left(
			\uk^i_{T b}
			-
			\frac{ \uk^+_{T b} }{ \uk^+_{T a} }
			\uk^i_{T a}  
			\right)
			%  e^{-i n \frac{ \uxzero^+ }{\dueoldap \uk^+_{T a}} } 
			\nonumber\\
			&
			\phantom{\exp}
			\prod_{b=1}^{a-1}
			(x_{b a}  - z)^{ -n \frac{\uk^+_{T b} }{\uk^+_{T a}} }
			\,
			\prod_{\rrr \TTT+1}^{N}
			(x_{a b}  + z)^{ -n \frac{\uk^+_{T b} }{\uk^+_{T a}} }
			,
		\end{align}
		and
		\begin{align}
			%%  \oldJ_{\snr n_1; \snt n_2}( \uk_{\sns},\, x_\SSS)
			\oldJ_{a, n_1;\, b, n_2}( \{x\};\, \{\uk^+\}
			%%x_a,\, \uk^a;\, x_b,\, \uk^b
			%%  ;\, \uxzero
			)
			=&
			\oint_{z_1=0, |z_1| > |z_2|} \frac{d z_1}{ 2\pi i}
			\oint_{z_2=0} \frac{d z_2}{ 2\pi i}
			\frac{
				1
				%%    e^{
					%%      -i
					%%      \frac{
						%%        n_{ 1}  \uxzero^+
						%%      }{\dueoldap \uk^+_{T a} }
					%%      -i
					%%      \frac{
						%%        n_{ 2} \uxzero^+
						%%      }{\dueoldap \uk^+_{T b} }
					%%    }
			}{
				z_1^{n_1} z_2^{n_2} (z_1 -z_2 +x_{a b}  )^2
			}
			\nonumber\\
			&
			\phantom{\exp}
			\prod_{c=1}^{a-1}
			(x_{c a}   - z_1)^{ -n_1 \frac{\uk^+_{T c} }{\uk^+_{T a}} }
			\prod_{c=a+1}^{M}
			(x_{a c}   + z_1)^{ -n_1 \frac{\uk^+_{T c} }{\uk^+_{T a}} }
			\nonumber\\
			&
			\phantom{\exp}
			\prod_{c=1}^{b-1}
			(x_{c b}    - z_2)^{ -n_2 \frac{\uk^+_{T c} }{\uk^+_{T b}} }
			\,
			\prod_{c=b+1}^{m}
			(x_{b c}    + z_2)^{ -n_2 \frac{\uk^+_{T c} }{\uk^+_{T b}} }
			.
		\end{align}

		Upon the same steps as in section \ref{sect:compaing single vertex},
		i.e. integration by part as in Eq.\eqref{eq:integrals in usual DDF vertex}
		and mapping Eq.\eqref{app:mapping framed usual}  we 
		get Eqs.\eqref{25},\eqref{26}.

		\section{Covariant polarizations from DDF-states}
		\label{Covpol}
		In open  bosonic string in the critical dimension $d=26$ the spectrum of the physical states in the covariant formalism is  obtained by applying on the lightest tachyon state with momentum $p$, i.e. $|0, \,p \rangle$, an arbitrary number of creation operators:
		\begin{eqnarray}
			\epsilon_{\mu_1\dots \mu_k} \,\alpha_{-n_1}^{\mu_1}\dots \alpha_{n_k}^{\mu_k}|0,\,p\rangle\label{1}
		\end{eqnarray}
		Their masses are given by:
		\begin{eqnarray}
			\alpha' m^2= N-1~~;~~N= \sum_{m=1}^\infty :\alpha_{-n}\cdot \alpha_n:
		\end{eqnarray}
		where $N$ is the level of the excited state.
		These states have positive norm and their polarization tensors $\epsilon_{\mu_1\dots \mu_k}$  are restricted by the so-called Virasoro conditions which have to be imposed, in the old covariant approach, to obtain physical states. However,  in the spectrum of the theory,  together with the positive norm states there are also other physical states, the {\em  Brower states},  having  zero-norm. They are obtained by solving the conditions:
		\begin{eqnarray}
			&&L_{-1}|\chi\rangle =0 ~~;~~L_m|\chi\rangle=0~~, \,m\geq 1~~;~~L_0|\chi\rangle=0\nonumber\\
			&&\left( L_{-2} +\frac{3}{2} L^2_{-1}\right)|\tilde{\chi}\rangle=0~~;~~ L_m| \tilde{\chi}\rangle=0~~,\,m\geq 1~~;~~(L_0+1)|\tilde{\chi}\rangle=0\label{3}
		\end{eqnarray}	
		At each level $N$ of the massive string spectrum, two types of physical null norm states must be considered alongside the positive norm states. Although these null-norm states do not contribute to scattering amplitudes and are often neglected, they play a crucial role in the theory. Specifically, Brower states, when added to the physical states described in Eq. \eqref{1}, induce transformations in the polarization tensors that correspond to residual on-shell "gauge" transformations permitted by the Virasoro conditions \cite{Lee:1989rc,Lee:1989rd}. These residual gauge transformations can be utilized to recast the covariant polarizations of the massive string states into forms that can be identified according to the representations of the Poincaré group.
		
		In the following subsections, we will derive the relationship between DDF states and covariant polarizations for the first massive string levels, showing the role of residual gauge transformations in this context.

		\subsection{Massive $N=2$ level}
		The positive norm states of the first $N=2$ massive level are collected by the expression:
		\begin{eqnarray}
			(T_\mu\alpha^\mu_{-2} +S_{\mu\nu}\,\alpha_{-1}^\mu\,\alpha_{-1}^\nu)|0,p\rangle~~;~~S_{\mu\nu}=S_{\nu\mu}\label{1.4}
		\end{eqnarray}
		The Virasoro constraints determine:
		\begin{eqnarray}
			\sqrt{2\alpha'} p^\mu\,S_{\mu\nu}+T_\nu=0~~;~~S^\mu_{~\mu}+2\sqrt{2\alpha'} p^\mu\,T_\mu=0\label{1.5}
		\end{eqnarray}		
		The norm null states of the level $N=2$
		\begin{eqnarray}
			&&(\xi\cdot \alpha_{-2}+\sqrt{2\alpha'} p\cdot \alpha_{-1}\,\xi\cdot \alpha_{-1})|0,p\rangle ~,~~p\cdot \xi=0\nonumber\\
			&& c\left[(\eta^{\mu\nu}+\alpha' \frac{(d+4)}{5}\,p_\mu p_\nu)\alpha_{-1}^\mu\,\alpha_{-1}^\nu +  \sqrt{2\alpha'} \frac{(d-1)}{5} p\cdot \alpha_{-2}\right]|0,p\rangle\nonumber\\
		\end{eqnarray} 
		when added to those written in Eq.\eqref{1.4}, change the polarization tensors as follows 
		\begin{eqnarray}
			S'_{\nu\mu}= S_{\nu\mu} +\sqrt{\frac{\alpha'}{2}}(p_\mu\xi_\nu+p_\nu\xi_\mu) +c\eta_{\mu\nu}+c\,\alpha'\frac{(d+4)}{5}\,p_\mu\,p_\nu ~~;~~T'_\nu=T_\nu+ c\,\sqrt{2\alpha'} \frac{(d-1)}{5} p_\nu+\xi_\nu\nonumber\\
		\end{eqnarray}
		One can easily verify that these transformations are the residual gauge transformations allowed by Virasoro's conditions. They depend on $d$-independent parameters, $\xi$ and $c$, and allow to gauge away the polarization vector $\epsilon_\nu=0$. The physical state is, therefore,  a massive spin-$2$ particle described by a symmetric and traceless  polarization tensor.
		This construction can be made explicit by choosing the gauge transformation parameter $\xi_\nu$ such that the transformed polarization tensor $\epsilon'_\nu$ is vanishing:
		\begin{eqnarray}
			\xi_\nu=- T_\nu - \sqrt{2\alpha'} \,c \frac{d-1}{5} p_\nu\qquad;\qquad p\cdot \xi=0\Rightarrow c= \sqrt{\frac{\alpha'}{2} } \frac{5\,p\cdot \epsilon}{(d-1)}\label{gcov}
		\end{eqnarray}
		Within this gauge transformation we have:
		\begin{eqnarray}\sqrt{2\alpha'} \,p^\mu\,S'_{\mu\nu}= T_\nu+\sqrt{2\alpha'} p^\mu\,S_{\mu\nu}=0~~;~~S^{'\mu}_{~\mu}=S^\mu_{~\mu} + 2 \sqrt{2\alpha'} \, p^\mu T_\mu=0\label{A.88}
		\end{eqnarray}
		where the identity to zero is a consequence of the Virasoro's conditions \eqref{1.5}. In the new gauge we have:
		\begin{eqnarray}
			T'_\mu=0~~;~~ S'_{\mu\nu}=S_{\mu\nu}-\sqrt{\frac{\alpha'}{2}}\left(p_\mu\,T_\nu+p_\nu\,T_\mu\right) -\frac{p\cdot T}{\sqrt{2} (d-1)}\Bigg( -5\eta_{\mu\nu} +\alpha' (d-6) p_\mu\,p_\nu\Bigg) \label{A.89}
		\end{eqnarray}

		It is also interesting to observe that making the peculiar choice of the DDF-polarization $\lambda^i=\lambda^i_{~i}=0$ with  $\epsilon_{(1)\mu}^i=\delta^i_\mu$ and $k=(k^-,0,\dots, 0)$,  we  get a covariant polarization which is transverse and traceless. Therefore, this choice of the DDF-polarizations  my provide a convenient partial parametrization of  $S'_{\mu\nu}$, i.e.:
		\begin{eqnarray}
			S'_{\mu\nu}&&\equiv\lambda_{ij}\,\varepsilon^i_{(1)\mu } \,\varepsilon_{(1)\nu}^j~~;~~\lambda^i_{~i}=0\label{s'2}
		\end{eqnarray}
		Eq. \eqref{s'2} doesn't parameterize the full degrees of freedom of the covariant two indices symmetric and traceless tensor, but as we saw in Section \ref{Amplitude}, for the states lying on the leading Regge trajectory, this parametrization yields the correct expression for the full amplitude.
		\subsection{Massive $N=3$ level}
		
		The massive string states at the level $N=3$ are obtained by acting on the lowest string states with  all the possible creation operators giving such a string level:  
		\begin{eqnarray}
			\left[\xi_\mu\alpha^\mu_{-3} +\xi_{\mu\nu}\, \alpha^\mu_{-2}\,\alpha_{-1}^\nu+\xi_{\mu\nu\rho} \,\alpha^\mu_{-1}\,\alpha^\nu_{-1} \alpha^\rho_{-1}\right]|0,\,p\rangle~~;~~\mu,\nu,\,\rho=0\dots d-1\label{42}
		\end{eqnarray}
		and imposing the Virasoro's condition:
		\begin{eqnarray}
			&&\overbracket{3\xi_\mu+\sqrt{2\alpha'} p^\nu\xi_{\mu\nu}=0}^{L_1=0}~~;~~\overbracket{\xi_{\mu\nu}+\xi_{\nu\mu}+ 3\sqrt{2\alpha'} \xi_{\mu\nu\rho}p^\rho=0}^{L_1=0}\nonumber\\
			&&\overbracket{2\sqrt{2\alpha'} p^\mu\xi_{\mu\nu} + 3\xi_\nu + 3 \eta^{\mu\rho}\xi_{\mu\rho\nu}=0}^{L_2=0}~~;~~\overbracket{3\sqrt{2\alpha'}p^\mu\xi_\mu+ 2\eta^{\mu\nu}\xi_{\mu\nu}=0}^{L_3=0} \label{2.28}
		\end{eqnarray}
		Together with these states with positive norm, we have also the physical norm null states:
		\begin{eqnarray}
			&&L_{-1}\left[\epsilon_\mu\alpha^\mu_{-2} +\epsilon_{\mu\nu}\alpha_{-1}^\mu\alpha_{-1}^\nu\right]|0,p\rangle~~;~~\Big[L_{-2}+\frac{3}{2} L_{-1}^2\,\epsilon_\mu^{(1)}\alpha^\mu_{-1}\Big]|0,p\rangle~~;~~p^2=-\frac{2}{\alpha'}
		\end{eqnarray}
		satisfying the conditions given in eq.\eqref{3}. These give imposes $p\cdot\epsilon_\mu^{(1)}=0$ and on $\epsilon_\mu$ and $\epsilon_{\nu\mu}$ the conditions given in Eq. \eqref{1.5}.
		We can no count the number of degrees of freedom of the states on such a level. The counting is correct only in $d=26$ therefore in the following we assume such a value for the space-time dimensions.    
		The states written in Eq. \eqref{42} together with Virasoro's conditions have 
		\begin{eqnarray}
			d +d^2 +\frac{1}{3!} d(d+1)(d+2)-d -\left[\frac{1}{2} d(d+1)-1\right]-d-1
		\end{eqnarray}
		degrees of freedom\footnote{Naively one could write that the conditions coming from the second constraint in Eq. \eqref{2.28} are those of a symmetric tensor. This is not correct because the condition coming from the  trace of such an equation is already taken in account by the other three Virasoro constraints.} The polarizations of the physical norm null states are the parameters of the    gauge transformations that can gauge away  $(d-1)+\frac{1}{2}(d-1)(d+2)$ components of the polarization tensor. These are the vector $\xi_\mu=0$, the symmetric part of the tensor $\xi_{\mu\nu}+\xi_{\nu\mu}=0$ and gauge away $d-1$-components of the three indices symmetric  tensor $\xi_{\mu\nu\rho}$. Therefore, the number of degrees of freedom after the gauge fixing are:
		\begin{eqnarray}
			&&d +d^2 +\frac{1}{3!} d(d+1)(d+2)-d -\left[\frac{1}{2} d(d+1)-1\right]-d-1-(d-1)-\frac{1}{2}(d-1)(d+2)\nonumber\\
			&&=\frac{1}{3!} (d-1) d(d+1)-(d-1) +\frac{1}{2} (d-1)(d-2)
		\end{eqnarray}
		This number corresponds to the degrees of freedom of  the   traceless three indices symmetric representation and the two-index antisymmetric representation  of the little group $SO(d-1)$, with $d=26$.  %The physical states turn out to be:
%		\begin{eqnarray}
%			\Big[\Big(\xi'_{\mu\nu} -\xi'_{\nu\mu}\Big)\alpha_{-1}^\mu\alpha_{-2}^\nu+\xi'_{\mu\nu\rho}\,\alpha_{-1}^\mu\alpha_{-1}^\nu\alpha_{-1}^\rho\Big]|0,p\rangle\label{cgauge3}
%		\end{eqnarray} 
%		where we have denoted the polarization with a prime because, similar to the case of the level $N=2$, they are the gauge-transformed versions of the initial polarizations.
%		In this gauge the \eqref{2.28} become:
%		\begin{eqnarray}
%			p^\mu\xi'_{\mu\nu}=-p^\mu\xi'_{\nu\mu}=0~~;~~p^\rho\xi'_{\rho\mu\nu}=0~~;~~\eta^{\mu\rho}\xi'_{\mu\rho\nu}=0\label{48}
%		\end{eqnarray}
		This physical state with the DDF-operator is given by the expression:
		\begin{eqnarray}
			\left[\lambda_i\,\epsilon^i_{(3)\mu}\, A_{-3}^\mu+ \lambda_{ij} \epsilon_{(2)\mu}^i\,\epsilon_{(1) \mu}^j\,A_{-2}^\mu\,A_{-1}^\nu+\lambda_{ijk}\epsilon_{(1) \mu}^i\,\epsilon_{(1) \nu}^j\epsilon_{(1) \rho}^k\,\alpha_{-1}^\mu\,\alpha_{-1}^\nu\,\alpha_{-1}^\rho\right]|0,p_T\rangle
		\end{eqnarray}
		The mapping between the $SO(d)$-covariant polarizations and $SO(d-2)$ DDF-polarization is obtained by evaluating:
		\begin{eqnarray}
			&&A_{-3}^i |0,p_T\rangle=\epsilon_{(3)\mu}^i\sum_{m=0}^n \frac{\alpha_{-m}^\mu}{(n-m)!} \,\frac{\partial^{n-m}}{\partial  z^{n-m} }e^{-n\sqrt{2\alpha'} \sum_{p=1}^\infty \frac{\alpha_{-p}\cdot k}{p}z^p}\Bigg|_{n=3} |0,p_T-3k\rangle\nonumber\\
			&&= \Big[\varepsilon_{(3)}\cdot \alpha_{-3} +9\alpha' \, \varepsilon_{(3)\sigma}^i\,k_\nu\,k_\rho \alpha_{-1}^\sigma\,\alpha_{-1}^\nu\,\alpha_{-1}^\rho -\frac{ 3}{2}\sqrt{2\alpha'}\left(2 \varepsilon_{(3)\rho}^i\,k_\nu+\varepsilon_{(3)\nu}^i\,k_\rho\right) \alpha_{-1}^\nu\,\alpha_{-2}^\rho\Big]|0,\,p\rangle\nonumber\\
			&&A_{-2}^i \,A_{-1}^j|0,p_T\rangle=\epsilon_{(2)\nu}^i \sum_{m=0}^{n=2}  \alpha_{-m}^\nu \frac{1}{(n-m)!}\frac{\partial^{n-m}}{\partial z^{n-m}} e^{-2\sqrt{2\alpha'} \sum_{p=1}^\infty \frac{\alpha_{-p}\cdot k}{p} z^p}\Bigg|_{z=0} \varepsilon^j_{(1)}\cdot \alpha_{-1}|0,\,p\rangle\nonumber\\
			&&+\varepsilon_{(2)}^i\cdot \varepsilon_{(1)}^j \,\frac{1}{3!}\frac{\partial^3}{\partial z^3}e^{-2\sqrt{2\alpha'} \sum_{p=1}^\infty \frac{\alpha_{-p}\cdot k}{p} z^p}\Bigg|_{z=0} | 0,\,p\rangle\nonumber\\
			&&=\epsilon_{(2)\nu}^i\Big[(\delta^\nu_\mu -2\alpha' p_T^\nu k_\mu)\alpha_{-2}^\mu -2 (\delta^\nu_\mu -2\alpha' p_T^\nu k_\mu)\alpha_{-1}^\mu\, \sqrt{2\alpha'}k\cdot \alpha_{-1}\Big] \varepsilon^j_{(1)}\cdot \alpha_{-1}|0,\,p\rangle\nonumber\\
			&&-\varepsilon_{(2)\nu}^i\,\varepsilon_{(1)\mu}^j\,\eta^{\mu\nu} \, \frac{2}{3}\Big[ \sqrt{2\alpha'}k\cdot \alpha_{-3} - 3\sqrt{2\alpha'} k \cdot\alpha_{-2}\,\sqrt{2\alpha'} k\cdot \alpha_{-1} + 2(\sqrt{2\alpha'}k\cdot \alpha_{-1})^3\Big]|0,\,p\rangle
		\end{eqnarray}
		and finally:
		\begin{eqnarray}
			&&A_{-1}^i\,A_{-1}^j\,A_{-1}^k|0,p_T\rangle=\varepsilon^i_{(1)}\cdot \alpha_{-1} \varepsilon_{(1)}^j\cdot \alpha_{-1}\,\varepsilon_{(1)}^k\cdot \alpha_{-1}|0,p_T-3k\rangle\nonumber\\
			&&+\frac{1}{2}\left[\varepsilon_{(1)}^i\cdot \varepsilon_{(1)}^j \,\varepsilon_{(1)}^k\cdot \alpha_{-1} +\varepsilon_{(1)}^i\cdot \varepsilon_{(1)}^k \,\varepsilon_{(1)}^j\cdot \alpha_{-1}]\right]\left( 2\alpha'(k\cdot \alpha_{-1})^2-\sqrt{2\alpha'} k\cdot \alpha_{-2}\right)|0,p_T-3k\rangle \nonumber\\
			&&-\varepsilon^j_{(1)}\cdot \varepsilon^k_{(1)}\sqrt{\frac{\alpha'}{2} }\varepsilon_{(1)}^i\cdot \alpha_{-1}  (k\cdot \alpha_{-2} -\sqrt{2\alpha'} (k\cdot \alpha_{-1})^2)|0,p_T-3k\rangle
		\end{eqnarray}
		The mapping with the covariant formalism gives:
		\begin{eqnarray}
			S_\mu&&= \lambda_i\varepsilon_{(3)\mu}^i
			-\frac{2}{3} \lambda_{ij} \varepsilon_{(2)}^i\cdot \varepsilon^j_{(1)}\sqrt{2\alpha'} k_\mu\nonumber\\
			S_{\rho\nu}&&=-3 \sqrt{2\alpha'}\lambda_i\varepsilon_{(3)\rho}^i\,k_\nu -\frac{3}{2} \sqrt{2\alpha'}\lambda_i\varepsilon_{(3)\nu}^i\,k_\rho+\lambda_{ij}\varepsilon_{(2)\rho}^i\,\varepsilon_{(1)\nu}^j+\lambda_{ij}  \varepsilon_{(2)}^i\cdot \varepsilon_{(1)}^j \,2\, (2\alpha')k_\rho\,k_\nu\nonumber\\
			&&-\sqrt{\frac{\alpha'}{2}}S_{ijk}( \varepsilon_{(1)}^i\cdot\varepsilon_{(1)}^j \varepsilon_{(1)\nu}^k + \varepsilon_{(1)}^i\cdot\varepsilon_{(1)}^k \varepsilon_{(1)\nu}^j+\varepsilon_{(1)}^j\cdot\varepsilon_{(1)}^k \varepsilon_{(1)\nu}^i)k_\rho \nonumber\\
			S_{\mu\nu\sigma}&&=\lambda_{ijk} \varepsilon^i_{(1)\sigma}\varepsilon^j_{(1)\nu}\varepsilon^k_{(1)\mu}+(2\alpha')\frac{3}{2} \left( \lambda_i\varepsilon_{(3)\sigma}^ik_\nu k_\mu+\lambda_i\varepsilon_{(3)\nu}^ik_\sigma k_\mu+\lambda_i\varepsilon_{(3)\mu}^ik_\nu k_\sigma \right)\nonumber\\
			&&-2\sqrt{2\alpha'}\frac{1}{6}\left(\lambda_{ij} \varepsilon^i_{(2)\sigma} \, 
			k_\nu\,\varepsilon^j_{(1)\mu}+\lambda_{ij} \varepsilon^i_{(2)\sigma} \, 
			k_\mu\,\varepsilon^j_{(1)\nu}+\lambda_{ij} \varepsilon^i_{(2)\nu} \, 
			k_\sigma\,\varepsilon^j_{(1)\mu}+\lambda_{ij} \varepsilon^i_{(2)\nu} \, 
			k_\mu\,\varepsilon^j_{(1)\sigma} \right.\nonumber\\ 
			&&\left.+\lambda_{ij} \varepsilon^i_{(2)\mu}\,k_\nu\,\varepsilon^j_{(1)\sigma}+\lambda_{ij} \varepsilon^i_{(2)\mu} \, 
			k_\sigma\,\varepsilon^j_{(1)\nu}\right)-\frac{4}{3} \lambda_{ij} \varepsilon_{(2)}^i \cdot \varepsilon_{(1)}^j(2\alpha')^\frac{3}{2}k_\sigma\,k_\nu\,k_\mu\nonumber\\
			&& +\alpha'\frac{1}{3}\lambda_{ijk}\left[ (\varepsilon_{(1)}^i\cdot \varepsilon_{(1)}^j\varepsilon_{(1)\sigma}^k+\varepsilon_{(1)}^i\cdot \varepsilon_{(1)}^k\varepsilon_{(1)\sigma}^j+\varepsilon_{(1)}^j\cdot \varepsilon_{(1)}^k\varepsilon_{(1)\sigma}^i)k_\nu k_\mu\right.\nonumber\\
			&&+(\varepsilon_{(1)}^i\cdot \varepsilon_{(1)}^j\varepsilon_{(1)\nu}^k+\varepsilon_{(1)}^i\cdot \varepsilon_{(1)}^k\varepsilon_{(1)\nu}^j+\varepsilon_{(1)}^j\cdot \varepsilon_{(1)}^k\varepsilon_{(1)\nu}^i)k_\sigma k_\mu\nonumber\\
			&&\left. +(\varepsilon_{(1)}^i\cdot \varepsilon_{(1)}^j\varepsilon_{(1)\mu}^k+\varepsilon_{(1)}^i\cdot \varepsilon_{(1)}^k\varepsilon_{(1)\mu}^j+\varepsilon_{(1)}^j\cdot \varepsilon_{(1)}^k\varepsilon_{(1)\mu}^i)k_\nu k_\sigma\right]\label{53}
		\end{eqnarray}
		One can verify that such tensors satisfy Virasoro's conditions given in Eq.\eqref{2.28}. 
		We observe that the DDF-parametrization of the covariant polarizations is not expressed in the gauge specified by the Equation \eqref{48}. Consequently, as in the case of the DDF-state at level n=2, a further gauge transformation is required to obtain  polarization tensors transverse  and in the case of the symmetric state also traceless.
		
		\subsection{Massive $N=4$ level}
		The most general DDF- state of the level $N=4$ is:
		\begin{eqnarray}
			&&|N=4\rangle= \left[\lambda_i\epsilon_{(4)\mu}^i\,A^\mu_{-4}+ \lambda_{ij} \epsilon_{(3)\mu}^i\, \epsilon_{(1)\nu}^j\,A_{-3}^\mu\,A_{-1}^\nu +S_{ij}\,\epsilon_{(2)\mu}^i\, \epsilon_{(2)\nu}^j\,A_{-2}^i\,A_{-2}^j\right.\nonumber\\
			&&\left.+ \lambda_{ijk}\, \epsilon_{(2)\mu_1}^i\, \epsilon_{(1)\mu_2}^j\,\epsilon_{(1)\mu_3}^k\,A_{-2}^{\mu_1}\,A_{-1}^{\mu_2}\,A_{-1}^{\mu_3}+ S_{ijkl}     \epsilon_{(1)\mu}^i\epsilon_{(1)\nu}^j\,\epsilon_{(1)\rho}^k\,\epsilon_{(1)\sigma}^l
			\,A_{-1}^\mu\,A_{-1}^\nu \, A_{-1}^\rho\,A_{-1}^\sigma \right]|0,\,p_T\rangle\nonumber\\
			&&\label {171L}
		\end{eqnarray} 
		The counting of the degrees of freedom is the same as in the light cone gauge. We have $(d-2)+(d-2)^2 + \frac{1}{2} (d-2)(d-1)+ \frac{1}{2} (d-2)(d-1)(d-2)+\frac{1}{4!} (d-2)(d-1)d(d+1)$ components, with $d=26$. These correspond to a symmetric, transverse and traceless rank-4 tensor 
		and a symmetric, transverse and traceless, two indices tensor, 
		a  transverse and traceless rank-3 tensor and 
		a scalar particle\cite{Blumenhagen:2013fgp}.
		
		The mapping between the covariant and DDF- polarization is obtained, as usual,  by acting with the DDF-operators on the lowest tachyon state. The first term in eq. \eqref{171L} gives  ($\tk=\sqrt{2\alpha'} k$):
		\begin{eqnarray}
			&&A_{-4}^i |0,p_T\rangle=\epsilon_{(3)\mu}^i\sum_{m=0}^n \frac{\alpha_{-m}^\mu}{(n-m)!} \,\frac{\partial^{n-m}}{\partial  z^{n-m} }e^{-n\sqrt{2\alpha'} \sum_{p=1}^\infty \frac{\alpha_{-p}\cdot k}{p}z^p}\Bigg|_{n=4} |0,p_T-4k\rangle\nonumber\\
			&&=\left[\varepsilon_{(4)}\cdot \alpha_{-4}
			-4 \varepsilon_{(4)}^b\, \tk^a\, \alpha_{-1\,a}\,\alpha_{-3b}-\frac{4}{3} \varepsilon_{(4)}^a \tk^b\alpha_{-1a }\alpha_{-3 b}\right.\nonumber\\
			&&-2 \varepsilon_{(4)}^a\tk^b \alpha_{-2a}\alpha_{-2b}+8\varepsilon_{(4)}^a\,\tk^b\,\tk^c \alpha_{-2a}\alpha_{-1b}\,\alpha_{-1c}+8\varepsilon_{(4)}^a\,\tk^b\,\tk^c\,
			\alpha_{-2a} \alpha_{-1b}\,\alpha_{-1c}\nonumber\\
			&&\left.-\frac{32}{3} \varepsilon_{(4)}^a\, \tk^b\,\tk^c\,\tk^d \alpha_{-1a}\alpha_{-1b}\,\alpha_{-1c}\alpha_{-1d}\right]|0,\,p_T-4 k\rangle
		\end{eqnarray}
		The next contribution is: 
		\begin{eqnarray}
			&&A_{-3}^i\,A_{-1}^j |0,\,p_T\rangle= \epsilon_{(3)\nu}^i\oint \frac{dz}{2\pi i}\,\frac{1}{x^4}\,\left[\sum_{n=0}^\infty \alpha_{-n}^\nu\, x^n+\frac{\alpha_1^\nu}{x}\right]\,e^{-3\sqrt{2\alpha'}\sum_{n=1}^\infty \frac{k\cdot \alpha_{-n}}{n} x^n}\varepsilon_{(n)}^j\cdot \alpha_{-1}|0,\,p_T-4k\rangle\nonumber\\
			&&= \left[\varepsilon_{(3)}^i\cdot \alpha_{-3}- 3\varepsilon_{(3)}^i \cdot \alpha_{-2} \,\tk \cdot\alpha_{-1}-\frac{3}{2}\tk\cdot \alpha_{-2} \,\varepsilon_{(3)}^i\cdot \alpha_{-1} +\frac{9}{2}\varepsilon_{(3)}^i\cdot \alpha_{-1}\, \tk\cdot \alpha_{-1} \tk \cdot \alpha_{-1} \right]\varepsilon_{(1)}^j \cdot \alpha_{-1}|0,\,p_T-4k\rangle\nonumber\\
			&&-\frac{3}{8} \varepsilon_{(3)}^i\cdot \varepsilon_{(1)}^j\left(2 \tk\cdot \alpha_{-4} -3(\tk \cdot \alpha_{-2})^2 -8\tk \cdot \alpha_{-3} \,\tk\cdot \alpha_{-1} +18\tk\cdot \alpha_{-2} (\tk \cdot \alpha_{-1})^2-9 (\tk\cdot \alpha_{-1})^4\right)|0,\,p_T-4k\rangle\nonumber\\
		\end{eqnarray}
		The action of the symmetric DDF-operators gives:
		\begin{eqnarray}
			&&A^i_{-2} \,A^j_{-2} |0,p_T\rangle= A_{-2}^i \,\left(\varepsilon_{(2)}^j\cdot  \alpha_{-2} -2 \varepsilon_{(2)}^j\cdot\alpha_{-1}\,\tk\cdot  \alpha_{-1}\right) |0,p_T-2k\rangle\nonumber\\
			&&=\left[\varepsilon_{(2)}^i\cdot \alpha_{-2} - 2 \varepsilon_{(2)}^i \cdot \alpha_{-1}\,\tk \alpha_{-1}\right] \left( \varepsilon_{(2)}^j\cdot\alpha_{-2} -2 \varepsilon_{(2)}^j\cdot\alpha_{-1}\,\tk\cdot  \alpha_{-1}\right) |0,\,p_T-4k\rangle\nonumber\\
			&&+\frac{4}{3} \varepsilon_{(2)}^i\cdot \varepsilon_{(2)}^j \left( \tk \cdot \alpha_{-3} - 3 \tk \cdot \alpha_{-2}\,\tk\cdot \alpha_{-1} + 2(\tk \cdot \alpha_{-1})^3\right) \tk \cdot \alpha_{-1}|0,\,p_T-4k\rangle\nonumber\\
			&&+\frac{1}{3} \varepsilon_{(2)}^i\cdot \varepsilon_{(2)}^j \left( -3 \tk \cdot \alpha_{-4} + 3(\tk \cdot \alpha_{-2})^2 + 8 \tk \cdot \alpha_{-3}\,\tk \cdot\alpha_{-1} -12 \tk \cdot \alpha_{-2}\,(\tk \cdot \alpha_{-1})^2+4(\tk \cdot \alpha_{-1})^4 \right)|0,\,p_T-4k\rangle\nonumber\\
		\end{eqnarray}
		The other term gives:
		\begin{eqnarray}
			&&A_{-2}^i \,A_{-1}^j\,A_{-1}^j|0,p_T\rangle = A_{-2}^i \,A_{-1}^j\,\varepsilon_{(1)}^k\cdot \alpha_{-1}|0,p_T-k\rangle \nonumber\\
			&&=\sum_{m=0}^2\epsilon^i_{(2)}\cdot \alpha_{-m} \frac{1}{(2-m)!}\frac{\partial^{2-m}}{\partial x^{2-m}}e^{-{\color{red} 2}\sum_{n=1}^\infty \frac{\tk \cdot \alpha_{-n}}{n} x^n}\Bigg|_{x=0}\left[\dots \right] |0,p_T-4k\rangle\nonumber\\
			&&+\frac{1}{3!}\left[\varepsilon_{(2)}^i\cdot \varepsilon_{(1)}^j\,\varepsilon_{(1)}^k\cdot \alpha_{-1}+\varepsilon_{(2)}^i\cdot \varepsilon_{(1)}^k\,\varepsilon_{(1)}^j\cdot \alpha_{-1}\right]\frac{\partial^3}{\partial x^3}e^{-{\color{red} 2}\sum_{n=0}^\infty \frac{\tk \cdot \alpha_{-n}}{n} x^n}\Bigg|_{x=0} |0,p_T-4k\rangle\nonumber\\
			&&=\left( \varepsilon_{(2)}^i \cdot \alpha_{-2} -2 \varepsilon_{(2)}^i \cdot \alpha_{-1}\,\tk \cdot \alpha_{-1}\right)\left[ \varepsilon_{(1)}^j \cdot \alpha_{-1} \,\varepsilon_{(1)}^k \cdot \alpha_{-1} -\frac{1}{2} \varepsilon_{(1)}^j\cdot \varepsilon_{(1)}^k\left( \tk\cdot \alpha_{-2} -(\tk \cdot \alpha_{-1})^2\right)\right]|0,p_T-4k\rangle\nonumber\\
			&&-{\frac{
					2}{3}}\left[\varepsilon_{(2)}^i\cdot \varepsilon_{(1)}^j\,\varepsilon_{(1)}^k\cdot \alpha_{-1}+\varepsilon_{(2)}^i\cdot \varepsilon_{(1)}^k\,\varepsilon_{(1)}^j\cdot \alpha_{-1}\right]{\left( \tk \cdot \alpha_{-3} - 3 \tk \cdot \alpha_{-2} \, k\cdot \alpha_{-1}+2(\tk \cdot \alpha_{-1})^3\right)}|0,p_T-4k\rangle\nonumber\\
		\end{eqnarray}
		The last quantity to compute is:
		\begin{eqnarray}
			&&A_{-1}^i\,A_{-1}^j\,A_{-1}^k \,A_{-i}^\gamma|0,\,p_T\rangle=A_{-1}^i\left[\varepsilon_{(1)}^j \cdot \alpha_{-1}\, \varepsilon_{(1)}^k \cdot \alpha_{-1} \, \varepsilon_{(1)}^\gamma \cdot \alpha_{-1}\right.\nonumber\\
			&&\left.+\frac{1}{2} \left( \varepsilon_{(1)}^j\cdot \varepsilon_{(1)} ^k \, \varepsilon_{(1)}^\gamma \cdot \alpha_{-1} +\varepsilon_{(1)}^j\cdot \varepsilon_{(1)}^\gamma\,\varepsilon_{(1)}^k \cdot \alpha_{(1)}+\varepsilon_{(1)}^k\cdot \varepsilon_{(1)}^\gamma\,\varepsilon_{(1)}^j\cdot \alpha_{-1}\right) \left((\tk \cdot \alpha_{-1})^2 -\tk \cdot\alpha_{-2}\right)\right]|0,p_T-3k\rangle \nonumber\\
			&&=\varepsilon_{(1)}^i\cdot \alpha_{-1}\,\left[\varepsilon_{(1)}^j \cdot \alpha_{-1}\, \varepsilon_{(1)}^k \cdot \alpha_{-1} \, \varepsilon_{(1)}^\gamma \cdot \alpha_{-1}\right.\nonumber\\
			&&\left.+\frac{1}{2} \left( \varepsilon_{(1)}^j\cdot \varepsilon_{(1)} ^k \, \varepsilon_{(1)}^\gamma \cdot \alpha_{-1} +\varepsilon_{(1)}^j\cdot \varepsilon_{(1)}^\gamma\,\varepsilon_{(1)}^k \cdot \alpha_{(1)}+\varepsilon_{(1)}^k\cdot \varepsilon_{(1)}^\gamma\,\varepsilon_{(1)}^j\cdot \alpha_{-1}\right) \left((\tk \cdot \alpha_{-1})^2 -\tk \cdot\alpha_{-2}\right)\right]|0,p\rangle\nonumber\\
			&&+\left[\varepsilon_{(1)}^j\cdot \varepsilon_{(1)}^i \,\varepsilon_{1}^k \cdot \alpha_{-1}\, \varepsilon_{(1)}^\gamma\cdot \alpha_{-1}+\varepsilon_{(1)}^i\cdot \varepsilon_{(1)}^k \,\varepsilon_{1}^j \cdot \alpha_{-1}\, \varepsilon_{(1)}^\gamma\cdot \alpha_{-1} +\varepsilon_{(1)}^i\cdot \varepsilon_{(1)}^\gamma \,\varepsilon_{1}^j \cdot \alpha_{-1}\, \varepsilon_{(1)}^k\cdot \alpha_{-1}\right.\nonumber\\
			&&\left.+\frac{1}{2} \left( \varepsilon_{(1)}^j\cdot \varepsilon_{(1)}^k \, \varepsilon_{(1)}^\gamma\cdot \varepsilon_{(1)}^i+ \varepsilon_{(1)}^j\cdot \varepsilon_{(1)}^\gamma \, \varepsilon_{(1)}^k\cdot \varepsilon_{(1)}^i+ \varepsilon_{(1)}^\gamma\cdot \varepsilon_{(1)}^k \, \varepsilon_{(1)}^j\cdot \varepsilon_{(1)}^i\right)\left( (\tk \cdot \alpha_{-1})^2-\tk \cdot \alpha_{-2}\right)\right]\nonumber\\
			&&\times{ \frac{1}{2}}\left((\tk \cdot \alpha_{-1})^2 -\tk \cdot \alpha_{-2}\right)|0,p\rangle
		\end{eqnarray}
		By equating Eq.\eqref{171L} with the covariant representation of the states of this level, as given in Eq. \eqref{69L}, we get the map between the covariant and DDF-polarizations:
		\begin{eqnarray}
			&&\xi_\mu= \lambda_i\varepsilon_{(4)\mu}^i -{\frac{3}{4}} \lambda_{ij}\varepsilon_{(3)}^i\cdot \varepsilon_{(1)}^j \tk_\mu -S_{ij} \varepsilon_{(2)}^i\cdot \varepsilon_{(2)}^j\, \tk_\mu \nonumber\\
			&&\xi_{\mu\nu}= - 4 \lambda_i \varepsilon_{(4)\mu}^i  \tk_\nu  -\frac{4}{3} \lambda_i\varepsilon_{(4)\nu }\,\tk_\mu^i+\lambda_{ij} \varepsilon_{(3)\mu}^i \,\varepsilon_{(1)\nu}^j+ 3\lambda_{ij} \varepsilon_{(3)}^i\cdot \varepsilon_{(1)}^j\, \tk_\mu\,\tk_\nu+4 S_{ij}\varepsilon_{(2)}^i\cdot\varepsilon_{(2)}^j\, \tk_\mu\,\tk_\nu\nonumber\\
			&&-{\frac{2}{3}} \lambda_{ijk}\left( \varepsilon_{(2)}^i\cdot \varepsilon_{(1)}^j\,\varepsilon_{(1)\nu}^k+\varepsilon_{(2)}^i\cdot \varepsilon_{(1)}^k\,\varepsilon_{(1))\nu}\right) \tk_\mu  \nonumber\\
			&&S_{\mu\nu}= -2 \lambda_i\varepsilon_{(4)\{\mu}^i\,\tk_{\nu\}}+\frac{9}{8} \lambda_{ij}\,\varepsilon_{(3)}^i \cdot \varepsilon_{(1)}^j\tk_\mu\,\tk_\nu +S_{ij} \varepsilon_{(2)\mu}^i \, \varepsilon_{(2)\nu}^j  +S_{ij} \,\varepsilon_{(2)}^i\cdot \varepsilon_{(2)}^j \,\tk_\mu\,\tk_\nu -\frac{1}{2} \lambda_{ijk} \varepsilon_{(2)\{\mu}^i\,\varepsilon_{(1)}^j\cdot \varepsilon_{(1)}^k \,\tk_{\nu\}} \nonumber\\
			&&+{\frac{1}{4}} S_{ijk\gamma} \left( \varepsilon_{(1)}^j\cdot \varepsilon_{(1)}^k \, \varepsilon_{(1)}^\gamma\cdot \varepsilon_{(1)}^i+ \varepsilon_{(1)}^j\cdot \varepsilon_{(1)}^\gamma \, \varepsilon_{(1)}^k\cdot \varepsilon_{(1)}^i+ \varepsilon_{(1)}^\gamma\cdot \varepsilon_{(1)}^k \, \varepsilon_{(1)}^j\cdot \varepsilon_{(1)}^i\right)\,\tk_\mu\,\tk_\nu\nonumber\\
			&&\xi_{\mu\nu\rho}= 8\lambda_i\varepsilon_{(4)\mu}^i\tk_\nu\tk_\rho+ 8\lambda_i\varepsilon_{(4)\{\nu}^i\tk_{\rho\}}\tk_\mu -\frac{3}{2}\lambda_{ij}  \tk_\mu \varepsilon_{(3)\{\nu}^i\,\varepsilon_{(1)\rho\}}^j - \frac{3}{2}\lambda_{ij} \varepsilon_{(3)\mu}^i  \,\left(\varepsilon_{(1)\nu}^j\tk_\rho+\varepsilon_{(1)\rho}^j\tk_\nu\right)\nonumber\\
			&& -\frac{27}{4} \lambda_{ij}\, \varepsilon_{(3)}^i \cdot \varepsilon_{(1)}^j\tk_\mu\tk_\nu\tk_\rho-2 S_{ij} \varepsilon_{(2)\mu}^i\,(\varepsilon_{(2)\nu}^j\tk_\rho+ \varepsilon_{(2)\rho}^j\tk_\nu)-8 S_{ij}\,\varepsilon_{(2)} ^i \cdot \varepsilon_{(2)}^j\ \tk_\mu\,\tk_\nu\,\tk_\rho    +\lambda_{ijk} \,\varepsilon_{(2)\mu}^i \varepsilon_{(1)\nu}^j\varepsilon^k_{(1)\rho} \nonumber\\
			&&+\frac{1}{2} \lambda_{ijk} \varepsilon_{(2)\mu}^i \varepsilon_{(1)}^i \cdot \varepsilon_{(1)}^k\tk_\nu\tk_\rho +\lambda_{ijk} \varepsilon_{(2)\{\nu}^i\tk_{\rho\}} \,\varepsilon_{(1)}^j \cdot \varepsilon_{(1)}^k \tk_\mu+{ 2} \lambda_{ijk}\left[\varepsilon_{(2)}^i\cdot \varepsilon_{(1)}^j\,\varepsilon_{(1)\rho}^k+\varepsilon_{(2)}^i\cdot \varepsilon_{(1)}^k\,\varepsilon_{(1)\rho}^j\right]  \tk_\mu\,\tk_\nu\nonumber\\
			&&  -\frac{1}{2}S_{ijk\gamma} \varepsilon_{(1)\nu}^i  \left( \varepsilon_{(1)}^j\cdot \varepsilon_{(1)} ^k \, \varepsilon_{(1)\rho}^\gamma +\varepsilon_{(1)}^j\cdot \varepsilon_{(1)}^\gamma\,\varepsilon_{(1)\rho}^k +\varepsilon_{(1)}^k\cdot \varepsilon_{(1)}^\gamma\,\varepsilon_{(1)\rho}^j\right) \tk_{\mu}\nonumber\\
			&&-\frac{1}{2} S_{ijk\gamma}\left(\varepsilon_{(1)}^j\cdot \varepsilon_{(1)}^i \,\varepsilon_{(1)\nu}^k\, \varepsilon_{(1)\rho}^\gamma+\varepsilon_{(1)}^i\cdot \varepsilon_{(1)}^k \,\varepsilon_{(1)\nu}^j \, \varepsilon_{(1)\rho}^\gamma +\varepsilon_{(1)}^i\cdot \varepsilon_{(1)}^\gamma \,\varepsilon_{(1)\nu}^j \, \varepsilon_{(1)\rho}^k\right)\tk_\mu \nonumber\\
			&&{-\frac{1}{2}S_{ijk\gamma} \left( \varepsilon_{(1)}^j\cdot \varepsilon_{(1)}^k \, \varepsilon_{(1)}^\gamma\cdot \varepsilon_{(1)}^i+ \varepsilon_{(1)}^j\cdot \varepsilon_{(1)}^\gamma \, \varepsilon_{(1)}^k\cdot \varepsilon_{(1)}^i+ \varepsilon_{(1)}^\gamma\cdot \varepsilon_{(1)}^k \, \varepsilon_{(1)}^j\cdot \varepsilon_{(1)}^i\right)\tk_\mu\,\tk_\nu\,\tk_\rho}\nonumber\\
			&&S_{\mu\nu\rho\gamma}= -\frac{32}{3} \lambda_i \,\varepsilon_{(4)\{\mu}^i\,\tk_\nu\tk_\rho\tk_{\gamma\}}+\frac{9}{2} \lambda_{ij} \varepsilon^i_{(3)\{\mu}\varepsilon_{(1)\nu}^j\,\tk_\rho\,\tk_{\gamma\}} +\frac{27}{8} \,\lambda_{ij}\varepsilon_{(3)}^i\cdot \varepsilon_{(1)}^j \tk_\mu\,\tk_\nu\,\tk_\rho\,\tk_\gamma +4  S_{ij}\varepsilon^i_{(2)\{\mu}\varepsilon_{(2)\nu}^j\tk_\rho\tk_{\gamma\}}\nonumber\\
			&&+4 S_{ij} \varepsilon_{(2)}^i\cdot \varepsilon^j_{(2)} \tk_\mu\tk_\nu\tk_\rho\tk_\gamma{-2\lambda_{ijk}\varepsilon_{(2)\{\mu}^i\, \varepsilon_{(1)\nu}^j\varepsilon_{(1)\rho}^k \tk_{\gamma\}}} -\lambda_{ijk} \varepsilon_{(2)\{\mu}^i \varepsilon_{(1)}^j \cdot \varepsilon_{(1)}^k \tk_\nu\tk_\rho\tk_{\gamma\}} \nonumber\\
			&&-{\frac{4}{3}} \lambda_{ijk} \left[\varepsilon_{(2)}^i\cdot \varepsilon^j_{(1)} \,\varepsilon_{(1)\{\mu}^k+\varepsilon_{(2)}^i \cdot \varepsilon_{(1)}^k \varepsilon_{(1)\{\mu}^j\right]\tk_\nu \tk_\rho\tk_{\gamma\}}+S_{ijkl} \varepsilon_{(1)\mu}^i\varepsilon_{(1)\nu}^j \varepsilon_{(1)\rho}^k\varepsilon_{(1)\gamma}^l\nonumber\\
			&&+\frac{1}{2} S_{ijkl}  \left( \varepsilon_{(1)}^j\cdot \varepsilon_{(1)} ^k \, \varepsilon_{(1)\{\mu}^l +\varepsilon_{(1)}^j\cdot \varepsilon_{(1)}^l\,\varepsilon_{(1)\{\mu}^k +\varepsilon_{(1)}^k\cdot \varepsilon_{(1)}^l\,\varepsilon_{(1)\{\mu}^j\right)\varepsilon^i_{(1)_\nu}\tk_\rho\tk_{\gamma\}}\nonumber\\
			&&+\frac{1}{2} S_{ijkl}\left[\varepsilon_{(1)}^j\cdot \varepsilon_{(1)}^i \,\varepsilon_{(1)\{\mu}^k \, \varepsilon_{(1)\nu}^l+\varepsilon_{(1)}^i\cdot \varepsilon_{(1)}^k \,\varepsilon_{(1)\{\mu}^j \, \varepsilon_{(1)\nu}^l+\varepsilon_{(1)}^i\cdot \varepsilon_{(1)}^l \,\varepsilon_{(1)\{\mu}^j \, \varepsilon_{(1)\nu}^k\right]\tk_{\rho}\tk_{\gamma\}}\nonumber\\
			&&+{\frac{1}{4}} S_{ijkl}\left( \varepsilon_{(1)}^j\cdot \varepsilon_{(1)}^k \, \varepsilon_{(1)}^l\cdot \varepsilon_{(1)}^i+ \varepsilon_{(1)}^j\cdot \varepsilon_{(1)}^l \, \varepsilon_{(1)}^k\cdot \varepsilon_{(1)}^i+ \varepsilon_{(1)}^l\cdot \varepsilon_{(1)}^k \, \varepsilon_{(1)}^j\cdot \varepsilon_{(1)}^i\right)\tk_{\mu}\tk_{\nu}\tk_{\rho}\tk_{\gamma}\nonumber\\
			\label{6.67a}
		\end{eqnarray}
		where we have denoted with $\star_{\{\mu_1}\dots\star_{\mu_n\}}=\frac{1}{n!} \left[\star_{\mu_1}\dots \star_{\mu_n}+{\rm perm.}\right]$.
		
		%% appendix on integrals
	
		\newcommand{\rud}{\rho_{21}}
		\newcommand{\rutre}{\rho_{31}}
		\newcommand{\rdt}{\rho_{32}}
		\newcommand{\rdu}{\rho_{12}}
		\newcommand{\rtd}{\rho_{23}}
		\newcommand{\rtu}{\rho_{13}}
		\newcommand{\ruc}{\rho_{c1}}
		
		\section{Correlation functions: single and double integrals }
		\label{app:threeIntegral}
		In this section, we compute the single and double integral appearing
		in the expression of the $N$-point amplitude in the case of three and
		four generic external coherent states.

		\subsection{Single three-point integrals}
		\label{three-point integral}
		We start by considering the single integral given in Eq.s \eqref{25}
		and \eqref{34}, here cited:
		\begin{eqnarray}
			{\cal  A}_{a,n_a}&&= \oint_{x_a} \frac{dz_a}{2\pi i} \,\frac{ \prod_{b >a}(z_a-x_b)^{-  n_a \rab } \,\prod_{b <a}(x_b-z_a)^{-  n_a\rab } }{(z_a-x_a)^{n_a}} \nonumber\\
			&&\times\left[\sum_{b\neq a }\frac{\sqrt{2\alpha'}\lambda^a_{n_a\,i} \, P^i_{ab}}{z_a-x_b}\right] \,,\nonumber\\
			\label{25A}
		\end{eqnarray} 
		in the case of three external DDF-states, i.e. $a,b=1,\,2,\,3$.
		The sum inside the integral is computed through the identity:
			\begin{eqnarray}
				\sum_{b\neq a=1}^3  \frac{P_{ab}^i}{z-x_b}= \left[\frac{1}{z-x_b}-\frac{1}{z-x_{a-1}}\right]P^i_{ab}~,
			\end{eqnarray}
			with the identification $x_3\equiv x_0$.
		The integral is easily evaluated by introducing the cross ratio which moves the pole from the Koba-Nielsen variable $x_a$ to the origin of the complex plane $z_a$. This choice is not unique,  and we provide details in the case of ${\cal  A}(\lambda^1_{n_1},\,x_1,\,k^1)$ where we define:
		\begin{eqnarray}
			y= \frac{(z_1-x_1)}{(z_1- x_2)}\frac{(x_3-x_2)}{(x_3-x_1)}%~~;~~y= \frac{(w-z_1)}{(w- z_3)}\frac{(z_2-z_3)}{z_2-z_1}~.
			\label{194A}
		\end{eqnarray}
		The integral turns out to be:
		\begin{eqnarray}
			{\cal  A}_{1,n_1}= && (x_{12}\,x_{13})^{-n_1}x_{12}^{-n_1\,\rud} \,x_{13}^{-n_1\,k_1\,\rutre}\,x_{23}^{n_1}\,
			\oint_0\frac{dy}{2\pi i} \frac{1}{y^{n_1}} (1-y)^{n_1\,\rud+n_1-1}
			\,\lambda_{n_a,i}\, P^i_{12}~.\nonumber\\                        
		\end{eqnarray}
		By observing that:
		\begin{eqnarray}
			\oint_0\frac{dy}{2\pi i} \frac{1}{y^{n_1}} (1-y)^{n_1\,\rud+n_1-1}&&
			=\frac{1}{(n_1-1)!}\frac{\partial^{n_1-1}}{\partial y^{n_1-1}}(1-y)^{ n_1\,\rud+n_1-1}\Bigg|_{y=0}\nonumber\\
			&&
			=(-1)^{n-1}\left( \begin{array}{c}
				n_1\,\rud+n_1-1\\
				n_1-1\end{array}\right)~,
		\end{eqnarray}
		we have\cite{Ademollo:1974kz}:
		\begin{eqnarray}
			&&{\cal  A}_{1,n_1}=\left(\frac{x_{23}}{x_{12}\,x_{13}}\right)^{n_1}x_{12}^{- \,n_1\,\rud} \,x_{13}^{-n_1\,\rud}\, (-1)^{n_1-1}\left( \begin{array}{c}
				n_1\,\rud+n_1-1\\
				n_1-1\end{array}\right)\, \lambda_{n_1,\,i}^1\,P^i_{12}~.\nonumber\\
		\end{eqnarray}
		Along the same line, one can compute the other two single integrals by getting:
		\begin{eqnarray}
			&&{\cal  A}_{2,n_2}=\left(\frac{x_{13}}{x_{23}\,x_{12}}\right)^{n_2}x_{23}^{- \,n_2\,\rdt} \,x_{12}^{-\,n_2\,\rdu}\, (-1)^{n_2-1}\left( \begin{array}{c}
				n_2\,\rdt+n_2-1\\
				n_2-1\end{array}\right)\, \lambda_{n_2,\,i}^2\,P_{23}^i\nonumber\\
		\end{eqnarray}
		and:
		\begin{eqnarray}
			&&{\cal  A}_{3,n_3}=\left(\frac{x_{12}}{x_{13}\,x_{2-3}}\right)^{n_3}x_{13}^{- \,n_3\,\rtu} \,x_{23}^{-n_3\,\rtd}\, (-1)^{n_3-1}\left( \begin{array}{c} n_3\,\rtu+n_3-1\\
				n_3-1\end{array}\right)\, \lambda_{n_3,\,i}^3\,P^i_{31}\nonumber\\
		\end{eqnarray}
		\subsection{ Double three-point integrals}
		\label{F.2}
		The second typology of integrals to compute are those defined in Eq.\eqref{26} with $x_a=x_b$, here  rewritten in the case $a=1$:
		\begin{eqnarray}
			{\cal  B}_{1;n_1,m_1}&&=\oint_{x_1} \frac{dz_1}{2\pi i}\frac{  \prod_{c >1}(z_1-x_c)^{-  n_1\,\ruc }}{ (w_1-x_1)^{n_1}}\, \oint_{x_1} \frac{dw_1}{2\pi i}\frac{ \prod_{c >1}(w_1-x_c)^{-  {m_1}\,\ruc } \,     }{(w_1-x_1)^{{m_1}}}\,\frac{1}{(z_1-w_1)^2}\,.\nonumber\\
			\label{26A}
		\end{eqnarray}
		To move the poles at the origin of the complex plane, we introduce for both variables for both variable $z_1$ and $w_1$ the same cross-ratio defined in \eqref{194A}. In the new variables, the integrals become:
		\begin{eqnarray}
			{\cal  B}_{1;n_1,m_1}&&= x_{12}^{-(n_1+{m_1})\,\rud}\,x_{13}^{- \,(n_1+{m_1})\,\rutre}\,\left(\frac{x_{23}}{x_{12}\,x_{13}}\right)^{{m_1}+\,n_1} \nonumber\\
			&&\times\oint_0 \frac{dy_1}{2\pi i} \frac{(1-y_1)^{-n_1\,\rutre}}{y_1^{n_1}}\,\oint_0 \frac{dy_2}{2\pi i} \frac{(1-y_2)^{-{m_1}\,\rdt}}{y_2^{{m_1}}}\,\,\frac{1}{(y_1-y_2)^2}~.\nonumber\\
		\end{eqnarray}
		We expand in $y_2/y_1$ getting:
		\begin{eqnarray}
			\frac{1}{(y_1-y_2)^2} = \frac{1}{y_1^2} \sum_{h=0}^\infty (h+1)\left( \frac{y_2}{y_1}\right)^h =\sum_{h=1}^\infty   h\,y_2^{h-1} \,y_1^{-h-1}~.\label{5.103}
		\end{eqnarray} 
		The two integrals can now be easily evaluated:
		\begin{eqnarray}
			&&{\cal  B}_{1;n_1,m_1}
			=(-1)^{n_1+m_1}\,x_{12}^{-(n_1+{m_1})\,\rud}\,x_{13}^{-(n_1+{m_1})\,\rutre}\,\left(\frac{x_{23}}{x_{12}\,x_{13}}\right)^{{m_1}+\,n_1}     \nonumber\\
			&&\times
			\, \sum_{h=1}^{{m_1}}  h\left(\begin{array}{c}
				- \,n_1\,\rutre\\
				n_1+h\end{array}\right)\,\left(\begin{array}{c}
				-{m_1}\,\rutre\\
				{m_1}-h\end{array}\right)~.\nonumber\\
		\end{eqnarray}
		By using the identity (A.7) of \cite{Ademollo:1974kz} here cited:
		\begin{eqnarray}
			\sum_{l=1}^n l\left( \begin{array}{c}
				-nx\\n-l\end{array}\right)\left( \begin{array}{c}
				-n'x\\n'+l\end{array}\right)=
			\left( \begin{array}{c} -nx\\n\end{array}\right) \left( \begin{array}{c} -n'x\\n'\end{array}\right)\frac{n\,n'(1+x)}{x(n+n')}~,
		\end{eqnarray}
		
		we get the identity:
		\begin{eqnarray}
			\oint_0\frac{dy_1}{2\pi i}\frac{(1-y)^{-nx} }{y_1^n}\oint_0\frac{dy_2}{2\pi i} \,\frac{(1-y_2)^{-n' x}}{y_2^{n'}}\,\frac{1}{(y_1-y_2)^2}=(-1)^{n+n'}\,\left( \begin{array}{c}
				-n\,x\\n\end{array}\right)\left(\begin{array}{c} -n'\,x\\n'\end{array}\right)\frac{n\,n'(1+x)}{x(n+n')}~.\nonumber\\
			&&\label{5.164}
		\end{eqnarray}
		We can therefore write:
		\begin{eqnarray}
			{\cal  B}_{1;n_1,m_1}&&=(-1)^{n_1+m_1}\,
			\left(\frac{x_{23}}{x_{12}\,x_{23} }\right)^{{m_1+n_1} }\,x_{12}^{-{ (n_1+m_1)}\,\rud} \,x_{13}^{-{(n_1+m_1)}\,\rutre}\nonumber\\
			&&\times  \,{\left(\begin{array}{c}-  n_1\,\rutre\\ n_1\end{array}\right)\,\left(\begin{array}{c}-  m_1\,\rutre\\ m_1\end{array}\right)}\,
			{\frac{n_1\,m_1}{n_1+m_1}}\, \frac{(1+\rud)}{\rutre}~.  \nonumber\\
			&&
		\end{eqnarray}
		Similarly, we can evaluate the other two integrals with the same topology by getting:
		\begin{eqnarray}
			{\cal  B}_{2;n_2,m_2}
			&&=(-1)^{n_2+m_2}\,\left(\frac{x_{13}}{x_{23}\,x_{12} }\right)^{{n_2+m_2}} \,x_{23}^{- {(n_2+m_2)}\,\rdt} \,x_{12}^{-{(n_2+m_2)}\,\rdu} \nonumber\\
			&&\times  {\,\left(\begin{array}{c}-  n_2\,\rdu\\ n_2\end{array}\right)\,\left(\begin{array}{c}-  m_2\,\rdu\\ m_2\end{array}\right)}\, {\frac{n_2\,m_2}{n_2+m_2}}\,\frac{(1+\rdu)}{\rdu}~,\nonumber\\
			&&\nonumber\\
			{\cal  B}_{3;n_3,m_3}
			&&=(-1)^{n_3+m_3}\,\left(\frac{x_{12}}{x_{13}\,x_{23} }\right)^{{m_3+n_3}} \,x_{13}^{- {(n_3+m_3)}\,\rtu} \,x_{23}^{- {(n_3+m_3)}\,\rtd} \nonumber\\
			&&\times \,{\left(\begin{array}{c}- n_3 \,\rtd\\ n_3\end{array}\right)\,\left(\begin{array}{c}-  m_3\, \rtd\\ m_3\end{array}\right)}\,\frac{ n_3\,m_3}{n_3+m_3}\,\frac{(1+\rtd)}{\rtd}~.
		\end{eqnarray}
		The last typology of integrals to be evaluated are the ${\cal B}$-integrals as defined in Eq.s  \eqref{26a} and \eqref{35} but 
		along two non-intersecting circles centered around two different Koba-Nielsen variables $x_a$ and $x_b$. We will give details in the case $a=1$ and $b=2$, for the others we only cite the result:
		\begin{eqnarray}
			{\cal  B}_{1,n_1;2,n_2}&&=\oint_{x_1} \frac{dz_1}{2\pi i} \frac{ (z_1-x_2)^{- n_1\,\rud}\, (z_1-x_3)^{- n_1\,\rutre}}{(z_1-x_1)^{n_1}}\nonumber\\
			&&\times \oint_{x_2}\frac{dz_2}{2\pi i}\frac{(z_2-x_3)^{-n_2\,\rdt}\,(x_1-z_2)^{- n_2\,\rdu}}{(z_2-x_2)^{n_2}} \,\frac{1}{(z_1-z_2)^2}~.\label{DI12}
		\end{eqnarray}
		This integrals are evaluated by introducing the cross-ratios
		\begin{eqnarray}
			y_1= \frac{(z_1-x_1)(x_2-x_3)}{(z_1-x_3)(x_1-x_2)}~~;~~y_2=\frac{(z_2-x_2)(x_1-x_3)}{(z_2-x_3)(x_1-x_2)}~,
		\end{eqnarray}
		obtaining:
		\begin{eqnarray}
			{\cal  B}_{1,n_1;2,n_2}&&=  x_{12}^{-n_1\,\rud}\,x_{13}^{- n_1\,\rutre} x_{23}^{-n_2\,\rdt} \,x_{12}^{- n_2\,\rdu}\left( \frac{x_{23}}{x_{12}\,x_{13}}\right)^{n_1}\,\left( \frac{x_{13}}{x_{23}\,x_{12}}\right)^{n_2}\nonumber\\
			&&\times\oint_0\frac{dy_1}{2\pi i}\, \frac{(1+y_1)^{-n_1\,\rud}}{y_1^{n_1}} \oint_{0} \frac{dy_2}{2\pi i} \, \frac{(1-y_2)^{-n_2\,\rdu}}{y_2^{n_2}}\frac{1}{(1+y_1-y_2)^2}~.\nonumber\\
		\end{eqnarray}
		By expanding for small $y_1$ we get:
		\begin{eqnarray}
			\frac{1}{(1+y_1-y_2)^2}= \frac{1}{y_1\,(1-y_2)}\,\sum_{h=1}^\infty h (-1)^{-h}\frac{y_1^h}{(1-y_2)^h}~,
		\end{eqnarray}
		and the integral turns out to be:
		\begin{eqnarray}
			{\cal  B}_{1,n_1;2,n_2}&&= x_{12}^{- n_1\,\rud}\,x_{13}^{- n_1\,\rutre} x_{23}^{-n_2\,\rdt} \,x_{12}^{-n_2\,\rdu}\,\left( \frac{x_{23}}{x_{12}\,x_{13}}\right)^{n_1}\,\left( \frac{x_{13}}{x_{23}\,x_{12}}\right)^{n_2}\nonumber\\
			&&\times\sum_{h=1}^{n_1} \,h\,(-1)^{-h}\oint_0\frac{dy_1}{2\pi i}\, \frac{(1+y_1)^{-n_1\,\rud}}{y_1^{n_1-h+1}} \oint_{0} \frac{dy_2}{2\pi i} \, \frac{(1-y_2)^{-n_2\,\rdu-h-1}}{y_2^{n_2}}\nonumber\\
			&&=x_{12}^{- n_1\,\rud}\,x_{13}^{-\rutre} x_{23}^{-n_2\,\rdt}\,x_{12}^{- n_2\,\rdu}\left( \frac{x_{23}}{z_{12}\,z_{13}}\right)^{n_1}\,\left( \frac{x_{13}}{x_{23}\,x_{12}}\right)^{n_2}\nonumber\\
			&&\times\sum_{h=1}^{n_1} \,h\,(-1)^{n_2 -h}\left( \begin{array}{c}
				- n_1\,\rud\\ n_1-h\end{array}\right) \left( \begin{array}{c}- n_2\,\rdu -h-1\\n_2-1\end{array}\right)~.\nonumber\\
		\end{eqnarray}
		We have chosen $k_a=(k_a^-,\,\vec{0})$, therefore, we get:
		\begin{eqnarray}
			2\alpha' \,k_a\cdot p_T^{(a)}=1 \Rightarrow k_a^-=-\frac{1}{2\alpha' (p_T^{(a)})^+}~.
		\end{eqnarray}
		This condition determines:
		\begin{eqnarray}
			\rab=2\alpha' \,k_a\cdot p_T^{(b)}= \frac{(p_T^{(b)})^+}{(p_T^{(a)})^+}~~;~~\rba=2\alpha' \,k_b\cdot p_T^{(a)}= \frac{(p_T^{(a)})^+}{(p_T^{(b)})^+}=\frac{1}{\rab}~.
		\end{eqnarray}
		With the help of the identity \cite{Ademollo:1974kz}:
		\begin{eqnarray}
			\sum_{h=1}^n (-1)^{n-h} h \,\left( \begin{array}{c} -n\, \omega \\n-h\end{array}\right) \left( \begin{array}{c} -\frac{n'}{\omega}-h-1\\n'-1\end{array}\right)= (-1)^{n-1}\, \left( \begin{array}{c} -n\omega \\n\end{array}\right) \,\left( \begin{array}{c}-\frac{n'}{\omega}\\ n' \end{array}\right) \frac{n\,n'(1+\omega)}{n \,\omega+n'} 
		\end{eqnarray}
		we arrive at the expression:
		\begin{eqnarray}
			{\cal  B}_{1,n_1;2,n_2}&&= x_{12}^{-2\alpha' n_1\,k_1\cdot p_2-2\alpha' n_2k_2\cdot p_1}\,x_{13}^{- n_1\,\rutre} x_{23}^{-n_2\,k_2\,\rdt}\,\left( \frac{x_{23}}{x_{12}\,x_{13}}\right)^{n_1}\,\left( \frac{x_{13}}{x_{23}\,x_{12}}\right)^{n_2}\nonumber\\
			&&\times\, (-1)^{n_2-1} \left( \begin{array}{c}- n_1 \rud\\n_1\end{array}\right)\left(\begin{array}{c}- \frac{n_2}{\rud}\\n_2\end{array}\right) \frac{n_1 \,n_2 (1+\rud)}{n_1\rud+n_2}~.  \nonumber\\
		\end{eqnarray}
		Similar calculation holds for the other two integrals:
		\begin{eqnarray}
			{\cal  B}_{1,n_1;3,n_3}&&=x_{13}^{- n_1\,\rutre- n_3\,\rtu}\,x_{12}^{-n_1\,\rud} \,x_{23}^{-n_3\,\rtd}\,\left( \frac{x_{12}}{x_{13}\,x_{23}}\right)^{n_3}\,\left( \frac{x_{23}}{x_{12}\,x_{13}}\right)^{n_1}\nonumber\\
			&&\times\, (-1)^{n_1-1} \left( \begin{array}{c}- n_1 \rtu\\ n_1\end{array}\right)\left(\begin{array}{c}- \frac{n_2}{\rtu}\\n_2\end{array}\right) \frac{n_1 \,n_3 (1+\rtu)}{n_3\rtu+n_1}~,   \nonumber\\
		\end{eqnarray}
		and:
		\begin{eqnarray}
			{\cal  B}_{2,n_2;3,n_3}
			&&=
			x_{23}^{-n_2\,\rdt-2\alpha' n_3\,\rtd}\, 
			x_{12}^{- n_2\,\rdu}\, x_{13}^{-n_3\,\rtu}\,
			\left( \frac{x_{12}}{x_{13}\,x_{23}}\right)^{n_3}\,
			\left( \frac{x_{23}}{x_{12}\,x_{13}}\right)^{n_1}
			\nonumber\\
			&&\times\, (-1)^{n_3-1} 
			\left( 
			\begin{array}{c}- n_1 \rho_{31}\\n_1\end{array}\right)
			\left(\begin{array}{c}- \frac{n_2}{\rtu}\\n_2\end{array}\right)
			\frac{n_1 \,n_3 (1+\rtu)}{n_3\rtu+n_1}~.  
			\nonumber\\
		\end{eqnarray}

		\subsection{N-point single pole integral}
		In this section, we extend the calculation of the three-point integral from Sec.\ref{three-point integral} to the general case involving $N$ coherent states. We begin by considering the following integral:
		\begin{eqnarray}
			&&{\cal A}_{n_a}^i(\{x\};\,\varepsilon_{(n_a)};\,\{\rho\};\{p\})
			= 
			\oint_{x_a} \frac{d z_a}{2\pi i} 
			\frac{\prod_{c=a+1}^N(z_a-x_c)^{-n_a {\rho_{c a}}}\,
				\prod_{c=1}^{a-1} (x_c-z_a)^{-n_a\rho_{c a}}}
			{(z_a-x_a)^{n_a}}
			\sum_{c\neq a}\frac{\varepsilon_{(n_a)}^i\cdot p_c}{z_a-x_c}\nonumber\\
			&&
		\end{eqnarray}
		Using momentum conservation to write $p_{a+1}=-\sum_{c\ne a+1} p_c$ and transversality $ \varepsilon \cdot p =0$ it can be written in the form:
		\begin{eqnarray}
			{\cal A}_{n_a}^i(\{x\};\,\varepsilon_{(n_a)};\,\{\rho\};\{p\})=\sum_{r\neq\{a-1,\,a,\,a+1\}=1}^{N} \varepsilon_a^i\cdot p_c\,(x_c -x_{a+1})I_c+\varepsilon_a^i\cdot p_{a-1}\,(x_{a-1} -x_{a+1})I_{a-1}\nonumber\\
			\label{C.10}
		\end{eqnarray}
		with:
		\begin{eqnarray}
			I_c=\oint_{x_a} \frac{dz_a}{2\pi i} \frac{(z-x)_{a;a+1}^{-n_a\rho_{a+1 a}}\,\prod_{b\neq\{ c, a-1, a,a+1\}}^N(z-x)_{a;b}^{-n_a\rho_{ba}}\,    (x-z)_{c;a}^{-n_a\rho_{ca}}\,     
				(z-x)_{a;a-1}^{-n_a\rho_{a-1a}} % \prod_{s\neq r=1}^{t-2} (x_s-z_t)^{-n_t\rho_{ts}}
			}{(z_a-x_a)^{n_a}\,(z_a-x_c)(z_a-x_{a+1})
			}\nonumber\\
			\label{I1r}
		\end{eqnarray}
		where we have introduced the notation:
		\begin{eqnarray}
			(z-x)_{a;b}\equiv \left\{\begin{array}{cc} z_a-x_b & a<b\\
				x_b-z_a& b<a
			\end{array}\right. \label{C.12}
		\end{eqnarray}
		and:
		\begin{eqnarray}
			I_{a-1}= \oint_{x_a} \frac{dz_a}{2\pi i} \frac{(z-x)_{a,a+1}^{-n_a\rho_{a+1a}}(x-z)_{a-1,a}^{-n_t\rho_{a-1a}}\,%\prod_{s=t+2}^N(z_t-x_s)^{-n_t\rho_{ts}}
				\,\prod_{c\neq \{a-1,a,a+1\}=1} (x-z)_{c;a}^{-n_t\rho_{ca}}}{(z_a-x_a)^{n_a} (z_a-x_{a+1}) (z_a-x_{a-1})}\nonumber\\
			\label{I1t}
		\end{eqnarray}
		The residue of the two integrals are:
		\begin{eqnarray}
			I_c=\frac{1}{(n_a-1)!}\frac{\partial^{n_a-1}}{\partial z_a^{n_a-1}}\left[
			\frac{(z-x)_{a;a+1}^{-n_a\rho_{a+1 a}}\,\prod_{b\neq\{ c, a-1, a,a+1\}}^N(z-x)_{a;b}^{-n_a\rho_{ba}}\,    (x-z)_{c;a}^{-n_a\rho_{ca}}\,     
				(z-x)_{a;a-1}^{-n_a\rho_{a-1a}} % \prod_{s\neq r=1}^{t-2} (x_s-z_t)^{-n_t\rho_{ts}}
			}{(z_a-x_a)^{n_a}\,(z_a-x_c)(z_a-x_{a+1})
			}\right]_{z_a=x_a}\nonumber\\   
			&&
		\end{eqnarray}
		and:
		\begin{eqnarray}
			I_{a-1}=\frac{1}{(n_a-1)!}\frac{\partial^{n_a-1}}{\partial z_a^{n_a-1}}\left[
			\frac{(z-x)_{a,a+1}^{-n_a\rho_{a+1a}}(x-z)_{a-1,a}^{-n_t\rho_{a-1a}}\,%\prod_{s=t+2}^N(z_t-x_s)^{-n_t\rho_{ts}}
				\,\prod_{c\neq \{a-1,a,a+1\}=1}^N (x-z)_{c;a}^{-n_t\rho_{ca}}}{(z_a-x_a)^{n_a}(z_a-x_{a-1})(z_a-x_{a+1})}\right]_{z_a=x_a}\nonumber\\
		\end{eqnarray}
		These residues can be explicitly computed by introducing the anharmonic ratios of the Koba-Nielsen variables associated with the external DDF legs. There are many possible anharmonic ratios,  we change variable defining:
		\begin{eqnarray}
			\Omega_{[{a\,a-1\,a+1}]}(z)
			\equiv\omega
			=\frac{(z-x_a)}{(z_a-x_{a+1})}\frac{(x_{a-1} -x_{a+1})}{(x_{a-1} -x_a)}
			\Rightarrow 
			z= 
			\frac{x_a (x_{a-1} -x_{a+1})-x_{a+1} \,\omega (x_{a-1}-x_{a})}
			{(x_{a-1} -x_{a+1})-\omega (x_{a-1} -x_a)}
			,\nonumber\\
			&&\label{264}
			\label{changevariable}
		\end{eqnarray}
		so that $\Omega(x_a)=\omega_a=0$, 
		$\Omega(x_{a-1})=\omega_{a-1}=1$ and  
		$\Omega(x_{a+1})=\omega_{a+1}=\infty$. 
		
		We start by evaluating the integral \eqref{I1r} with the change of variable proposed in Eq. \eqref{changevariable}, getting:
		\begin{eqnarray}
			&&I_c=\frac{(x_{a-1}-x_a)}{(x_c-x_a)(x_{a-1} -x_{a+1})}\left( \frac{x_{a-1}-x_{a+1}}{(x_{a-1} -x_a)(x_a-x_{a+1})}\right)^{n_a}\,(x_a-x_c)^{-n_a\rho_{ca}}\nonumber\\
			&&\times \prod_{b\neq \{a,c\}}|x_b-x_a|^{-n_t\rho_{ba }}\,\oint_0 \frac{d\omega}{2\pi i} \, \frac{1}{\omega^{n_a}}\, (1- \omega)^{-n_a \rho_{a-1a}}\nonumber\\
			&&\times\, \prod_{b\neq\{\,a-1,\, a,\,a+1\}=1}^N \left[ 1 - \omega \frac{(x_{a+1}-x_b)(x_{a-1} -x_a)}{(x_a-x_b)(x_{a-1} -x_{a+1})}\right]^{-n_a \rho_{ba}} \,\left[ 1- \omega \frac{(x_c-x_{a+1}) (x_{a-1}-x_a)}{(x_c -x_a)(x_{a-1} -x_{a+1})}\right]^{-1}\nonumber\\
			&&\label{C.15}
		\end{eqnarray}
		
		We can now use the binomial expansion, getting:
		\begin{eqnarray}
			\oint_0 \frac{d
				\omega}{2\pi i} \, \frac{1}{\omega^{n_a}}\, (1- \omega)^{-n_a \rho_{a-1a}}\, &&\prod_{ b\neq\{a-1,\, a,\,a+1\}=1}^N
			\sum_{k_b=0}^\infty \left( \begin{array}{c}
				-n_a \rho_{ba}-\delta_{bc}\\k_b\end{array}\right) (-1)^{k_b}\, \omega^{k_b}\nonumber\\
				&&\times  \frac{(x_{a+1}-x_b)^{k_b}(x_{a-1} -x_a)^{k_b}}{(x_a-x_b)^{k_b}(x_{a-1} -x_a)^{k_b}}\nonumber\\
	\end{eqnarray}
The integral is different from zero when:
\begin{eqnarray}
n_a-1-\sum_{b\neq\{ a-1,a,a+1\}}k_b\geq 0 \Rightarrow k_b=\{1,\dots,\,n_a-1\}\,, 
\end{eqnarray}
since the minimal value of $\sum_{b\neq\{a-1,a,a+1\}} k_b=0$.  %$k_s= n_t-N-2$. 
Let's observe that:
\begin{eqnarray}
&&\prod_{b\neq\{a-1,\, a,\,a+1\}=\bf{1}}^N\, \sum_{{k_b}=0}^\infty\left( \begin{array}{c}
	-n_a \rho_{ts}-\delta_{rs}\\k_b\end{array}\right) (-1)^{k_s}\, y^{k_s} \frac{(x_{a+1}-x_b)^{k_b}(x_{a-1} -x_a)^,{k_b}}{(x_a-x_b)^{k_s}(x_{a-1} -x_a)^{k_b}}\nonumber\\
&&= \sum_{{\bf K}_{(t-1,t\,\,t+1)}=\bf{0}}^\infty\prod_{b\neq\{a-1,\, a,\,a+1\}=1}^N\left( \begin{array}{c}
	-n_a \rho_{ba}-\delta_{cb}\\k_b\end{array}\right) (-1)^{k_b}\, y^{k_b} \frac{(x_{a+1}-x_b)^{k_b}(x_{a-1} -x_a)^{k_b}}{(x_a-x_b)^{k_b}(x_{a-1} -x_a)^{k_b}}\nonumber\\
\label{268}
\end{eqnarray}
Here, ${\bf K}_{(t-1,t\,\,t+1)}= \{ k_b  \}$ is a vector with $N-3$  components, labeled with the index $b=\{1,\dots, N\}\backslash\{a-1,\,a,a+1\}$.

The integral becomes:
\begin{eqnarray}
&&I_c=\frac{(x_{a-1}-x_a)}{(x_c-x_a)(x_{a-1} -x_{a+1})}\left( \frac{x_{a-1}-x_{a+1}}{(x_{a-1} -x_a)(x_a-x_{a+1})}\right)^{n_t}\,\prod_{b\neq \{a\}}|x_b-x_a|^{-n_t\rho_{ba}}\nonumber\\
&&\sum_{\vec{K}_{(a-1,a\,\,a+1)}=\bf{0}}^{(n_a-1)\bf{1}}\prod_{b\neq\{a-1,\, a,\,a+1\}=1}^N\left( \begin{array}{c}
	-n_a \rho_{ba}-\delta_{bc}\\k_b\end{array}\right)\, \frac{(x_{a+1}-x_b)^{k_b}(x_{a-1} -x_a)^{k_b}}{(x_a-x_b)^{k_b}(x_{a-1} -x_{a+1})^{k_b}}\delta_{n_a-1-\sum_{s\neq \{a,a-1,\,a+1\}}k_s\geq 0}
\nonumber\\
&&\times\,
(-1)^{n_a-1} \left( \begin{array}{c}
	-n_a \rho_{a-1a}\\
	n-\sum_{b\neq \{a-1,a,a+1\}}k_b-1\end{array}\right)
\end{eqnarray}
The other integral turns out to be:
\begin{eqnarray}
I_{a-1}=&& \frac{(x_{a-1} -x_a)(x_{a-1}-x_{a+1}) (x_a-x_{a+1})}{(x_{a-1} -x_a)^{n_a} (x_a-x_{a+1})^{n_a}}\, \frac{(x_{a-1} -x_{a+1})^{-n_t\rho_{a+1 a}}}{(x_{a-1} -x_{a+1})}\, \frac{|x_a -x_{a+1}|^{-n_a\rho_{a+1 a}}}{(x_a-x_{a+1})}\nonumber\\
&&\times \frac{|x_{a-1} -x_a|^{-n_a\rho_{a-1 a}}}{(x_{a} -x_{a-1})}\, \frac{(x_{a-1} -x_{a+1})^{-n_a \rho_{a-1 a}}}{(x_{a-1} -x_{a+1})} \prod_{b\neq\{a-1, a,a+1\}=1}^N |x_a -x_b|^{-n_a \rho_{ba} }(x_{a-1} -x_{a+1})^{-n_a\rho_{ba}}\nonumber\\
&& \times \oint_0 \frac{d\omega}{2\pi i} \, \frac{1}{\omega^{n_a}}\, (1- \omega)^{-n_t \rho_{a-1a }-1}\, \prod_{b\neq \{a-1,\,a,\,a+1\}=1}^N \left[ 1 - \omega \frac{(x_{a+1}-x_b)(x_{a-1} -x_a) }{(x_a-x_b)(x_{a-1} -x_{b+1})}\right]^{-n_a \rho_{ba}} \nonumber\\
\label{C.19}
\end{eqnarray} 
with:
\begin{eqnarray}
&&\oint_0 \frac{d\omega}{2\pi i} \, \frac{1}{\omega^{n_a}}\, (1- \omega)^{-n_a \rho_{a-1a}-1}\, \prod_{b\neq \{a-1,\,a,\,a+1\}=1}^N \left[ 1 - \omega \frac{(x_{a+1}-x_b)(x_{a-1} -x_a)}{(x_a-x_b)(x_{a-1} -x_{a+1})}\right]^{-n_a \rho_{ba}}\nonumber\\
&&=\sum_{{\bf K}_{a-1,a,a+1}=\bf{0} }^{(n_a-1)\bf{1}}  \prod_{b\neq \{a-1,\,a,\,a+1\}=1}^N (-1)^{n_a-1}\left( \begin{array}{c} -n_a\rho_{ba}\\ k_b\end{array}\right)\frac{(x_{a+1}-x_b)^{k_b}(x_{a-1} -x_a)^{k_b}}{(x_a-x_b)^{k_b}(x_{a-1} -x_a)^{k_b}}\nonumber\\
&&\times 
\left( \begin{array}{c} -n_t\rho_{a-1 a } -1\\n_a-\sum_{s\neq \{t-1,t,t+1\}}k_b-1\end{array}\right)\delta_{n_a-1-\sum_{b\neq \{a,a-1,\,a+1\}}k_b\geq 0}
\nonumber\\
\end{eqnarray} 
We finally get:
\begin{eqnarray}
&&I_{a-1}= -\frac{1}{x_{a-1}-x_{a+1}}\left( \frac{(x_{a-1}-x_{a+1})}{(x_{a-1}-x_a)(x_a-x_{a+1})}\right)^{n_a}\prod_{b\neq a}|x_b-x_a|^{-n_t\rho_{ba}}\nonumber\\
&&\times\oint_0 \frac{d\omega}{2\pi i} \, \frac{1}{\omega^{n_a}}\, (1- y)^{-n_a \rho_{a-1 a}-1}\, \prod_{b\neq \{a-1,\,a,\,a+1\}=1}^N \left[ 1 - \omega \frac{(x_{a+1}-x_b)(x_{a-1} -x_a)}{(x_a-x_b)(x_{a-1} -x_{a+1})}\right]^{-n_a \rho_{ba}}\nonumber\\
&&=-\frac{(-1)^{n_a-1} }{x_{a-1}-x_{a+1}}\left( \frac{(x_{a-1}-x_{a+1})}{(x_{a-1}-x_a)(x_a-x_{a+1})}\right)^{n_a}\,\prod_{b\neq a}|x_b-x_a|^{-n_a\rho_{ba}}\nonumber\\
&&\times \sum_{\vec{K}_{t-1,t,t+1}=\bf{0} }^\infty  \prod_{b\neq \{a-1,\,a,\,a+1\}=1}^N \left( \begin{array}{c} -n_a\rho_{ba}\\ k_a\end{array}\right)\frac{(x_{a+1}-x_b)^{k_b}(x_{a-1} -x_a)^{k_b}}{(x_a-x_b)^{k_b}(x_{a-1} -x_{a+1})^{k_b}}\nonumber\\
&&\times 
\left( \begin{array}{c} -n_a\rho_{a-1 a} -1\\n_a-\sum_{b\neq \{a-1,a,a+1\}}k_b-1\end{array}\right)\delta_{n_a-1-\sum_{b\neq \{a-1,a,\,a+1\}}k_s\geq 0}
\nonumber\\
&&
\end{eqnarray}
This integral is in agreement with the case $N=3$ discussed in the previous subsection.
\subsection{Double Integrals} 
The first kind of double integrals we evaluate are those around two circles, ${\cal C}_a$ and ${\cal C}_b$ centered at the Koba Nielsen variables  $x_a$ and $x_b$ respectively, with  $a\neq b$ and ${\cal C}_a \bigcap {\cal C}_b=0$. These are denoted with:
\begin{eqnarray}
{\cal B}_{a,n_a;b,n_b}&&=\oint_{x_a}\frac{dz_a}{2\pi i} \frac{ \prod_{c=a+1}^N(z_a -x_c)^{-n_a\rho_{ca}}\,\prod_{c=1}^{a-1} (x_c- z_a)^{-n_a\rho_{ca}}}{(z_a-x_a)^{n_a}}\nonumber\\
&&\times\, \oint_{x_b} \frac{dz_b}{2\pi i} \frac{ \prod_{c=b+1}^N(z_b -x_c)^{-n_b\rho_{bc}}\,\prod_{c=1}^{b-1} (x_c- z_b)^{-n_b\rho_{cb}}}{(z_b-x_b)^{n_b}}\,\frac{1}{(z_a-z_b)^2}\nonumber\\
\end{eqnarray}
The result of the integration is:
\begin{eqnarray}
{\cal B}_{a,n_a;b,n_b}=\frac{1}{(n_a-1)!}\frac{1}{(n_b-1)!}
\frac{\partial^{n_a-1}}{\partial z_a^{n_a-1}}\frac{\partial^{n_b-1}}{\partial z_b^{n_b-1}}\left[\frac{ \prod_{c\neq a}^N(z -x)^{-n_a\rho_{ca}}_{a;c}}{(z_a-x_a)^{n_a}}
%\right.\nonumber\\
%&&\left.\times
\, \frac{\prod_{c\neq b} (z-x)_{b;c}^{-n_b\rho_{cb}}}{(z_b-x_b)^{n_b}}\,\frac{1}{(z_a-z_b)^2}\right]_{
z_a=x_a;z_b=x_b}\nonumber\\
\end{eqnarray}
The residues are explicitly evaluated by introducing the anharmonic ratios which map the variables $x_{a;b}$ to the origin of the complex plane:
\begin{eqnarray}
\omega_a=\frac{(z_a-x_a)(x_b-x_c)}{(z_a-x_c)(x_b-x_a)}~~;~~\omega_b=\frac{(z_b-x_b)(x_a-x_c)}{(z_b-x_c)(x_a-x_b)}
\end{eqnarray}
with $x_c$, $c\neq \{a,\,b\}$,  a third arbitrary Koba-Nielsen variable. After some algebra, we get:
\begin{eqnarray}
&&{\cal B}_{a,n_a;b,n_b}=(-1)^{n_b+1} \left( \frac{(x_c-x_b)}{(x_c-x_a)\, (x_a-x_b)}\right)^{n_a}\left(\frac{(x_c-x_a)}{(x_c-x_a)(x_a-x_b)}\right)^{n_b}\prod_{s\neq a}|x_s-x_a|^{-n_a\rho_{sa}}\nonumber\\
&&\times\,\prod_{s\neq b}|x_s-x_b|^{-n_b\rho_{sb}} \oint_0\frac{d\omega_a}{2\pi i} \, \,\frac{1}{\omega_a^{n_a}}\,\prod_{s\neq \{a,b,c\}}\left[ 1+\frac{(x_c-x_s)(x_b-x_a)}{(x_a-x_s)(x_c-x_b)}\omega_1\right]^{-n_a \rho_{sa}} (1-\omega_a)^{-n_a\rho_{ba}}\nonumber\\
&&\times \oint_0\frac{d \omega_b}{2\pi i} \, \frac{1}{ \omega_b^{n_b}}\,
\prod_{s\neq \{a,b,c\}}\left[1+\frac{(x_c-x_s)(x_a-x_b)}{(x_b-x_s)(x_c-x_a)}\omega_b\right]^{-n_b\rho_{sb}}(1-\omega_b)^{-n_b\rho_{ab}}\,
\frac{1}{(1-\omega_a-\omega_b)^2}\label{276}
\end{eqnarray}
We now use the expansion\footnote{We are expanding for small $\omega_a$; however, we could equivalently expand for small $\omega_b$. In the case $N=3$, as shown in Appendix \ref{F.2}, the result of the integration is symmetric under the exchange of $a\leftrightarrow b$.}:
\begin{eqnarray}
\frac{1}{(1-\omega_a-\omega_b)^2}= \frac{1}{\omega_a(1-\omega_b)}\sum_{l=1}^\infty \frac{l\,\omega_a^l}{(1-\omega_b)^l}\label{C.47}
\end{eqnarray}
and (with a similar binomial expansion in $\omega_a$):
\begin{eqnarray}
&&\prod_{s\neq \{a,b,c\}=1}^N \left[1 +\frac{(x_c-x_s)(x_b-x_a)}{(x_b-x_s)(x_c-x_a)}\omega_b\right ]^{-n_{b}\rho_{sb}}= \prod_{s\neq \{b,a,c\}=1}^N\sum_{k_s=0}^\infty \left( \begin{array}{c} -n_b \rho_{sb}\\ k_s\end{array}\right)\frac{(x_c-x_s)^{k_s}(x_b-x_a)^{k_s}}{(x_b-x_s)^{k_s}(x_c-x_a)^{k_s}}\omega_b^{k_s}\nonumber\\
&&=\sum_{{\bf K}_{\{a,b,c\}}={\bf 0}}^\infty\prod_{s\neq \{a,c,c\}=1}^N\left( \begin{array}{c} -n_b \rho_{sb}\\ k_s\end{array}\right)\frac{(x_c-x_s)^{k_s}(x_b-x_a)^{k_s}}{(x_b-x_s)^{k_s}(x_t-x_a)^{k_s}}\omega_b^{k_s}\label{278}
\end{eqnarray}
where the notation ${\bf K}_{\{a,b,c\}}$ is  defined after Eq. \eqref{268}. By substituting the binomial expansion from Eq. \eqref{278}  into Eq. \eqref{276}, the integrals are easily evaluated, yielding:
\begin{eqnarray}
&&{\cal B}_{a,n_a;\,b;n_b}=(-1)^{n_a}\left(\frac{(x_c-x_b)}{(x_c -x_a)\, (x_a-x_b)}\right) ^{n_a} \left(\frac{(x_c-x_a)}{(x_c-x_b)\, (x_a-x_b)}\right)^{n_b}\nonumber\\
&&\times \prod_{s\neq a=1}^N |x_a-x_s|^{-n_a\rho_{sa} }\,\prod_{s\neq b=1}^N |x_b-x_s|^{-n_b\rho_{sb} } \nonumber\\
&&\times \sum_{h=1}^{n_a} h\,\sum_{{\bf K}_{\{a,b,t\}}={\bf 0}}^\infty \sum_{{\bf L}_{\{a,b,t\}}={\bf 0}}^\infty\,(-1)^{-h+n_b-\sum_{s\neq\{ a,b,t\}}l_{s}+n_a-\sum_{s\neq\{ a,b,t\}}k_{s}}\nonumber\\
&&\times \prod_{s\neq\{ a,b,t\}=1}^N \left( \begin{array}{c} -n_a\rho_{sa}\\k_{s}\end{array}\right)\,\frac{(x_c-x_s)^{k_{s}}(x_b-x_a)^{k_{s}}}{(x_a-x_s)^{k_{s}}(x_t-x_b)^{k_{s}}}\nonumber\\
&&\times \,\prod_{s\neq\{a, b,t\}=1}^N \left( \begin{array}{c} -n_b\rho_{sb}\\ l_{s}\end{array}\right) \,\frac{(x_c-x_s)^{l_{s}}(x_a-x_b)^{l_{s}}}{(x_b-x_s)^{l_{s}}(x_c-x_a)^{l_{s}}}\,\left(\begin{array}{c} -n_a\rho_{ba}\\n_a-h-\sum_{s\neq a}k_{s} \end{array}\right)\left(\begin{array}{c} -n_b\rho_{ab}-h-1\\n_b-\sum_{s\neq b}l_{s}-1\end{array}\right)\nonumber\\
&&\times \delta_{n_a-h-\sum_{s\neq a}k_{s}\geq 0}\, \delta_{n_b-\sum_{s\neq b}l_{s}-1\geq0}\nonumber\\
\end{eqnarray}
This result for $N=3$ agrees with the integrals given in Appendix\eqref{F.2} and the case $N=3$  is also consistent with  \cite{Ademollo:1974kz}.

Similarly, we now proceed to evaluate the double integrals, where the regions of integration are circles centered around the same Koba Nielsen variable,$x_a$: 
\begin{eqnarray}
&&{\cal B}_{a;n_a,m_a}=\oint_{x_a}\frac{dz_a}{2\pi i} \frac{ \prod_{c=a+1}^N(z_a -x_c)^{-n_a\rho_{ca}}\,\prod_{c=1}^{a-1} (x_c- z_a)^{-n_a\rho_{ca}}}{(z_a-x_a)^{n_a}}\nonumber\\
&& \oint_{x_a} \frac{dw_a}{2\pi i} \frac{ \prod_{c=a+1}^N(w_a -x_c)^{-m_a\rho_{ca}}\,\prod_{c=1}^{a-1} (x_c- w_a)^{-m_a\rho_{ca}}}{(w_a-x_a)^{m_a}}\,\frac{1}{(z_a-w_a)^2}\, ,
\end{eqnarray}
and we may take $|z_a-x_a|>|w_a-x_a|$.  In this case, the change of variables is the same as proposed in Eq. \eqref{264}, here cited:
\begin{eqnarray}
\omega_a^1=\frac{(z_a-x_a)(x_{a_2}-x_{a_1})}{(z_a-x_{a_1})(x_{a_2}-x_a)}~~;~~\omega_a^2=\frac{(w_a-x_a)(x_{a_2}-x_{a_1})}{(w_a-x_{a_1})(x_{a_2}-x_a)}
\end{eqnarray}
Here, $x_{a_1}$ and $x_{a_2}$  are two arbitrary Koba Nielsen variables.   In the new variables, the integrals take the form:
\begin{eqnarray}
&&{\cal B}_{a;n_a,m_a}=\left[ \frac{(x_{a_1}-x_{a_2})}{(x_{a_1}-x_a)(x_a-x_{a_2})}\right]^{n_a+m_a} \prod_{c\neq a}|x_a-x_c|^{-(n_a+m_a)\rho_{ca}}\nonumber\\
&&\times \oint_0\frac{d\omega_a^1}{2\pi i}\, \frac{(1-\omega_a^1)^{-n_a\rho_{a_2 a}}}{(\omega_a^1)^{n_a}}\prod_{c\neq \{a,\,a_1,\,a_2\}=1}^N\left[1+\frac{(x_{a_2}-x_{a})(x_a-x_c)}{(x_{a_1}-x_{a_2})(x_a-x_c)}\omega_a^1\right]^{-n_a\rho_{ca}} \nonumber\\
&&\times \oint_0\frac{dy_2}{2\pi i}\, \frac{(1-\omega_a^2)^{-m_a\rho_{a_2a}}}{(\omega_a^2)^{m_a}}\,\prod_{c\neq \{ a,\,a_1,\,a_2\}}\left[1+\frac{(x_{a_2}-x_{a})(x_{a_1}-x_c)}{(x_{a_1}-x_{a_2})(x_a-x_c)}\omega_a^2\right]^{-m_a\rho_{ca}}\,
\frac{1}{(\omega_a^1-\omega_a^2)^2}\nonumber\\
&&
\end{eqnarray} 
The integrals can be evaluated after performing the binomial expansions, yielding:
\begin{eqnarray}
&&{\cal B}_{a;n_a,m_a}=(-1)^{n_a+m_a}
\left[ \frac{(x_{a_1}-x_{a_2})}{(x_{a_1}-x_a)(x_a-x_{a_2})}\right]^{n_a+m_a} \prod_{c\neq a}|x_a-x_c|^{-(n_a+m_a)\rho_{ca}}\nonumber\\
&&\times\,\, \sum_{h=1}^{m_a} h\,\sum_{{\bf K}_{\{a,a_1,a_2\}}={\bf 0}}^\infty(-1)^{\sum_{c\neq \{a,a_1,a_2\}} k_c}\prod_{c\neq\{a,a_1,a_2\}=1}^\infty \,\left(\begin{array}{c} -n_a\rho_{ca}\\k_c\end{array}\right)\,\frac{(x_{a_2}-x_{a})^{k_c}(x_{a_1}-x_c)^{k_c}}{(x_{a_1}-x_{a_2})^{k_c}(x_a-x_c)^{k_c}}\nonumber\\
&&\times \,\sum_{{\bf L}_{\{a,a_1,a_2\}}={\bf 0}}^\infty\prod_{b\neq\{a,a_1,a_2\}=1}^\infty (-1)^{\sum_{c\neq \{a,a_1,a_2\} }l_c}\,\left(\begin{array}{c} -m_a\rho_{ba}\\l_b\end{array}\right)\,\frac{(x_{a_2}-x_{a})^{l_b}(x_{a_1}-x_b)^{l_b}}{(x_{a_1}-x_{a_2})^{l_b}(x_a-x_c)^{l_b}}\nonumber\\
&&\times \left(\begin{array}{c} -n_a\rho_{a_2a}\\n_a+h+\sum_{c\neq \{a,a_1,a_2\}=1}k_c\end{array}\right)\,\left(\begin{array}{c} -m_a\rho_{a_2a}\\ m_a-h+\sum_{b\neq \{a,a_1,a_2\}l_b}\end{array}\right)\delta_{n_a+h+\sum_{c\neq \{a,a_1,a_2\}k_c\geq 0}}\nonumber\\
&&\times\delta_{ m_a-h+\sum_{b\neq \{a,a_1,a_2\}l_b\geq 0}}\nonumber\\
&&
\end{eqnarray}
Again, in the case $N=3$ where there is only the sum over the integer $h$, the expression matches the result in Appendix\eqref{F.2} and is consistent with  Ref.\cite{Ademollo:1974kz}.

		\renewcommand{\xt}{{x_a}}
		\renewcommand{\xr}{{x_d}}
		\renewcommand{\xs}{{x_b}}
		\newcommand{\xu}{{x_c}}
		
		\newcommand{\TT}{{a}}
		\renewcommand{\SS}{{b}}
		\newcommand{\UU}{{c}}
		\newcommand{\RR}{{d}}

		\subsection{Single pole integral rewritten}
		We want now to compute the same integral as in the previous section in a slightly different way
		\begin{align}
			\oldI_{\TT, n}^i
			=&
			\oint_{z=\xt}
			\frac{d z}{ 2\pi i}
			U_{[\TT]}^{-n}(z)
			\,
			\sum_{\SS=1,\, \SS\ne\TT}
			\frac{ \varepsilon_\TT^i \cdot p_\SS }{ ( z -\xs ) }
			,
		\end{align}
		with 
		\begin{align}
			U_{[\TT]}(z)
			&=
			(z-\xt)\,
			\hat U_{[\TT]}(z)
			=
			\prod_{\RR=1}^{\TT-1}
			(\xr - z)^{  \rho_{\RR \TT} }\,
			(z-\xt)\,
			\prod_{\RR=\TT+1}^{N}
			(-\xr + z)^{ \rho_{\RR \TT} }
			.
		\end{align}
		To this purpose we rewrite
		\begin{align}
			%  \left( \Pi_{[\TT]} p_\SS \right)^i
			\varepsilon_\TT^i \cdot p_\SS
			=
			\epsilon^i \cdot p_\SS
			-
			\frac{k_\SS \cdot p_\SS}{k_\TT \cdot p_\TT}
			\epsilon^i \cdot p_\TT
			,
		\end{align}
		and notice $\left( \varepsilon_{\TT} \cdot p_\TT \right)^i=0$
		so that we can use the momentum conservation
		$\sum_\SS p_\SS^\mu=0$ to write
		\begin{align}
			\sum_{\SS}   \left( \varepsilon_{\TT} \cdot p_\SS \right)^i=
			\sum_{\SS\ne\TT}   \left( \varepsilon_{\TT} \cdot p_\SS \right)^i=0
			.
		\end{align}
		This allows to get
		\begin{align}
			\left( \varepsilon_{\TT} \cdot p_{\TT-1} \right)^i
			=-
			\sum_{\SS\ne\TT, \TT-1}   \left( \varepsilon_{\TT} \cdot p_\SS \right)^i
			,
		\end{align}
		and after substituting back 
		we can now compute
		\begin{align}
			\oldI_{\TT, n}^i
			=&
			\sum_{\SS\ne \TT, \TT-1}
			\varepsilon^i \cdot p_\SS
			\hat   \oldI_{\TT,\, n ;\, \SS}
			=
			\sum_{\SS\ne \TT, \TT-1}
			\left(
			\epsilon^i \cdot p_\SS
			-
			\frac{k_\SS \cdot p_\SS}{k_\TT \cdot p_\TT}
			\epsilon^i \cdot p_\TT
			\right)
			\hat   \oldI_{\TT,\, n ;\, \SS}
			\nonumber\\
			=&
			%  \sdap
			\sum_{\SS\ne \TT, \TT-1}
			\left(
			\uk^i_{T \sN {\SS} }
			- \frac{ \uk^+_{T \sN {\SS} } }{ \uk^+_{T \sN  \TT} } \uk^i_{T \sN \TT}
			\right)
			\hat   \oldI_{\TT,\, n ;\, \SS}
			,
		\end{align}
		where we have defined
		\begin{align}
			\hat \oldI_{\TT,\, n;\, \SS}
			=&
			\oint_{z=\xt}
			\frac{d z}{ 2\pi i}
			\frac{ \xs -x_{\TT-1} }{ ( z -\xs ) (z -x_{\TT-1} ) }
			U_{[\TT]}^{-n}(z)
			\nonumber\\
			=&
			\oint_{z=\xt}
			\frac{d z}{ 2\pi i}
			\frac{ \xs -x_{\TT-1} }{ ( z -\xs ) (z -x_{\TT-1} ) }
			\left(
			\prod_{\RR=1}^{\TT-1}
			(x_{\RR} -z)^{ \rho_{\RR \TT} }
			\,
			(z -\xt)
			\,
			\prod_{\RR=\TT+1}^{N}
			(x_{\RR}  + z)^{ \rho_{\RR \TT} }
			\right)^{-n}
			,~~~~
			\SS\ne \TT, \TT-1
			,
		\end{align}
		The latter expression has an improved  $z\rightarrow \infty$ limit as
		$\frac{1}{z^2} + O \left(\frac{1}{z^3} \right)$

		It is then natural to introduce the anharmonic ratio
		\begin{align}
			\Omega_{[\TT\, \TT-1\, \TT+1]}(z)
			=&
			\frac{
				(z-x_{\TT}) (x_{\TT-1}-x_{\TT+1})
			}{
				(z-x_{\TT+1}) (x_{\TT-1}-x_{\TT})
			}
			,
			\nonumber\\
			&
			\Omega_{[\TT\, \TT-1\, \TT+1]}(x_\TT)=0,~~~
			\Omega_{[\TT\, \TT-1\, \TT+1]}(x_{\TT-1})=1,~~~
			\Omega_{[\TT\, \TT-1\, \TT+1]}(x_{\TT+1})=\infty
			,
		\end{align}
		where $\TT$ is defined $mod\, N$.
		In the following we use for $\RR\ne \TT+1$
		\begin{equation}
			\Omega_{[\TT\, \TT-1\, \TT+1]}(x_\RR)= \omega_\RR
			,~~~~
			\Omega_{[\TT\, \TT-1\, \TT+1]}(z)=\Omega
		\end{equation}
		to lighten the notation.
		We point out that $\omega_{\TT-1}=1$ but it will be anyhow shown
		in the net formulas
		for the same reason of having a lighter notation.

		Notice that there are two cases:
		\begin{enumerate}
			\item $\SS\ne \TT+1, \TT, \TT-1$,
			\item $\SS=\TT+1\ne \TT, \TT-1$,  
		\end{enumerate}
		since in the second case $\omega_\SS= \infty$.
		This latter case has to be treated slightly differently.
		For $N=3$ since $\SS\ne \TT, \TT-1$ it happens that $\SS=\TT+1$ and so
		we are always in the second case.

		It is then easy to verify that
		\begin{align}
			\frac{ (\xs -x_{\TT-1})\, dz }{ ( z -\xs ) (z -x_{\TT-1} ) }
			=&
			%% {{ \left(x_{\TT-1}-x_{\SS}\right)\, \left(x_{\TT+1}-x_{\TT}\right)\,
					%%   d \Omega}
				%%   \over
				%%   {\left(\Omega-1\right)\,
					%%     \left(\Omega\,\left(x_{\TT}-x_{\TT-1}\right)\,
					%%     \left(x_{\TT+1}-x_{\SS}\right)-\left(x_{\TT}-x_{\SS}\right)\,
					%%     \left(x_{\TT+1}-x_{\TT-1}\right)
					%%     \right)}}
			\frac{ (x_{\TT-1} -\xs) (\xt -x_{\TT+1}) }{ (x_{\TT} -\xs) (x_{\TT-1} -x_{\TT+1})}
			\frac{1}{1 -\frac{\Omega}{\omega_\SS}}
			\frac{d \Omega}{1 -\frac{\Omega}{\omega_{\TT-1} }},
			\nonumber\\
			=&
			\frac{ (x_{\TT-1} -\xs) (\xt -x_{\TT+1}) }{ (x_{\TT} -\xs) (x_{\TT-1} -x_{\TT+1})}
			\frac{d \Omega}{1 -\frac{\Omega}{\omega_{\TT-1} }},
			~~~~\mbox{for }\, \SS=\TT+1\, \mbox{ as follows from }
			\omega_\SS\rightarrow \infty,
			\nonumber\\
			z-x_{\RR}
			=&
			D\,
			(x_{\TT} - x_{\RR})\, (x_{\TT+1} - x_{\TT-1})\,
			\left( 1 - \frac{\Omega}{\omega_\RR} \right),
			\RR\ne \TT, \TT+1,
			\nonumber\\
			z-x_{\TT}
			=&
			-
			D\,
			(x_{\TT} - x_{\TT-1})\, (x_{\TT+1} - x_{\TT})\,
			\Omega
			,
			\nonumber\\
			z-x_{\TT+1}
			=&
			D\,
			(x_{\TT} - x_{\TT+1})\, (x_{\TT+1} - x_{\TT-1})
			,
		\end{align}
		with
		\begin{align}
			D=&
			\frac{1}{(x_{\TT+1}-x_{\TT-1}) - (x_{\TT}-x_{\TT-1}) \Omega}
			,
		\end{align}
		and
		\begin{align}
			\frac{z-\xr}{z-\xt}
			=&
			\frac{(\xt-x_\RR) (x_{\TT-1}-x_{\TT+1})}{(\xt-x_{\TT+1})
				(x_{\TT-1}-x_{\TT})}
			\left( 1- \frac{\Omega}{\omega_\RR} \right)
			\frac{1}{\Omega}
			,
		\end{align}
		whose constant coefficient is positive when $\RR>\TT$ and negative for 
		$\RR<\TT$ as it should be for having no ambiguities in the following
		expression for $  U_{[\TT]}(z)$.

		We can then easily get
		\begin{align}
			U_{[\TT]}(z)
			&=
			\prod_{\RR=1,\, \RR\ne\TT}^N (x_{\RR; \TT})^{ \rho_{\RR \TT} }\,
			\frac{ (x_{\TT-1}-\xt) (\xt-x_{\TT+1}) }{ x_{\TT-1}-x_{\TT+1} }\,
			\prod_{\RR=1,\, \RR\ne\TT, \TT+1}^N
			\left( 1 - \frac{\Omega}{\omega_\RR} \right)^{\rho_{\RR \TT}}\,
			{\Omega}
			\nonumber\\
			&=
			\left. \left(
			\frac{d U_{[\TT]}}{d z} / \frac{d\Omega}{d z}
			\right)\right|_{z=\xt}
			\prod_{\RR=1,\, \RR\ne\TT,\TT+1}^N
			\left( 1 - \frac{\Omega}{\omega_\RR} \right)^{\rho_{\RR \TT}}\,
			{\Omega}
			,
		\end{align}
		with
		\begin{align}
			\left. \left( \frac{d U_{[\TT]}}{d z}   \right)\right|_{z=\xt}
			=&
			\hat U_{[\TT]}( \xt)
			=
			\prod_{\RR=1,\, \RR\ne\TT}^N (x_{\RR; \TT})^{ \rho_{\RR \TT} }
			,
			\nonumber\\
			\left. \left(
			\frac{d\Omega}{d z}
			\right)\right|_{z=\xt}
			=&
			\frac{ x_{\TT-1}-x_{\TT+1} }{ (x_{\TT-1}-\xt) (\xt-x_{\TT+1}) }
			,
		\end{align}
		and we have used $\sum_\RR \rho_{\RR\TT}=0$ and $\omega_{\TT+1}=\infty$.
		Notice that the rewriting of the constant factor can be obtained
		either by direct computation or by noticing that it is
		$\frac{d U_{[\TT]}}{d \Omega}\Bigg|_{\Omega=0}$ since
		$U_{[\TT]}(z) = z\, \hat U_{[\TT]}(z)$.

		We can now assemble all and get (remember $\RR\ne \TT, \TT-1$)
		\begin{align}
			\hat \oldI_{\TT,\, n;\, \SS}
			=&
			\frac{ (x_{\TT-1} -\xs) (\xt -x_{\TT+1}) }{ (x_{\TT} -\xs) (x_{\TT-1} -x_{\TT+1})}
			\left[
			\left. \left(
			\frac{d U_{[\TT]}}{d z} / \frac{d\Omega}{d z}
			\right)\right|_{z=\xt}
			\right]^{-n}
			\nonumber\\
			&
			\times
			\oint_{\Omega=0}
			\frac{d \Omega}{ 2\pi i \Omega^n }\,
			\frac{1}{1 -\frac{\Omega}{\omega_\SS}}
			\frac{1}{1 -\frac{\Omega}{\omega_{\TT-1} }}  ,
			\,
			\prod_{\RR=1,\, \RR\ne\TT, \TT+1}^N
			\left( 1 - \frac{\Omega}{\omega_\RR} \right)^{-n \rho_{\RR \TT}}\,
			,
		\end{align}
		expanding the binomials we get for $\SS\ne \TT+1$
		\begin{align}
			=&
			\frac{ (x_{\TT-1} -\xs) (\xt -x_{\TT+1}) }{ (x_{\TT} -\xs) (x_{\TT-1} -x_{\TT+1})}
			\left[
			\left. \left(
			\frac{d U_{[\TT]}}{d z} / \frac{d\Omega}{d z}
			\right)\right|_{z=\xt}
			\right]^{-n}
			\nonumber\\
			&
			\times
			\oint_{\Omega=0}
			\frac{d \Omega}{ 2\pi i \Omega^n }\,
			\prod_{\RR=1,\, \RR\ne\TT, \TT+1, \TT-1, \SS}^N
			\left[
			\sum_{l_\RR=1}^N
			{ -n \rho_{\RR \TT} \choose l_\RR }
			\left(- \frac{\Omega}{\omega_\RR} \right)^{l_\RR}
			\right]
			\nonumber\\
			&
			\times
			\left[
			\sum_{l_\SS=1}^N
			{ -n \rho_{\SS \TT} -1 \choose l_\SS }
			\left(- \frac{\Omega}{\omega_\SS} \right)^{l_\SS}
			\right]
			\left[
			\sum_{l_{\TT-1}=1}^N
			{ -n \rho_{\TT-1\, \TT} -1 \choose l_{\TT-1 } }
			\left(- \Omega \right)^{l_{\TT-1}}
			\right]
		\end{align}

		\begin{align}
			=&
			(-1)^{n-1}
			\frac{ (x_{\TT-1} -\xs) (\xt -x_{\TT+1}) }{ (x_{\TT} -\xs) (x_{\TT-1} -x_{\TT+1})}
			\left[
			\left. \left(
			\frac{d U_{[\TT]}}{d z} / \frac{d\Omega}{d z}
			\right)\right|_{z=\xt}
			\right]^{-n}
			\nonumber\\
			\times
			&
			\sum_{ \{l_\RR\}_{\RR=1,\, \RR \ne \TT, \TT-1, \TT+1} }
			\prod_{\RR\ne \TT, \TT-1, \TT+1, \SS }
			{ -n \rho_{\RR \TT} \choose l_\RR }
			\,
			{ -n \rho_{\SS\, \TT} -1 \choose l_\SS }
			\,
			{ -n \rho_{\TT-1\, \TT} -1 \choose n -1 - \sum_\RR' l_\RR}
			\,
			\prod_{\RR \ne \TT, \TT-1, \TT+1} \frac{1}{\omega_{\RR}^{l_\RR}}
			\,
			\delta_{\sum_{\RR}' l_\RR \le  n-1}
			,
			\label{eq:single_I_s_not_tp1}
		\end{align}
		where $\sum_{\RR}'=\sum_{\RR=1,\, \RR\ne \TT, \TT-1, \TT+1}^N$.
		
		For $\SS= \TT+1$ we can expand again the binomials but the terms
		dropping the terms with $\frac{1}{\omega_\SS}=0$ (or equivalently
		setting $l_\SS=0$) to get
		% SECOND CASE
		\begin{align}
			\oldI_{\TT,\, n;\, \SS=\TT+1}=&
			\frac{ (x_{\TT-1} -\xs) (\xt -x_{\TT+1}) }{ (x_{\TT} -\xs) (x_{\TT-1} -x_{\TT+1})}
			\left[
			\left. \left(
			\frac{d U_{[\TT]}}{d z} / \frac{d\Omega}{d z}
			\right)\right|_{z=\xt}
			\right]^{-n}
			\nonumber\\
			&
			\times
			\oint_{\Omega=0}
			\frac{d \Omega}{ 2\pi i \Omega^n }\,
			\prod_{\RR=1,\, \RR\ne\TT, \TT+1, \TT-1}^N
			\left[
			\sum_{l_\RR=1}^N
			{ -n \rho_{\RR \TT} \choose l_\RR }
			\left(- \frac{\Omega}{\omega_\RR} \right)^{l_\RR}
			\right]
			\nonumber\\
			&
			\times
			%% \left[
			%%   \sum_{l_\SS=1}^N
			%%   { -n \rho_{\SS \TT} -1 \choose l_\SS }
			%%   \left(- \frac{\Omega}{\omega_\SS} \right)^{l_\SS}
			%%   \right]
			%
			\left[
			\sum_{l_{\TT-1}=1}^N
			{ -n \rho_{\TT-1\, \TT} -1 \choose l_{\TT-1 } }
			\left(- \Omega \right)^{l_{\TT-1}}
			\right]
		\end{align}

		\begin{align}
			=&
			(-1)^{n-1}
			%
			%%  \frac{ (x_{\TT-1} -x_{\TT-1}) }{  (x_{\TT-1} -x_{\TT+1})}
			%  
			\left[
			\left. \left(
			\frac{d U_{[\TT]}}{d z} / \frac{d\Omega}{d z}
			\right)\right|_{z=\xt}
			\right]^{-n}
			\nonumber\\
			\times
			&
			\sum_{ \{l_\RR\}_{\RR=1,\, \RR \ne \TT, \TT-1, \TT+1} }
			\prod_{\RR\ne \TT, \TT-1, \TT+1 }
			{ -n \rho_{\RR \TT} \choose l_\RR }
			%% \,
			%% { -n \rho_{\SS\, \TT} -1 \choose l_\SS }
			\,
			{ -n \rho_{\TT-1\, \TT} -1 \choose n -1 - \sum_\RR' l_\RR}
			\,
			\prod_{\RR \ne \TT, \TT-1, \TT+1} \frac{1}{\omega_{\RR}^{l_\RR}}
			\,
			\delta_{\sum_{\RR}' l_\RR \le  n-1}
			,
			\label{eq:single_I_s_tp1}
		\end{align}
		where still $\sum_{\RR}'=\sum_{\RR=1,\, \RR\ne \TT, \TT-1, \TT+1}^N$.
		
		Notice that the prefactor which does not depend of any anharmonic ratio
		should combine with other contributions in order to get an expression
		which depends on anharmonic ratios.
		
		For example for $N=3$, $\TT=1$ we have $\SS=\TT+1=2$ and we are in the
		second case of eq. \ref{eq:single_I_s_tp1}.
		There are no $\RR$ left and no $1/\omega_\RR$ factors
		then $\delta_{\sum_{\RR}' l_\RR +l_\SS +l_{\TT-1},  n-1}$
		becomes simply $\delta_{l_{\TT-1=3}, n-1}$.
		Finally
		\begin{align}
			\hat \oldI_{\TT=1,\, n;\, \SS=2}
			=&
			(-1)^{n-1}
			\,
			\left(
			\prod_{\RR\ne\TT} x_{\RR; \TT}^{\rho_{\RR \TT}}
			\,
			\frac{x_{\TT-1\, \TT} x_{\TT\, \TT+1}}{ x_{\TT-1\, \TT+1} }
			\right)^{-n}
			\,
			\sum_{l_3}
			{ -n \rho_{\TT-1=3\, \TT=1} -1 \choose l_3 }\,
			(-\omega_{ 3} )^{l_2}\,
			\delta_{l_3, n-1}
			\nonumber\\
			=&
			(-1)^{n-1}
			\,
			\left(
			\prod_{\RR\ne\TT} x_{\RR; \TT}^{\rho_{\RR \TT}}
			\,
			\frac{x_{\TT-1\, \TT} x_{\TT\, \TT+1}}{ x_{\TT-1\, \TT+1} }
			\right)^{-n}
			{ -n \rho_{3 1} -1 \choose n-1 }\,
			.
		\end{align}
		
		This result matches the one from direct computation upon the use of
		$\sum_\RR \rho_{\RR \TT}=0$
		\begin{align}
			\nonumber\\
			\hat   I_{\TT, n; \RR=\TT+1 }
			&=
			(-1)^{n-1}
			\frac{1}{ (n-1)! } \prod_{k=1}^{n-1}( n \rho_{\TT+1\, \TT} +k)   
			%%   \nonumber\\
			%% &
			\times
			\left(
			\frac{  x_{\TT ;  \TT+1} }{ x_{\TT ; \TT-1} }
			\right)^{ -n  \rho_{\TT+1\, \TT}}
			\,
			\left(
			\frac{  x_{\TT ;  \TT+1} }{ x_{\TT-1 ; \TT+1} }
			\right)^{ -n }
			.
			\label{eq:single integal N3 igor2}
		\end{align}
		
		In the $N=4$ case we can consider $\TT=2$ then $\SS\ne \TT,\TT-1$ so
		$\SS=3,4$.
		If we consider $\SS=4$ we are in the first case.
		The sum over $(l_\RR)_{\RR=1,\, \RR \ne \TT, \TT-1, \TT+1}$ reduces to $(l_4)$.
		The product $\prod_{\RR\ne \TT, \TT-1, \TT+1, \SS }$ is simply $1$
		then we get
		\begin{align}
			\oldI_{\TT=2,\, n;\, \SS=4}
			=&
			(-1)^{n-1}
			\left( \prod_{\RR\ne2} x_{\RR; 2}^{\rho_{\RR 2}}
			\,
			\frac{x_{12} x_{23} }{x_{13}} \right)^{-n}
			\frac{x_{14} x_{23}}{ x_{24} x_{13} }
			\sum_{l_4=0}^{n-1}
			{-n \rho_{4 2} -1 \choose l_4}
			{-n \rho_{1 2} -1 \choose n -1 -l_4}
			\frac{1}{\omega_4^{l_4}}
			,
		\end{align}
		with $\omega_4=  \frac{x_{13} x_{24}}{ x_{12} x_{34} }$.
		
		Actually for $N=4$ since $\SS\ne \TT,\TT-1$ we can only have
		the ``special'' case $\SS=\TT+1$
		and the ``generic'' case $\SS=\TT+2$,
		so we can write the ``special'' case
		\begin{align}
			\oldI_{\TT,\, n;\, \SS=\TT+1}
			=&
			(-1)^{n-1}
			\left( \prod_{\RR\ne\TT} x_{\RR; \TT}^{\rho_{\RR \TT}}
			\,
			\frac{x_{\TT\, \TT+1} x_{\TT-1\, \TT} }{x_{\TT-1\, \TT+1}}
			\right)^{-n}
			\sum_{l_{\TT+2}=0}^{n-1}
			{-n \rho_{{\TT+2}\, \TT} \choose l_{\TT+2}}
			{-n \rho_{\TT-1\, \TT} -1 \choose n -1 -l_{\TT+2}}
			\frac{1}{\omega_{\TT+2}^{l_{\TT+2}}}
			,
		\end{align}
		with
		$\omega_{\TT+2}
		=\frac{x_{\TT+2\, \TT} x_{\TT-1\, \TT+1} }
		{x_{\TT+2\, \TT+1} x_{\TT-1\, \TT}}$
		and the very similar ``generic'' case
		(the major difference with the ``special'' case is
		$-n \rho_{{\TT+2}\, \TT} \rightarrow -n \rho_{{\TT+2}\, \TT} -1$)
		\begin{align}
			\oldI_{\TT,\, n;\, \SS=\TT+2}
			=&
			(-1)^{n-1}
			\left( \prod_{\RR\ne\TT} x_{\RR; \TT}^{\rho_{\RR \TT}}
			\,
			\frac{x_{\TT\, \TT+1} x_{\TT-1\, \TT} }{x_{\TT-1\, \TT+1}}
			\right)^{-n}
			\frac{x_{\TT-1\, \TT+2} x_{\TT\, \TT+1} }{x_{\TT\, \TT+2}}
			\sum_{l_{\TT+2}=0}^{n-1}
			{-n \rho_{{\TT+2}\, \TT} -1 \choose l_{\TT+2}}
			{-n \rho_{\TT-1\, \TT} -1 \choose n -1 -l_{\TT+2}}
			\frac{1}{\omega_{\TT+2}^{l_{\TT+2}}}
			,
		\end{align}
		with again
		$\omega_{\TT+2}
		=\frac{x_{\TT+2\, \TT} x_{\TT-1\, \TT+1} }
		{x_{\TT+2\, \TT+1} x_{\TT-1\, \TT}}$.

		\subsection{Double integrals with $\TT=\SS$ rewritten}
		Let us write the following integral in a way that it is expressed
		as much as possible using anharmonic ratios
		\begin{align}
			\oldJ_{\TT n_1, \TT n_2}
			=&
			\oint_{z_1=\xt, |z_1-\xt| > |z_2-\xt|}
			\oint_{z_2=\xt} \frac{d z_2}{ 2\pi i}
			\frac{d z_1}{ 2\pi i}
			\frac{ 1  }{(z_1 -z_2)^2 }
			U_{[\TT]}^{-n_1}(z_1)\, U_{[\TT]}^{-n_1}(z_2)\, 
			\nonumber\\
			=&
			\oint_{z_1=\xt, |z_1-\xt| > |z_2-\xt|}
			\oint_{z_2=\xt} \frac{d z_2}{ 2\pi i}
			\frac{d z_1}{ 2\pi i}
			\frac{ 1  }{(z_1 -z_2)^2 }
			\frac{ 1  }{(z_1 - \xt)^{n_1} }
			\frac{ 1  }{(z_2 - \xt)^{n_2} }
			\nonumber\\
			&
			\times
			\prod_{\RR=1}^{\TT-1}
			(x_{\RR} - z_1)^{ -n_1 \rho_{\RR \TT} }
			\prod_{\RR=t+1}^{N}
			(- x_{\RR} + z_1)^{ -n_1 \rho_{\RR \TT} }
			\nonumber\\
			&
			\times
			\prod_{\RR=1}^{\TT-1}
			(x_{\RR} - z_2)^{ -n_2 \rho_{\RR \TT}}
			\prod_{\RR=t+1}^{N}
			(- x_{\RR} + z_2)^{ -n_2 \rho_{\RR \TT} }
			\nonumber\\
			,
		\end{align}
		with, as before
		\begin{align}
			U_{[\TT]}(z)
			&=
			(z -\xt)\, \hat U_{[\TT]}(z)
			=
			\prod_{\RR=1}^{\TT-1}
			(\xr - z)^{  \rho_{\RR \TT} }\,
			(z-\xt)\,
			\prod_{\RR=\TT+1}^{N}
			(-\xr + z)^{ \rho_{\RR \TT} }
			.
		\end{align}
		
		It is then natural to introduce the anharmonic ratio
		\begin{align}
			\Omega_{[\TT\, \TT-1\, \TT+1]}(z)
			=&
			\frac{
				(z-x_{\TT}) (x_{\TT-1}-x_{\TT+1})
			}{
				(z-x_{\TT+1}) (x_{\TT-1}-x_{\TT})
			}
			,
			\nonumber\\
			&
			\Omega_{[\TT\, \TT-1\, \TT+1]}(x_\TT)=0,~~~
			\Omega_{[\TT\, \TT-1\, \TT+1]}(x_{\TT-1})=1,~~~
			\Omega_{[\TT\, \TT-1\, \TT+1]}(x_{\TT+1})=\infty
			,
		\end{align}
		where $\TT$ is defined $mod\, N$.
		In the following we use for $\RR\ne \TT,\TT+1$
		\begin{equation}
			\Omega_{[\TT\, \TT-1\, \TT+1]}(x_\RR)= \omega_\RR
			,~~~~
			\Omega_{[\TT\, \TT-1\, \TT+1]}(z_1)=\Omega_1
			,~~~~
			\Omega_{[\TT\, \TT-1\, \TT+1]}(z_2)=\Omega_2
			,
		\end{equation}
		to lighten the notation.
		Notice that $\omega_{\TT-1}=1$ but it will be shown in the net formulas
		for the same reason of having a lighter notation.
		
		It is then easy to verify that
		\begin{align}
			\frac{ d z_1\, d z_2  }{(z_1 -z_2)^2 }
			=&
			\frac{ d \Omega_1\, d \Omega_2  }{(\Omega_1 -\Omega_2)^2 }
			,
			\nonumber\\
			z-x_{\RR}
			=&
			D\,
			(x_{\TT} - x_{\RR})\, (x_{\TT+1} - x_{\TT-1})\,
			\left( 1 - \frac{\Omega}{\omega_\RR} \right),
			\RR\ne \TT, \TT+1,
			\nonumber\\
			z-x_{\TT}
			=&
			-
			D\,
			(x_{\TT} - x_{\TT-1})\, (x_{\TT+1} - x_{\TT})\,
			\Omega
			,
			\nonumber\\
			z-x_{\TT+1}
			=&
			D\,
			(x_{\TT} - x_{\TT+1})\, (x_{\TT+1} - x_{\TT-1})
			,
		\end{align}
		with
		\begin{align}
			D=&
			\frac{1}{(x_{\TT+1}-x_{\TT-1}) - (x_{\TT}-x_{\TT-1}) \Omega}
			,
		\end{align}
		and
		\begin{align}
			\frac{z-\xr}{z-\xt}
			=&
			\frac{(\xt-x_\RR) (x_{\TT-1}-x_{\TT+1})}{(\xt-x_{\TT+1})
				(x_{\TT-1}-x_{\TT})}
			\left( 1- \frac{\Omega}{\omega_\RR} \right)
			\frac{1}{\Omega}
			,
		\end{align}
		whose constant coefficient is positive when $\RR>\TT$ and negative for 
		$\RR<\TT$ as it should be for having no ambiguities in the following
		expression for $  U_{[\TT]}(z)$.
		
		We can then easily get
		\begin{align}
			U_{[\TT]}(z)
			&=
			\prod_{\RR=1,\, \RR\ne\TT}^N (x_{\RR; \TT})^{ \rho_{\RR \TT} }\,
			\frac{ (x_{\TT-1}-\xt) (\xt-x_{\TT+1}) }{ x_{\TT-1}-x_{\TT+1} }\,
			\prod_{\RR=1,\, \RR\ne\TT, \TT+1}^N
			\left( 1 - \frac{\Omega}{\omega_\RR} \right)^{\rho_{\RR \TT}}\,
			{\Omega}
			\nonumber\\
			&=
			\left. \left(
			\frac{d U_{[\TT]}}{d z} / \frac{d\Omega}{d z}
			\right)\right|_{z=\xt}
			\prod_{\RR=1,\, \RR\ne\TT,\TT+1}^N
			\left( 1 - \frac{\Omega}{\omega_\RR} \right)^{\rho_{\RR \TT}}\,
			{\Omega}
			,
		\end{align}
		with
		\begin{align}
			\left. \left( \frac{d U_{[\TT]}}{d z}   \right)\right|_{z=\xt}
			=&
			\hat U_{[\TT]}( \xt)
			=
			\prod_{\RR=1,\, \RR\ne\TT}^N (x_{\RR; \TT})^{ \rho_{\RR \TT} }
			,
			\nonumber\\
			\left. \left(
			\frac{d\Omega}{d z}
			\right)\right|_{z=\xt}
			=&
			\frac{ x_{\TT-1}-x_{\TT+1} }{ (x_{\TT-1}-\xt) (\xt-x_{\TT+1}) }
			,
		\end{align}
		and we have used $\sum_\RR \rho_{\RR\TT}=0$ ($\rho_{\TT\TT}=1$) and
		$\omega_{\TT+1}=\infty$ .
		Notice that the rewriting of the constant factor can be obtained
		either by direct computation or by noticing that it is
		$\frac{d U_{[\TT]}}{d \Omega}\Bigg|_{\Omega=0}$ since
		$U_{[\TT]}(z) = (z -\xt)\, \hat U_{[\TT]}(z)$.

		The integral of interest can be written as
		\begin{align}
			\oldJ_{\TT n_1, \TT n_2}
			=
			\left(
			\left. \left(
			\frac{d U_{[\TT]}}{d z} / \frac{d\Omega}{d z}
			\right)\right|_{z=\xt}
			\right)^{-(n_1+n_2) }
			&
			\oint_{\Omega_1=0, |\Omega_1| > |\Omega_2|}
			\oint_{\Omega_2=0} \frac{d \Omega_2}{ 2\pi i}
			\frac{d \Omega_1}{ 2\pi i}
			\frac{ 1  }{(\Omega_1 -\Omega_2)^2 }
			\frac{ 1  }{\Omega_1^{n_1} }
			\frac{ 1  }{\Omega_2^{n_2} }
			\nonumber\\
			&
			\prod_{\RR=1,\, \RR\ne \TT, \TT+1}^N
			\left[
			\left( 1-\frac{\Omega_1}{\omega_\RR} \right)^{-n_1 \rho_{\RR \TT} }
			\left( 1-\frac{\Omega_2}{\omega_\RR} \right)^{-n_1 \rho_{\RR \TT} }
			\right]
			\nonumber\\
			=
			\left(
			\left. \left(
			\frac{d U_{[\TT]}}{d z} / \frac{d\Omega}{d z}
			\right)\right|_{z=\xt}
			\right)^{-(n_1+n_2) }
			&   
			\sum_{k=1}^{n_2}
			k\, 
			\oint_{\Omega_1=0}
			\frac{d \Omega_1}{ 2\pi i}
			\frac{ 1  }{\Omega_1^{n_1 +k +1} }
			\prod_{\RR=1,\, \RR\ne \TT, \TT+1}^N
			\left( 1-\frac{\Omega_1}{\omega_\RR} \right)^{-n_1 \rho_{\RR \TT} }
			\nonumber\\
			&
			\oint_{\Omega_2=0}
			\frac{d \Omega_2}{ 2\pi i}
			\frac{ 1  }{\Omega_2^{n_2 -k +1} }
			\prod_{\RR=1,\, \RR\ne \TT, \TT+1}^N
			\left( 1-\frac{\Omega_2}{\omega_\RR} \right)^{-n_2 \rho_{\RR \TT} }
			,
		\end{align}
		and evaluated.
		In particular it is symmetric in $n_1$ and $n_2$ so that we can choose
		$n_2 \le n_1$ and get
		\begin{align}
			\oldJ_{\TT n_1, \TT n_2}
			=&
			\left(
			\prod_{\RR=1,\, \RR\ne\TT}^N (x_{\RR; \TT})^{ \rho_{\RR \TT} }\,
			\frac{ (x_{\TT-1}-\xt) (\xt-x_{\TT+1}) }{ x_{\TT-1}-x_{\TT+1} }\,
			\right)^{-(n_1+n_2) }
			\nonumber\\
			\sum_{k=1}^{n_2} k\,
			&
			\sum_{ (l_{1 \RR})_{\RR\ne \TT,\TT+1}\in \N^{N-2}}
			\prod_{\RR=1,\, \RR\ne \TT,\TT+1}^N       { -n_1 \rho_{\RR \TT} \choose l_{1 \RR} }
			\sum_{ (l_{2 \RR})_{\RR\ne \TT,\TT+1}\in \N^{N-2}}
			\prod_{\RR=1,\, \RR\ne \TT,\TT+1}^N { -n_2 \rho_{\RR \TT} \choose
				l_{2 \RR} }
			\nonumber\\
			&
			\prod_{\RR=1,\, \RR\ne \TT,\TT+1}^N \frac{(-1)^{n_1+n_2}}{\omega_\RR^{l_{1 \RR}+l_{2 \RR}}}
			\delta_{\sum''_\RR l_{1 \RR}, n_1+k}\,
			\delta_{\sum''_\RR l_{2 \RR}+k, n_2}\,
			,
		\end{align}
		where
		$\sum''_\RR= \sum_{\RR=1,\,\RR\ne \TT, \TT+1}^N$.
		
		Let us see the explicit $N=4$ case.
		We consider first $\TT=4$ so we get the natural ``gauge fixing''
		$\omega_4=0$, $\omega_3=1$ and $\omega_1=\infty$.
		We are left with $\omega_2\equiv \omega$ as the only variable.
		The previous general expression boils down to
		\begin{align}
			\oldJ_{4 n_1,\, 4 n_2}
			=&
			(-1)^{n_1+n_2}\,
			\left( \prod_{\RR=1}^3 x_{\RR 4}^{\rho_{\RR 4}}
			\frac{x_{34} x_{14}}{x_{13}} \right)^{-(n_1+n_2)}
			\nonumber\\
			&
			\sum_{k=1}^{n_2} k\,
			\sum_{(l_{12},\, l_{13})\in\N^2}
			\sum_{(l_{22},\, l_{23})\in\N^2}
			{ -n_1\, \rho_{24} \choose l_{12} }   { -n_1\, \rho_{34} \choose l_{13} }
			{ -n_2\, \rho_{24} \choose l_{22} }   { -n_2\, \rho_{34} \choose l_{23} }
			\frac{1}{ \omega^{l_{12}+l_{22}} }
			\nonumber\\
			&
			\delta_{l_{12}+l_{13}, n_1+k}\,
			\delta_{l_{22}+l_{23}+k, n_2}
			,
		\end{align}
		and solving for $l_{13}$ and $l_{23}$
		\begin{align}
			\oldJ_{4 n_1,\, 4 n_2}
			=&
			(-1)^{n_1+n_2}\,
			\left( \prod_{\RR=1}^3 x_{\RR 4}^{\rho_{\RR 4}}
			\frac{x_{34} x_{14}}{x_{13}} \right)^{-(n_1+n_2)}
			\nonumber\\
			&
			\sum_{k=1}^{n_2} k\,
			\sum_{l_{12}=0}^{n_1+k}
			\sum_{l_{22}=0}^{n_2-k}
			{ -n_1\, \rho_{24} \choose l_{12} }   { -n_1\, \rho_{34} \choose n_1+k-l_{12} }
			{ -n_2\, \rho_{24} \choose l_{22} }   { -n_2\, \rho_{34} \choose n_2-k-l_{22} }
			\frac{1}{ \omega^{l_{12}+l_{22}} }
			.
		\end{align}
		For actually computing amplitudes it is computationally better to fix
		$x_4=0$, $x_3=1$, $x_2=x$ and $x_1=x_\infty$ so that
		$\omega= \frac{x (x_\infty -1)}{x_\infty-x}$ and we get
		\begin{align}
			\oldJ_{4 n_1,\, 4 n_2}
			=&
			(-1)^{n_1+n_2}
			\left( \hat U_{[4]}(0)\, \frac{x_\infty }{x_\infty-1} \right)^{-(n_1+n_2)}
			\nonumber\\
			&
			\sum_{k=1}^{n_2} k\,
			\sum_{l_{12}=0}^{n_1+k}
			\sum_{l_{22}=0}^{n_2-k}
			{ -n_1\, \rho_{24} \choose l_{12} }   { -n_1\, \rho_{34} \choose n_1+k-l_{12} }
			{ -n_2\, \rho_{24} \choose l_{22} }   { -n_2\, \rho_{34} \choose n_2-k-l_{22} }
			\left( \frac{x_\infty-x}{x (x_\infty -1)} \right) ^{l_{12}+l_{22}}
			,
			\nonumber\\
			\hat U_{[4]}(0)
			=&
			x_\infty^{\rho_{14}}\, x^{\rho_{24}}
			=
			x_\infty^{-1-\rho_{24}- \rho_{34}}\, x^{\rho_{24}}
			.
		\end{align}

		In a similar way for $\TT=1$ we have $\omega_{[1] \TT=1}=0$,
		$\omega_{[1] \TT=4}=1$ and $\omega_{[1] \TT=2}=\infty$,
		the possible values in the sums are $\RR=3,4$.
		The only non trivial $\omega$ is $\omega_{[1]3}= \frac{\omega}{\omega-1}
		\Rightarrow \frac{x (x_\infty-1)}{(x-1) x_\infty}$
		and we get
		\begin{align}
			\oldJ_{1 n_1,\, 1 n_2}
			=&
			(-1)^{n_1+n_2}\,
			\left( \prod_{\RR=2}^4 x_{1 \RR }^{\rho_{\RR 1}}
			\frac{x_{14} x_{12}}{x_{24}} \right)^{-(n_1+n_2)}
			\nonumber\\
			&
			\sum_{k=1}^{n_2} k\,
			\sum_{l_{13}=0}^{n_1+k}
			\sum_{l_{23}=0}^{n_2-k}
			{ -n_1\, \rho_{31} \choose l_{13} }   { -n_1\, \rho_{41} \choose n_1+k-l_{13} }
			{ -n_2\, \rho_{31} \choose l_{23} }   { -n_2\, \rho_{41} \choose n_2-k-l_{23} }
			\frac{1}{ \omega_{[1] 3}^{l_{13}+l_{23}} }
			.
		\end{align}

		In a similar way for $\TT=2$ we have $\omega_{[2] \TT=2}=0$,
		$\omega_{[2] \TT=1}=1$ and $\omega_{[2] \TT=3}=\infty$,
		the possible values in the sums are $\RR=1,4$.
		The only non trivial $\omega$ is $\omega_{[2] 4}= \omega
		\Rightarrow \frac{x (x_\infty-1)}{(x_\infty-x)}$
		and we get
		\begin{align}
			\oldJ_{2 n_1,\, 2 n_2}
			=&
			(-1)^{n_1+n_2}\,
			\left( \prod_{\RR=1,\,\RR\ne2 }^4 x_{2; \RR }^{\rho_{\RR 2}}
			\frac{x_{14} x_{12}}{x_{24}} \right)^{-(n_1+n_2)}
			\nonumber\\
			&
			\sum_{k=1}^{n_2} k\,
			\sum_{l_{14}=0}^{n_1+k}
			\sum_{l_{24}=0}^{n_2-k}
			{ -n_1\, \rho_{42} \choose l_{14} }   { -n_1\, \rho_{12} \choose n_1+k-l_{14} }
			{ -n_2\, \rho_{42} \choose l_{24} }   { -n_2\, \rho_{12} \choose n_2-k-l_{24} }
			\frac{1}{ \omega_{[2] 4}^{l_{14}+l_{24}} }
			,
		\end{align}
		where $\omega_4={\omega}$.
		
		In a similar way for $\TT=3$ we have $\omega_{[3] \TT=3}=0$,
		$\omega_{[3] \TT=2}=1$ and $\omega_{[3] \TT=4}=\infty$,
		the possible values in the sums are $\RR=1,2$.
		The only non trivial $\omega$ is $\omega_{[3] 1}= \frac{\omega}{\omega-1}
		\Rightarrow \frac{x (x_\infty-1)}{x_\infty (x-1)}$
		and we get
		\begin{align}
			\oldJ_{3 n_1,\, 3 n_2}
			=&
			(-1)^{n_1+n_2}\,
			\left( \prod_{\RR=1,\, \RR\ne3}^4 x_{\RR; 3}^{\rho_{\RR 3}}
			\frac{x_{23} x_{34}}{x_{24}} \right)^{-(n_1+n_2)}
			\nonumber\\
			&
			\sum_{k=1}^{n_2} k\,
			\sum_{l_{11}=0}^{n_1+k}
			\sum_{l_{21}=0}^{n_2-k}
			{ -n_1\, \rho_{13} \choose l_{11} }   { -n_1\, \rho_{23} \choose n_1+k-l_{11} }
			{ -n_2\, \rho_{13} \choose l_{21} }   { -n_2\, \rho_{23} \choose n_2-k-l_{21} }
			\frac{1}{ \omega_{[3] 1}^{l_{11}+l_{21}} }
			.
		\end{align}

		\subsection{Double integrals with $\TT\ne\SS$
  rewritten
			% : 3rd try and simpler
		}
		
		Let us write the integral we want to express using as much as possible
		anharmonic ratios as
		\begin{align}
			\oldJ_{\TT n_1, \SS n_2}
			=&
			\oint_{z_1=\xt}
			\oint_{z_2=\xs} \frac{d z_2}{ 2\pi i}
			\frac{d z_1}{ 2\pi i}
			\frac{ 1  }{(z_1 -z_2)^2 }
			U_{[\TT]}^{-n_1}(z_1)\, U_{[\SS]}^{-n_2}(z_2)\, 
			\nonumber\\
			=&
			\oint_{z_1=\xt}
			\oint_{z_2=\xs} \frac{d z_2}{ 2\pi i}
			\frac{d z_1}{ 2\pi i}
			\frac{ 1  }{(z_1 -z_2)^2 }
			\frac{ 1  }{(z_1 - \xt)^{n_1} }
			\frac{ 1  }{(z_2 - \xs)^{n_2} }
			\nonumber\\
			&
			\times
			\prod_{\RR=1}^{\TT-1}
			(\xr - z_1)^{ -n_1 \rho_{\RR \TT} }
			\prod_{\RR=\TT+1}^{N}
			(-\xr + z_1)^{ -n_1 \rho_{\RR \TT} }
			\nonumber\\
			&
			\times
			\prod_{\RR=1}^{\SS-1}
			(\xr - z_2)^{ -n_2 \rho_{\RR \SS}}
			\prod_{r=t+1}^{N}
			(-\xr + z_2)^{ -n_2 \rho_{\RR \SS} }
			,
		\end{align}
		with 
		\begin{align}
			U_{[\TT]}(z_1)
			&=
			\prod_{\RR=1}^{\TT-1}
			(\xr - z_1)^{  \rho_{\RR \TT} }\,
			(z-\xt)\,
			\prod_{\RR=\TT+1}^{N}
			(-\xr + z_1)^{ \rho_{\RR \TT} }
			.
		\end{align}
		
		It is then natural to introduce the anharmonic ratio
		\begin{align}
			\Omega_{[\SS\, \UU\, \TT]}(z)
			=&
			\frac{
				(z-\xt) (\xu-\xs)
			}{
				(z-\xs) (\xu-\xt)
			}
			,
			\nonumber\\
			&
			\Omega_{[\SS\, \UU\, \TT]}(\xs)=0,~~~
			\Omega_{[\SS\, \UU\, \TT]}(\xu)=1,~~~
			\Omega_{[\SS\, \UU\, \TT]}(\xt)=\infty
			,
		\end{align}
		where $s, t, u$ are defined $mod\, N$.
		In the following we use (for $r\ne t,s$)
		\begin{equation}
			\Omega_{[\SS\, \UU\, \TT]}(\xr)= \omega_r
			,~~~~
			\Omega_{[\SS\, \UU\, \TT]}(z_1)= \Omega_1
			,~~~~
			\Omega_{[\SS\, \UU\, \TT]}(z_2)= \Omega_2
			,
		\end{equation}
		to lighten the notation.
		Notice that $\omega_{\UU}=1$ but it will be shown in the next formulas
		for the same reason of having more compact expressions.
		
		It is then easy to verify that
		\begin{align}
			\oint_{z_1=0}\, \oint_{z_2=0}
			\frac{ d z_1\, d z_2  }{(z_1 -z_2)^2 } \dots
			=&
			\oint_{\Omega_1=\infty} \oint_{\Omega_2=0}
			\frac{ d \Omega_1\, d \Omega_2  }{(\Omega_1 -\Omega_2)^2 } \dots
			,
		\end{align}
		and
		\begin{align}
			z-\xr
			=&
			\left\{
			\begin{array}{l c}
				(\xs - \xr )\,
				\frac{
					1 - \frac{\Omega}{\omega_\RR}
				}{
					1 + \Omega \frac{\xu-\xs}{\xt-\xu}
				}
				&
				\mbox{good for } \Omega\sim 0
				\\
				(\xt-\xr)
				\frac{
					1 - \frac{\omega_\RR}{\Omega}
				}{
					1 + \frac{1}{\Omega} \frac{\xt-\xu}{\xu-\xs}
				}
				&
				\mbox{good for } \Omega\sim \infty
			\end{array}
			\right.
			~~~
			\RR\ne \SS,\TT
			,
		\end{align}
		where it is worth noticing that the $\Omega$ independent factors are
		positive for $\RR>\SS$ in the first line and $\RR>\TT$ in the second lie
		so that they can be factorized out from $(z-\xr)^\rho$.
		And similarly for $\xr-z$.
		
		We can also discuss how the integration paths are mapped.
		The naive expectation is right $|\Omega_1| > |\Omega_2|$.
		This can be seen by considering the paths
		$|z_1-\xt|=R_\TT$ and $|z_2-\xs|=R_\SS$ with $R_\TT>R_\SS$.
		
		We need also the special cases
		\begin{align}
			z-\xt
			=&
			\left\{
			\begin{array}{l c}
				(\xs - \xt )\,
				\frac{
					1 
				}{
					1 + \Omega \frac{\xu-\xs}{\xt-\xu}
				}
				&
				\mbox{good for } \Omega\sim 0
				\\
				-\frac{ (\xt-\xs) (\xt-\xu) }{ \xu-\xs }
				\frac{1}{\Omega}
				\frac{
					1 
				}{
					1 + \frac{1}{\Omega} \frac{\xt-\xu}{\xu-\xs}
				}
				&
				\mbox{good for } \Omega\sim \infty
			\end{array}
			\right.
			,
			\nonumber\\
			z-\xs
			=&
			\left\{
			\begin{array}{l c}
				\frac{ (\xt-\xs) (\xu-\xs) }{ \xt-\xu }
				\frac{
					\Omega
				}{
					1 + \Omega \frac{\xu-\xs}{\xt-\xu}
				}
				&
				\mbox{good for } \Omega\sim 0
				\\
				(\xt-\xs)
				\frac{
					1 
				}{
					1 + \frac{1}{\Omega} \frac{\xt-\xu}{\xu-\xs}
				}
				&
				\mbox{good for } \Omega\sim \infty
			\end{array}
			\right.
			.
		\end{align}
		Again the signs of the prefactors are the right ones for avoiding
		issues when exponentiating to a generic real $\rho$.

		We can now easily compute
		\begin{align}
			U_{[\SS]}(z_2)
			%% =&
			%%    \prod_{\RR=1}^{\SS-1}
			%%    (\xr - z_2)^{  \rho_{\RR \SS}}\,
			%%    (z-\xs)\,
			%%  \prod_{r=t+1}^{N}
			%%  (-\xr + z_2)^{ \rho_{\RR \SS} }
			%%  \nonumber\\
			&=
			\prod_{\RR=1,\, \RR\ne\SS}^N (x_{\RR; \SS})^{ \rho_{\RR \SS} }\,
			\frac{ (\xt-\xs) (\xu-\xs) }{ \xt-\xu }\,
			\prod_{\RR=1,\, \RR\ne\SS,\TT}^N\left( 1 - \frac{\Omega_2}{\omega_\RR} \right)^{\rho_{\RR \SS} }\,
			\Omega_2
			\nonumber\\
			&=
			\left. \left(
			\frac{d U_{[\SS]}}{d z_2} / \frac{d \Omega_{2}}{d z_2}
			\right)\right|_{z_2=\xs}
			\prod_{\RR=1,\, \RR\ne\SS,\TT}^N\left( 1 - \frac{\Omega_2}{\omega_\RR} \right)^{\rho_{\RR \SS} }\,
			\Omega_2
			,
		\end{align}
		and
		\begin{align}
			U_{[\TT]}(z_1)
			%% =&
			%%    \prod_{\RR=1}^{\TT-1}
			%%   (\xr - z_1)^{ -n_1 \rho_{\RR \TT} }
			%%   \prod_{\RR=\TT+1}^{N}
			%%   (-\xr + z_1)^{ -n_1 \rho_{\RR \TT} }
			%%   \nonumber\\
			=&
			-
			\prod_{\RR=1,\, \RR\ne\RR}^N (x_{\RR; \TT})^{ \rho_{\RR \TT} }\,
			\frac{ (\xs-\xt) (\xu-\xt) }{ \xu-\xs }\,
			\prod_{\RR=1,\, \RR\ne\SS,\TT}^N\left( 1 - \frac{\omega_\RR}{\Omega_1} \right)^{\rho_{\RR \TT} }\,
			\frac{1}{\Omega_1}
			\nonumber\\
			&=
			\left. \left(
			\frac{d U_{[\TT]}}{d z_1} / \frac{d (1/\Omega_{1}) }{d z_1}
			\right)\right|_{z_1=\xt}
			\prod_{\RR=1,\, \RR\ne\SS,\TT}^N\left( 1 - \frac{\omega_\RR}{\Omega_1} \right)^{\rho_{\RR \TT} }\,
			\frac{1}{\Omega_1}
			,
		\end{align}
		with
		\begin{align}
			\left. \frac{d \Omega_{2}}{d z_2} \right|_{z_2=\xs}
			=&
			\frac{ \xt-\xu }{ (\xt-\xs) (\xu-\xs) }
			,
			\nonumber\\
			\left.  \frac{d (1/\Omega_{1}) }{d z_1} \right|_{z_1=\xt}
			=&
			\frac{ \xu-\xs }{ (\xs-\xt) (\xu-\xt) }
			.
		\end{align}
		We have also used $\sum_\RR \rho_{\RR\TT}= \sum_\RR \rho_{\RR\SS}=0$.
		Notice that the rewriting of the constant factors can be obtained
		either by direct computation or by noticing that they are
		$\frac{d U_{[\SS]}}{d \Omega_2}\Bigg|_{\Omega_2=0}$
		and
		$\frac{d U_{[\TT]}}{d (1/\Omega_1)}\Bigg|_{\Omega_1=\infty}$ respectively
		
		We can then proceed to the evaluation of the integral
		\begin{align}
			\oldJ_{\TT n_1, \SS n_2}
			=&
			\left(
			-
			\prod_{\RR=1,\, \RR\ne\TT}^N (x_{\RR; \TT})^{ \rho_{\RR \TT} }\,
			\frac{ (\xs-\xt) (\xu-\xt) }{ \xu-\xs }
			\right)^{-n_1}\,
			\left(
			\prod_{\RR=1,\, \RR\ne\SS}^N (x_{\RR; \SS})^{ \rho_{\RR \SS} }\,
			\frac{ (\xt-\xs) (\xu-\xt) }{ \xt-\xu }
			\right)^{-n_2}
			\nonumber\\
			&
			\oint_{\Omega_1=\infty} \oint_{\Omega_2=0}
			\frac{ d \Omega_1\, d \Omega_2  }{
				(2\pi i)^2 (\Omega_1 -\Omega_2)^2 } 
			\Omega_1^{n_1}
			\frac{1}{\Omega_2^{n_2}}\,
			%%  \nonumber\\
			%%  %
			%%  &
			\prod_{\RR=1,\, \RR\ne\SS,\TT}^N
			\left( 1 - \frac{\omega_\RR}{\Omega_1} \right)^{-n_1 \rho_{\RR\TT}}\,
			\prod_{\RR=1,\, \RR\ne\SS,\TT}^N
			\left( 1 - \frac{\Omega_2}{\omega_\RR} \right)^{-n_2 \rho_{\RR\SS}}
			\nonumber\\
			=
			\left(\dots\right)^{-n_1}\,
			\left(\dots\right)^{-n_2}\,
			\nonumber\\
			\sum_{k=1}^{n_2}\, k\,\Bigg\{
			&
			\oint_{\Omega_1=\infty}
			\frac{ d \Omega_1  }{ 2\pi i} \frac{1}{\Omega_1^{k+1-n_1} } 
			\prod_{\RR=1,\, \RR\ne\SS,\TT}^N
			\left( 1 - \frac{\omega_\RR}{\Omega_1} \right)^{-n_1 \rho_{\RR\TT}}\,
			%  \nonumber\\
			%  &
			\oint_{\Omega_2=0}
			\frac{ d \Omega_2  }{ 2\pi i} \frac{1}{\Omega_2^{n_2+1-k} } 
			\prod_{\RR=1,\, \RR\ne\SS,\TT}^N
			\left( 1 - \frac{\Omega_2}{\omega_\RR} \right)^{-n_2 \rho_{\RR\SS}}  
			.
		\end{align}
		Then we can expand the binomials and perform the integrations with
		result
		\begin{align}
			\oldJ_{\TT n_1, \SS n_2}
			=&
			\left(
			-
			\prod_{\RR=1,\, \RR\ne\TT}^N (x_{\RR; \TT})^{ \rho_{\RR \TT} }\,
			\frac{ (\xs-\xt) (\xu-\xt) }{ \xu-\xs }
			\right)^{-n_1}\,
			\left(
			\prod_{\RR=1,\, \RR\ne\SS}^N (x_{\RR; \SS})^{ \rho_{\RR \SS} }\,
			\frac{ (\xt-\xs) (\xu-\xs) }{ \xt-\xu }
			\right)^{-n_2}
			\nonumber\\
			&
			\sum_{k=1}^{min(n_1,\,n_2)} k\,
			\Bigg\{
			%% \nonumber\\
			%% \times
			%% &
			(-1)
			\sum_{(l_{1 \RR})_{\RR\ne\SS, \TT, \UU}\in\N^{N-3}}
			\prod_{\RR=1, \RR\ne\SS, \TT}^N
			\left[ {-n_1 \rho_{\RR \TT} \choose l_{1 \RR} }\,
			(-\omega_\RR)^{l_{1 \RR}} \right]\,
			\delta_{\sum'_\RR l_{1 \RR}+ l_{1 \UU}, n_1-k}
			\nonumber\\
			\times
			&
			\sum_{(l_{2 \RR})_{\RR\ne\SS, \TT, \UU}\in\N^{N-3}}
			\prod_{\RR=1, \RR\ne\SS, \TT,}^N
			\left[ {-n_1 \rho_{\RR \SS} \choose l_{2 \RR} }\,
			(\frac{-1}{\omega_\RR})^{l_{2 \RR}} \right]\,
			\delta_{\sum'_\RR l_{2 \RR}+ l_{2 \UU}, n_2-k}
			\Bigg\}
			,
		\end{align}
		where the factor $(-1)$ is due to the integration around $\Omega_1=\infty$.
		In a more compact form
		\begin{align}
			\oldJ_{\TT n_1, \SS n_2}
			=&
			(-1)^{n_2+1}
			\left(
			\prod_{\RR=1,\, \RR\ne\TT}^N (x_{\RR; \TT})^{ \rho_{\RR \TT} }\,
			\frac{ (\xs-\xt) (\xu-\xt) }{ \xu-\xs }
			\right)^{-n_1}\,
			\left(
			\prod_{\RR=1,\, \RR\ne\SS}^N (x_{\RR; \SS})^{ \rho_{\RR \SS} }\,
			\frac{ (\xt-\xs) (\xu-\xs) }{ \xt-\xu }
			\right)^{-n_2}
			\nonumber\\
			\sum_{k=1}^{min(n_1,\,n_2)} k\,
			\Bigg\{
			&
			%% \nonumber\\
			%% \times
			%% &
			\sum_{(l_{1 \RR})_{\RR\ne\SS, \TT, \UU}\in\N^{N-3}}
			\prod_{\RR=1, \RR\ne\SS, \TT, \UU}^N
			\left[ {-n_1 \rho_{\RR \TT} \choose l_{1 \RR} }\,
			\omega_\RR^{l_{1 \RR}} \right]\,
			{-n_1 \rho_{\UU \TT} \choose n_1-k
				%                            -\sum_{\RR=1, \RR\ne\SS, \TT, \UU}^N l_{1\RR} }\,
			-\sum_{\RR}' l_{1\RR} }\,
		\delta_{\sum'_\RR l_{1 \RR}\le n_1-k}
		\nonumber\\
		\times
		&
		\sum_{(l_{2 \RR})_{\RR\ne\SS, \TT, \UU}\in\N^{N-3}}
		\prod_{\RR=1, \RR\ne\SS, \TT, \UU}^N
		\left[ {-n_1 \rho_{\RR \SS} \choose l_{2 \RR} }\,
		(\frac{1}{\omega_\RR})^{l_{2 \RR}} \right]\,
		{-n_1 \rho_{\UU \SS} \choose n_2-k
			-\sum_{\RR}' l_{2 \RR} }\,
		\delta_{\sum'_\RR l_{2 \RR}\le n_2-k}
		\Bigg\}
		,
	\end{align} 
where $\sum'_\RR= \sum_{\RR=1,\, \RR\ne \SS,\TT,\UU}^N$.

		%%%%%%%%%%%%%%%%%%%%%%%%%%%%%%%%%%%%%%%%%%%%%%%%%%%%%%%%%%%%%%%%%%%%%%
		%%%%%%%%%%%%%%%%%%%%%%%%%%%%%%%%%%%%%%%%%%%%%%%%%%%%%%%%%%%%%%%%%%%%%%
		%%%%%%%%%%%%%%%%%%%%%%%%%%%%%%%%%%%%%%%%%%%%%%%%%%%%%%%%%%%%%%%%%%%%%%

		\printbibliography
		
	\end{document}